\documentclass{article}
\usepackage[margin=1in]{geometry}
\usepackage{amsmath,amssymb,mathrsfs,bm} 
\usepackage{graphicx}
\usepackage{titling}

\usepackage[labelformat=simple]{subcaption}

\usepackage[title]{appendix}
\usepackage{enumitem}
\usepackage{float}

\usepackage{dcolumn}
\usepackage{siunitx}
\usepackage{pifont}

\usepackage{multirow}

\usepackage[]{natbib} 
\numberwithin{equation}{section}

\newcommand{\rd}{\mathrm{d}}
\newcommand{\ri}{\mathrm{i}}
\newcommand{\SM}{{\color{brown}the supplementary materials}}

\usepackage{hyperref}
\hypersetup{
    colorlinks=true,
    linkcolor=blue,
    filecolor=magenta,
    urlcolor=cyan,
    citecolor=blue,
    breaklinks=true
}

\usepackage{tikz,etoolbox}
\newrobustcmd*{\mycirc}{\tikz{\filldraw[draw=cyan,fill=cyan!100] (0,0) circle [radius=0.075cm];}}
\newrobustcmd*{\mytriang}{\tikz{\filldraw[draw=magenta,fill=magenta!100] (-0.1,0) --(0.1cm,0.0) -- (0,0.15cm) -- (-0.1,0);}}

\allowdisplaybreaks

\title{Reduced modelling and global instability of finite-Reynolds-number flow in compliant rectangular channels} 

\author{Xiaojia Wang, Ivan C. Christov\\[1mm]
{\normalsize\textit{School of Mechanical Engineering, Purdue University, West Lafayette, Indiana, 47907, USA}}}

\date{\today}

\begin{document}

\maketitle

\tableofcontents

\clearpage

\begin{abstract}
Experiments have shown that flow in compliant microchannels can become unstable at a much lower Reynolds number than the corresponding flow in a rigid conduit. Therefore, it has been suggested that the wall's elastic compliance can be exploited towards new modalities of microscale mixing. While previous studies mainly focused on the local instability induced by the fluid--structure interactions (FSIs) in the system, we derive a  one-dimensional (1D) model to study the FSI's effect on the global instability. The proposed 1D FSI model is tailored to long, shallow rectangular microchannels with a deformable top wall, similar to the experiments. Going beyond the usual lubrication flows analyzed in these geometries, we include finite fluid inertia and couple the reduced flow equations to a novel reduced 1D wall deformation equation. Although a quantitative comparison to previous experiments is difficult, the behaviors of the proposed model show qualitative agreement with the experimental observations, and capture several key effects. Specifically, we find the critical conditions under which the inflated base state of the 1D FSI model is linearly unstable to infinitesimal perturbations. The critical Reynolds numbers predicted are in agreement with experimental observations. The unstable modes are highly oscillatory, with frequencies close to the natural frequency of the wall, suggesting that the observed instabilities are resonance phenomena. Furthermore, during the start-up from an undeformed initial state, self-sustained oscillations can be triggered by FSI. Our modeling framework can be applied to other microfluidic systems with similar geometric scale separation under different operating conditions.
\end{abstract}

\section{Introduction}
\label{sec:intro}

Soft materials, such as elastomers, are used to fabricate microfluidic devices \citep{SFB14}. Consequently, fluid--structure interactions (FSIs) between the compliant walls and the fluids conveyed within such microconduits have emerged as a fundamental mechanics problem to understand \citep{C21}. Previous studies have focused on the steady, inertialess flow regime. In this regime, by leveraging the FSIs, a myriad of applications to microfluidics have been proposed, such as: pressure sensors \citep{HHM02, OYE13}, strain sensors \citep{Dhong18}, microrheometers with increased sensitivity \citep{SLVYW21}, and passive technique for profiling microchannels' shape \citep{KCCWC21}. More recently, microfluidic systems have also begun to access inertial flow regimes up to a Reynolds number $Re \simeq 10^2$ \citep{DITT07}. Although this range of $Re$ is low compared to the well-documented flow-instability $Re$ for flows in rigid conduits, flows in compliant microconduits, surprisingly, have been observed to go unstable. Dye stream experiments in a rectangular microchannel with a soft bottom wall by \citet{VK13} showed that the stream begins to {oscillate} at $Re\approx 178$ and can break up at $Re\approx 200$. Other experimental studies confirmed the existence of this phenomenon, in both channels and tubes \citep{KS79, VK12, NS15, KB16}. \citet{VK13} suggested that the instabilities observed are induced by FSIs. Importantly, the resulting unstable flows increased the mixing efficiency by several orders of magnitude, compared to a stable steady flow. This observation has important implications for new strategies of harnessing FSI-induced instabilities to enhance mixing at the microscale, which is notoriously challenging \citep{OW04,K13}. \citet{VK13} thus introduced the terms ``ultrafast mixing'' and ``soft-wall turbulence'' to refer these novel phenomena. However, it should be noted that FSI-induced unstable flows are fundamentally different from the usual wall-bounded turbulent flows at high $Re$ \citep{SK15,SK17}.

Importantly, here, the low-$Re$ flows of interest are \emph{not} such that $Re\to 0$. The flows of interest can be up to $Re\simeq 10^2$. {The flow conduits in microfluidics are often manufactured to be long and slender, with a small aspect ratio $\epsilon\ll1$ (here, defined as the ratio of radius to length for a tube, or height to length for a channel). In this context, the ``reduced'' Reynolds number  $\hat{Re}=\epsilon Re$ is the relevant quantity to assess inertial effects in the flow because $\hat{Re}$ is the coefficient of the inertial terms in the suitably scaled Navier--Stokes equations (as we will show in \S~\ref{sec:fluid}). So,} the observed instabilities in soft microconduits typically occur at $\hat{Re}$ up to $\mathcal{O}(1)$. However, for $\hat{Re}=\mathcal{O}(1)$, the flow can neither be considered as inertialess nor as inviscid, and we will demonstrate that there exists a balance between the fluid inertia, the dominant pressure gradient and viscous forces in the flow.

However, even at low $Re$, analyzing the instability of pressure-driven flows in compliant conduits is far more challenging than that in rigid conduits. One key challenge is that, due to FSI, the base state is not the classical unidirectional flow solution for a rigid conduit (\textit{e.g.}, Poiseuille or Hagen--Poiseuille flow). At steady state, a compliant channel will deform due to the hydrodynamic pressure within, and this deformation will, in turn, influence the velocity and pressure fields in the flow \citep{GEGJ06,C21}. Since the pressure decreases along the flow-wise direction in a pressure-driven flow, the deformation is not uniform, with larger deformation near the inlet and smaller deformation near the outlet. This non-flat shape of the deformed channel was indeed observed in the experiments by \citet{VK13}, however its two-way coupled nature to the flow was not captured in previous stability models. Importantly, the coupling between the flow and the solid deformation gives rise to a \emph{non-constant} pressure gradient in the streamwise direction, leading to a \emph{nonlinear} relationship between the flow rate and the pressure drop \citep{GEGJ06,SLHLUB09,CCSS17}. Only a global stability analysis can take this spatially varying non-flat base state into account. However, global analyses are difficult to perform for three-dimensional (3D) FSI problems. To this end, in the present work, we undertake reduced-order modeling.

\begin{table}[t]
    \centering
    \begin{tabular}{l@{\hskip 2em}l@{\hskip 2em}l@{\hskip 2em}l@{\hskip 2em}l}
    \hline\hline
       &  Flat base? & $\hat{Re}=\epsilon Re$  & Instability type &  Method  \\
    \hline
  \textbf{Kumaran family} & \\[1mm]
  \citet{K95} & Yes & $\mathcal{O}(1)$ & Local & OS \\
  \citet{GK05} & No & $\mathcal{O}(1)$ & Local & OS \\
  \citet{GS09} & Yes & $\mathcal{O}(1)$ & Local & OS \\
  \citet{VK13} & No & $\mathcal{O}(1)$ & Local & OS \\
  \citet{VK15} & No & $\mathcal{O}(1)$ & Local & OS \\[1mm]
    \hline
   \textbf{Collapsible tubes} & \\[1mm]
   \citet{J90, J92} & No & $\gg 1$ & Global & RM \\
   \citet{LP96, LP98} & No & $\gg 1$ & Global & Num. \\
   \citet{JH03} & No & $\gg 1$ & Global & Asym. \& Num.\\
   \citet{LCLP08} & No & $\gg 1$ & Global & Num. \\
   \citet{SWJ09} & Yes & $\gg 1$ & Global \& Local & RM \\
   \citet{SHWJ10} & Yes & $\gg 1$ & Global \& Local & OS \& MLEE\\ 
   \citet{HB10}  & Yes & $\gg 1$ & Global & Num. \\
   \citet{LLC12} & No & $\gg 1$ & Global & Num. \\
   \citet{XBJ13, XBJ14} & Yes & $\gg 1$ & Global & RM \\
   \citet{PPP14} & No & $\gg 1$ & Global & RM \\
   \citet{WLS21} & No & $\gg 1$ & Global & Num.\\[1mm]
   \hline
   \textbf{Present work} & No & $\mathcal{O}(1)$ & Global & RM \\
    \hline\hline
    \end{tabular}
    \caption{Comparison of selected previous studies on instability of pressure-driven flows in complaint conduits. In the last column, `OS' stands for Orr--Sommerfeld-type stability analysis; `RM' stands for reduced modeling; `Num.' specifically stands for 2D two-way coupled FSI simulations; `Asym.' stands for asymptotic analysis; and `MLEE' stands for matched local eigenfunction expansion method.}
    \label{tab:prev_refs}
\end{table}

Previous studies on instabilities due to microscale FSIs analyzed the problem from the local perspective. {For convenience, we term this line of research as the ``Kumaran family,''} and a list of representative studies is compared/contrasted in table~\ref{tab:prev_refs}. The Kumaran family studies typically derive a modified Orr--Sommerfeld equation by perturbing the fluid--solid interface with infinitesimal disturbances. Early studies neglected the effect of FSIs on the base state by taking the base flow to be the unidirectional one in a rigid conduit \citep{K95, GS09}. Recent work by \citet{VK13,VK15} sought to improve the previous linear stability analyses by incorporating the effect of nonuniform deformation of the conduit wall. However, instead of being derived from the governing equations of two-way coupled FSI problem, the deformed shape of the channel was imaged experimentally and then reconstructed for use in computational fluid dynamics (CFD) simulations. By assuming steady flow, the simulated velocity profile and the pressure distribution were taken as the base state and ``imported'' into the linear stability analysis. Nevertheless, to arrive at an Orr--Sommerfeld equation, \citet{VK13,VK15} had to assume that the variation of the channel deformation along the streamwise direction is so slow that the flow is nearly parallel. Thus, long-wave perturbations cannot be considered, and the analysis is strictly local. Notably, the local unstable modes were predicted to arise at $Re\lesssim 100$, both in compliant channels \citep{GK05,VK13} and in compliant tubes \citep{VK15}. However, it is difficult to reach a unified understanding from the current state of this literature because the explanation regarding the onset of instability is different for each {situation}. For instance, considering a neo-Hookean material as the compliant wall (instead of a linearly elastic one) modifies the linear stability of the flow in a compliant tube \citep{GS09}. Different formulations of the linear stability analysis can also lead to completely different conclusions \citep{PS19}. The most recent advances and perspectives following this line of research are thoroughly reviewed by \citet{K21}.

To fill these knowledge gaps, in this work, we analyze the global stability of microscale flows undergoing FSI. Specifically, we address the effect of the non-flat deformed base state, and we construct a reduced (global) model to make the analysis possible. The idea of reduced modeling is inspired by the research program on collapsible tubes; representative prior work is compared/contrasted in table~\ref{tab:prev_refs}. Although collapsible tubes research focuses on inertial flows with $\hat{Re}\gg 1$, and is not concerned with flows at the microscale, this research program has demonstrated the power of reduced modeling. Compared to complex two-way-coupled unsteady FSI simulations, reduced mathematical models are better suited for exploring the (potentially large) parameter space of such FSI problems. Reduced models can also aid the mathematical analysis and thus promote the understanding of the instability mechanisms. Although early one-dimensional (1D) reduced models \citep{S77} incorporated \textit{ad hoc} assumptions, such as an empirical tube law for deformation and an energy loss term for flow separation, these models surprisingly provided good qualitative agreement with experimental observations \citep{JP89} and predicted the expected complex oscillations \citep{J90, J92}. More recently, \citet{PPP14} constructed a 1D model based on the so-called interactive boundary layer theory and predicted oscillations induced by wall inertia. Meanwhile, \citet{SWJ09} invoked the long-wave approximation and built a 1D model to study the global and local instabilities in collapsible tubes. This model was then used extensively to investigate the effect of the pre-tension of the soft wall \citep{SHWJ10}, the effect of the length of a downstream rigid segment \citep{XBJ13, XBJ14}, and the model was also applied to understand retinal venous pulsation \citep{SF19}.

Along these lines, in this work, we derive a 1D FSI model inspired, in several ways, by the 1D model of \citet{SWJ09}. However, instead of considering $\hat{Re}\gg 1$, we focus on $\hat{Re}$ up to $\mathcal{O}(1)$, consistent with the microchannel experiments of \citet{VK13}. The new model admits a non-flat fluid--solid interface at steady state, resulting from the nonlinear pressure distribution within the channel. We conduct a global stability analysis to properly take this spatially-varying base state into account, complementary to the local stability analyses in the Kumaran family of studies. With the finite fluid inertia and non-flat base state accounted for, our 1D FSI model is the first reduced model that addresses the global stability of pressure-driven flow in a compliant microchannel. 

\begin{figure}[t]
    \centering
    \begin{subfigure}[]{0.7\textwidth}
    \centering
    \includegraphics[width=\textwidth]{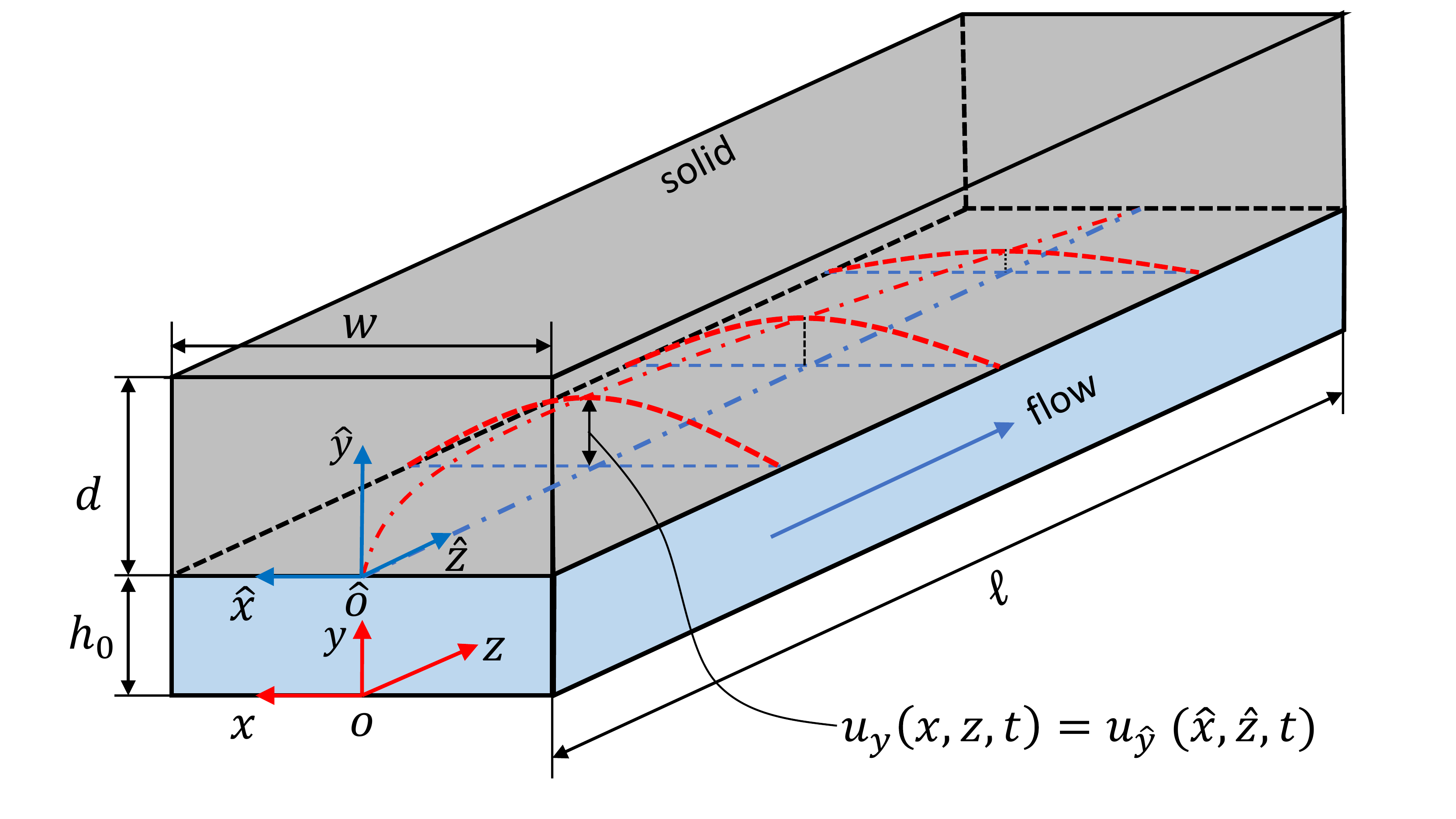}
    \subcaption{}
    \label{subfig:gen1d-geo-3d}
    \end{subfigure}
    \begin{subfigure}[]{0.55\textwidth}
    \centering
    \includegraphics[width=\textwidth]{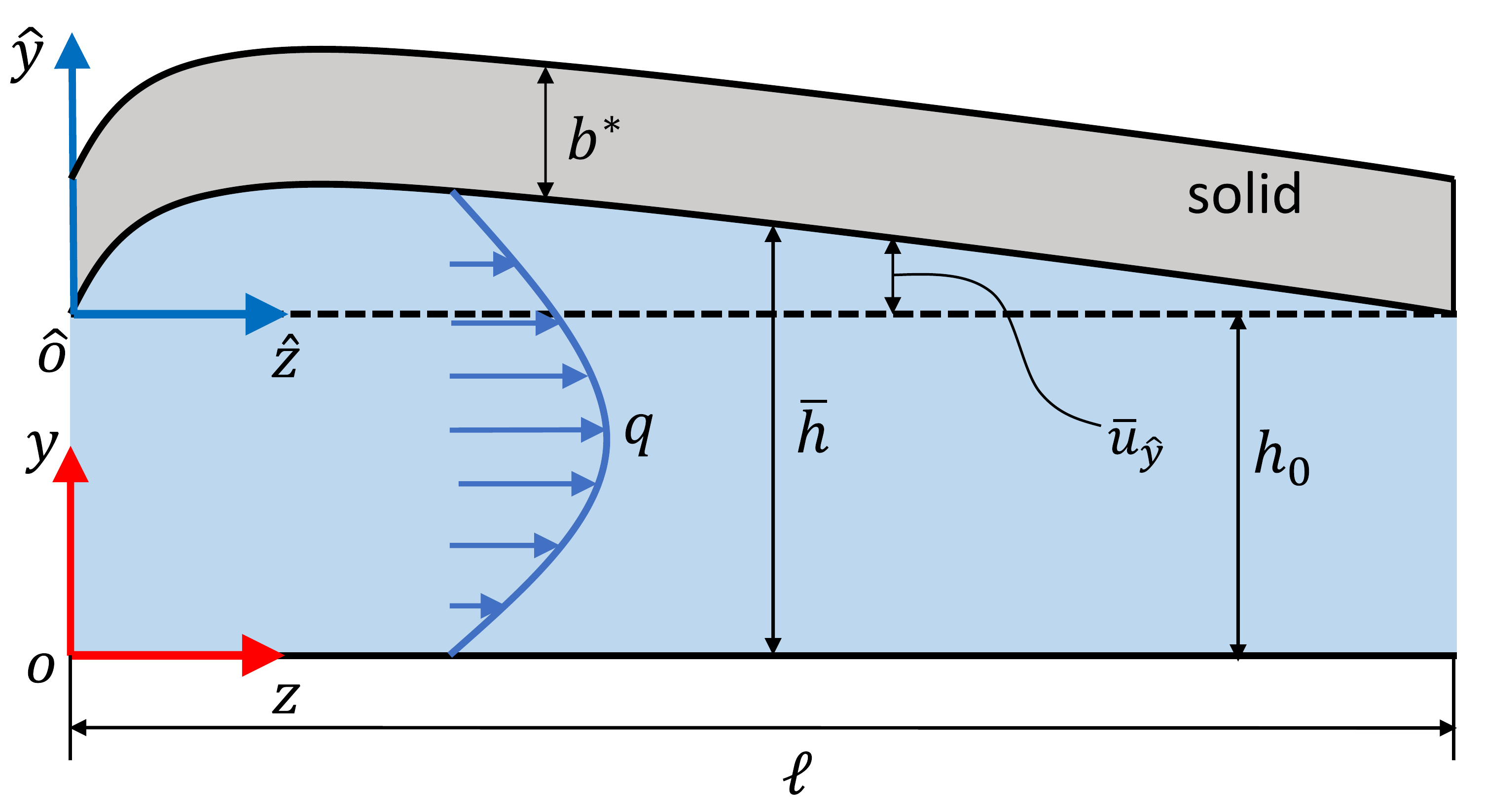}
    \subcaption{}
    \label{subfig:gen1d-geo-2d}
    \end{subfigure}
    \caption{(\textit{a}) Schematic of the 3D geometry of the compliant microchannel with a deformable top wall \citep{WC21}, with key dimensional variables labeled. The red dash-dotted curve and the red dashed curves sketch the deformed fluid--solid interface at the mid-plane ($x=0$), and the typical cross-sectional deformation profiles at the interface at different flow-wise locations, respectively. (\textit{b}) Schematic of the configuration of the reduced 2D problem obtained by width-averaging the interface displacement.}
    \label{gen1d-geo}
\end{figure}

To this end, this paper is organized as follows. We introduce the configuration of the microchannel in \S~\ref{sec:prob-statement}. In \S~\ref{sec:fluid}, we invoke the lubrication approximation to simplify the governing equations of the internal flow. Assuming linear elasticity, we extract the dominant mechanism in the wall deformation through a scaling argument, leveraging the slenderness of the wall (\S~\ref{subsec:solid-O1}). Then in \S~\ref{subsec:eff_height}, the obtained displacement field is averaged over the spanwise direction ($\hat{x}$), reducing the three-dimensional (3D) system (figure~\ref{subfig:gen1d-geo-3d}) to a two-dimensional (2D) one (figure~\ref{subfig:gen1d-geo-2d}). The ultimate solid model is 1D, obtained by introducing weak inertia and {modeling} the weak deformation-induced tension (\S~\ref{subsec:weak_effects}). In \S~\ref{sec:coupling},  we couple the 1D solid model with the depth-averaged Navier--Stokes equations to achieve a reduced 1D FSI model relating the wall deformation to the flow rate and the pressure in the flow. We analyze the base state of the 1D FSI model in \S~\ref{sec:steady}. Then in \S~\ref{sec:linear-stability}, we conduct a global linear stability analysis based on the nonuniform inflated base state, the results of which are validated against unsteady numerical simulations of the 1D FSI model (\S~\ref{subsec:dynamic1}). The dynamics of the 1D FSI model is also simulated by taking the undeformed state as the initial condition (\S~\ref{subsec:dynamic2}). Finally, in \S~\ref{sec:discussion}, we conclude with a summary of the key results of our study, and their broader context.

\section{Problem statement}
\label{sec:prob-statement}

Consider a microchannel, as shown in figure~\ref{subfig:gen1d-geo-3d}, with undeformed height of $h_0$, width of $w$, and length of $\ell$. The microchannel is assumed to be long and shallow so that $h_0\ll w\ll \ell$. Introducing the dimensionless aspect ratios $\epsilon=h_0/\ell$ and $\delta=h_0/w$, then we have $\epsilon\ll\delta\ll 1$. In reality, the three walls of the channel can be made rigid with a soft wall bonded on top, as the geometry considered in \citep{CCSS17, SC18}. Alternatively, the top and side walls are soft and bonded to a rigid bottom wall, as was the case in \citep{WC19}. In either case, the deformation of the top wall is dominant. Therefore, in our modeling, the deformation of the side walls is neglected. Further, we denote the thickness of the top wall by $d$. To make the model general, at this stage, we do not specify the magnitude $d$ compared to the other dimensions, but we do require that $d\ll \ell$. As the fluid is pushed through the microchannel, from the inlet to the outlet, the hydrodynamic pressure will deform the fluid--solid interface at the top wall. The displacement of the interface is denoted by $u_y(x,z,t)$. Finally, since the microchannel is often restricted from moving at the inlet ($z=0$) and the outlet ($z=\ell$) planes by external connections (or the outlet is open to ambient gauge pressure, thus has negligible deformation), we assume zero displacement of the fluid--solid interface at both ends ($z=0,\ell$).

For convenience, we introduce two coordinate systems. As shown in figure~\ref{subfig:gen1d-geo-3d}, the $o_{xyz}$ coordinate system is located at the bottom wall of the microchannel, with its origin set at the center of the inlet. The $\hat{o}_{\hat{x} \hat{y} \hat{z}}$ coordinate system is the $o_{xyz}$ system translated along $y$ by $h_0$, thus its origin is located at the undeformed fluid--solid interface. Specifically, we have $x = \hat{x}$, $y = \hat{y}+h_0$ and $z = \hat{z}$.

\section{Fluid mechanics problem formulation}
\label{sec:fluid}

\subsection{Scaling and identification of the dominant effects}

Assume the working fluid is incompressible and Newtonian, with a density of $\rho_f$ and dynamic viscosity of $\mu$. With the displacement of the fluid--solid interface denoted as $u_y(x,z,t)$, the deformed channel height can be written as $h(x,z,t) = h_0 +u_y(x,z,t)$. Then, the deformed configuration of the fluid domain is $\{ (x,y,z) \;|\; -w/2\leq x \leq +w/2,\, 0\leq y\leq h(x,z,t),\, 0\leq z\leq \ell\}$. Further, we assume that $h(x,z,t)\ll w\ll \ell$, \textit{i.e.}, the slenderness and shallowness assumptions on the conduit hold true even after its deformation. The former assumption is important because it allows us to use $h_0$ as the scale for $y$. 

Under these assumptions, the governing equations are the unsteady incompressible Navier--Stokes equations, which take the form:
\begin{subequations}\label{iNS}
\begin{align}
  \underbrace{\frac{\partial v_x}{\partial x}}_{\mathcal{O}(1)} + \underbrace{\frac{\partial v_y}{\partial y}}_{\mathcal{O}(1)} + \underbrace{\frac{\partial v_z}{\partial z}}_{\mathcal{O}(1)} &= 0, \label{COM}\displaybreak[3]\\
  \underbrace{\frac{\partial v_x}{\partial t}}_{\mathcal{O}\left(\epsilon^2\delta^{-2}\hat{Re}\right)} + \underbrace{v_x\frac{\partial v_x}{\partial x}}_{\mathcal{O}\left(\epsilon^2\delta^{-2}\hat{Re}\right)} + \underbrace{v_y\frac{\partial v_x}{\partial y}}_{\mathcal{O}\left(\epsilon^2\delta^{-2}\hat{Re}\right)}  + \underbrace{v_z\frac{\partial v_x}{\partial z}}_{\mathcal{O}\left(\epsilon^2\delta^{-2}\hat{Re}\right)} &= -\underbrace{\frac{1}{\rho_f}\frac{\partial p}{\partial x}}_{\mathcal{O}(1)} + \underbrace{\frac{\mu}{\rho_f}\frac{\partial^2 v_x}{\partial x^2}}_{\mathcal{O}(\epsilon^2)} + \underbrace{\frac{\mu}{\rho_f}\frac{\partial^2 v_x}{\partial y^2}}_{\mathcal{O}\left(\epsilon^2\delta^{-2}\right)} + \underbrace{\frac{\mu}{\rho_f}\frac{\partial^2 v_x}{\partial z^2}}_{\mathcal{O}\left(\epsilon^4\delta^{-2}\right)}, \label{COLM-x}\\
  \underbrace{\frac{\partial v_y}{\partial t}}_{\mathcal{O}(\epsilon^2\hat{Re})} + \underbrace{v_x\frac{\partial v_y}{\partial x}}_{\mathcal{O}(\epsilon^2\hat{Re})} + \underbrace{v_y\frac{\partial v_y}{\partial y}}_{\mathcal{O}(\epsilon^2\hat{Re})}  + \underbrace{v_z\frac{\partial v_y}{\partial z}}_{\mathcal{O}(\epsilon^2\hat{Re})}  &= -\underbrace{\frac{1}{\rho_f}\frac{\partial p}{\partial y}}_{\mathcal{O}(1)} + \underbrace{\frac{\mu}{\rho_f}\frac{\partial^2 v_y}{\partial x^2}}_{\mathcal{O}(\epsilon^2\delta^2)} + \underbrace{\frac{\mu}{\rho_f}\frac{\partial^2 v_y}{\partial y^2}}_{\mathcal{O}\left(\epsilon^2\right)} + \underbrace{\frac{\mu}{\rho_f}\frac{\partial^2 v_y}{\partial z^2}}_{\mathcal{O}\left(\epsilon^4\right)}, \label{COLM-y}\\
  \underbrace{\frac{\partial v_z}{\partial t}}_{\mathcal{O}(\hat{Re})} + \underbrace{v_x\frac{\partial v_z}{\partial x}}_{\mathcal{O}(\hat{Re})} + \underbrace{v_y\frac{\partial v_z}{\partial y}}_{\mathcal{O}(\hat{Re})}  + \underbrace{v_z\frac{\partial v_z}{\partial z}}_{\mathcal{O}(\hat{Re})}  &=  -\underbrace{\frac{1}{\rho_f}\frac{\partial p}{\partial z}}_{\mathcal{O}(1)} + \underbrace{\frac{\mu}{\rho_f}\frac{\partial^2 v_z}{\partial x^2}}_{\mathcal{O}(\delta^2)} + \underbrace{\frac{\mu}{\rho_f}\frac{\partial^2 v_z}{\partial y^2}}_{\mathcal{O}\left(1\right)} + \underbrace{\frac{\mu}{\rho_f}\frac{\partial^2 v_z}{\partial z^2}}_{\mathcal{O}\left(\epsilon^2\right)}, \label{COLM-z}
\end{align}
\end{subequations}
with the order of magnitude of each term listed underneath, based on the scales from table~\ref{tab:fluid-scales}.

\begin{table}
    \centering
    \begin{tabular}{l l l l l l l l l l} 
       \hline\hline
      Var. & $t$ &  $x$ or $\hat{x}$ & $y$ & $\hat{y}$ & $z$ or $\hat{z}$ & $v_x$  & $v_y$ & $v_z$ & $p$\\
      \hline
      Scale  & $\mathcal{T}_f$  & $w$ & $h_0$ & $d$ & $\ell$ & $\epsilon \mathcal{V}_c/\delta$ & $\epsilon\mathcal{V}_c$ & $\mathcal{V}_c$ & $\mathcal{P}_c$\\
      \hline\hline
    \end{tabular}
    \caption{The scales for the variables in the incompressible Navier--Stokes equations \eqref{iNS}.}
    \label{tab:fluid-scales}
\end{table}

In table~\ref{tab:fluid-scales}, $\mathcal{V}_c$ is the characteristic velocity scale. Specifically, to ensure the conservation of mass of equation \eqref{COM}, $\epsilon\mathcal{V}_c/\delta$, $\epsilon\mathcal{V}_c$ and $\mathcal{V}_c$ are chosen to be the characteristic scales for the velocity components $v_x$, $v_y$ and $v_z$, respectively. Also, as is standard for low-Reynolds-number flow, to achieve a balance between the pressure and the viscous stresses in equation \eqref{COLM-z}, the characteristic pressure scale, $\mathcal{P}_c$, and $\mathcal{V}_c$ are related by $\mathcal{P}_c=\mu \mathcal{V}_c \ell/h_0^2$. If the volumetric flow rate, $q$, at the inlet is fixed, we can choose $\mathcal{V}_c = q/(wh_0)$, then $\mathcal{P}_c = \mu q \ell /(wh_0^3)$. However, if the pressure drop, $\Delta p = p|_{z=0}- p|_{z=\ell}$, is prescribed, $\mathcal{P}_c = \Delta p$ and, accordingly, $\mathcal{V}_c = \Delta p h_0^2/(\mu\ell)$. The Reynolds number is defined as $Re=\rho_f\mathcal{V}_c h_0/{\mu}$. However, {owing to the shallowness and slenderness of the fluid domain ($\epsilon\ll \delta \ll 1$), the inertial effects in the $z$-momentum equation~\eqref{COLM-z} are more dominant than in the other two momentum equations. In this scaling, $\hat{Re}=\epsilon Re$ emerges as the sole prefactor of the inertial terms in equation~\eqref{COLM-z} (rather than $Re$), thus we say that the reduced Reynolds number $\hat{Re}$ is more suitable for quantifying the inertia of this flow.} Finally, $\mathcal{T}_f$ is taken to be the characteristic time scale for axial advection (the dominant flow direction): $\mathcal{T}_f=\ell /\mathcal{V}_c$.

\subsection{Reduction: lubrication approximation}
\label{subsec:fluid-reduction}

Recall that we are interested {in flow in a shallow and slender microchannel ($h_0\ll w\ll \ell$) such that $\epsilon=h_0/\ell\ll \delta=h_0/w \ll 1$}. Based on the discussion above, it is clear that the dominant balance of terms occurs in the $z$-momentum equation \eqref{COLM-z}. Only the pressure terms are left in equations \eqref{COLM-x} and \eqref{COLM-y}, indicating that, at the leading order in $\epsilon$ and $\delta$, the hydrodynamic pressure $p$ is \emph{only} a function of the streamwise location $z$, as in the classic lubrication approximation \citep{panton,W06_book}. More importantly, this argument is true even at finite Reynolds number, \textit{i.e.}, $\hat{Re}= \mathcal{O}(1)$, which is typical of the microfluidic experiments that we compare to \citep{VK13}. Specifically, with $\hat{Re}= \mathcal{O}(1)$, the dominant balance in the flow-wise  momentum equation \eqref{COLM-z} occurs between fluid inertia, the pressure gradient, and viscous forces. The same balance was employed by \citet{IWC20} to derive a 1D FSI model from the 2D Navies--Stokes equations (but under different assumptions on the solid mechanics problem).

Examining further the right-hand side of equation \eqref{COLM-z}, the balance of forces at the leading order indicates that $\partial p/\partial z \sim \partial \tau_{yz}/\partial y$ because the shear stress is $\tau_{yz} \sim \mu \partial v_z/\partial y$. Introducing $\mathcal{S}_c$ as the characteristic scale for $\tau_{yz}$ and substituting the other scales from table~\ref{tab:fluid-scales}, this balance suggests that $\mathcal{P}_c/\ell = \mathcal{S}_c/h_0$, leading to $\mathcal{S}_c=(h_0/\ell) \mathcal{P}_c=\epsilon \mathcal{P}_c$. For $\epsilon\ll 1$, we conclude that $\tau_{yz}\ll p$. Hence, at the leading order in $\epsilon$ and $\delta$, $p(z)$ is the only flow-induced load exerted on the fluid--solid interface.

\section{Solid mechanics problem formulation}
\label{sec:solid}

\subsection{Scaling and identification of the dominant effects}
\label{subsec:solid-O1}

\begin{table}
    \centering
    \begin{tabular}{l l l l l l l l l l l l l l}
       \hline\hline
      Var. & $t$ & $\hat{x}$ & $\hat{y}$ & $\hat{z}$ & $u_{\hat{x}}$  & $u_{\hat{y}}$ or $u_y$ & $u_{\hat{z}}$ & $\sigma_{\hat{x}\hat{x}}$ & $\sigma_{\hat{x}\hat{y}}$ & $\sigma_{\hat{x}\hat{z}}$ & $\sigma_{\hat{y}\hat{y}}$ & $\sigma_{\hat{y}\hat{z}}$ & $\sigma_{\hat{z}\hat{z}}$\\
      \hline
      Scale & $\mathcal{T}_f$ & $w$ & $d$ & $\ell$ & $\mathcal{U}_{c,x}$ & $\mathcal{U}_c$ & $\mathcal{U}_{c,z}$ & $\mathcal{D}_{\hat{x}\hat{x}}$ & $\mathcal{D}_{\hat{x}\hat{y}}$ & $\mathcal{D}_{\hat{x}\hat{z}}$ & $\mathcal{P}_c$ & $\epsilon\mathcal{P}_c$ & $\mathcal{D}_{\hat{z}\hat{z}}$\\
      \hline\hline
    \end{tabular}
    \caption{The scales for the variables in the linear elastodynamics equations \eqref{cauchy-eq}.}
    \label{tab:solid-scales}
\end{table}

For the solid mechanics problem, it is more convenient to use the $\hat{o}_{\hat{x} \hat{y} \hat{z}}$ coordinate system, where we denote the displacement of the fluid--solid interface by $u_{\hat{y}}$, as shown in figure~\ref{subfig:gen1d-geo-3d}. We consider the case in which the maximum of $\hat{u}_y$ is small compared with the smallest dimension of the solid, so that the small-strain theory of linear elasticity is applicable. Specifically, if the wall is ``thick,'' meaning $w \lesssim d \ll \ell$, we require that $\hat{u}_y \ll w$. However, if the wall is ``thin,'' meaning $d\lesssim w \ll \ell$, we require that $\hat{u}_y \ll d$ \citep{WC21}. 

The following discussion proceeds along the lines of \citet{WC19}. However, here, we provide a more general derivation for the reader's convenience. First, using the scales from table~\ref{tab:solid-scales}, the balance between the Cauchy stresses and the solid inertia within the wall, neglecting any body forces, is:
\begin{subequations}\label{cauchy-eq}
\begin{align}
     \underbrace{\rho_s\frac{\partial^2 u_{\hat{x}}}{\partial t^2}}_{\mathcal{O}(\rho_s \mathcal{U}_{c,x}/\mathcal{T}_f^2) } + \underbrace{\frac{\partial \sigma_{\hat{x} \hat{x}}}{\partial \hat{x}}}_{\mathcal{O}({\mathcal{D}_{\hat{x} \hat{x}}}/{w})} + \underbrace{\frac{\partial \sigma_{\hat{x} \hat{y}}}{\partial \hat{y}}}_{\mathcal{O}({\mathcal{D}_{\hat{x} \hat{y}}}/{d})} + \underbrace{\frac{\partial \sigma_{\hat{x} \hat{z}}}{\partial \hat{z}}}_{\mathcal{O}({\mathcal{D}_{\hat{x} \hat{z}}}/{\ell})} &= 0,\label{cauchy-x} \\
     \underbrace{\rho_s\frac{\partial^2 u_{\hat{y}}}{\partial t^2}}_{\mathcal{O}(\rho_s\mathcal{U}_c/\mathcal{T}_f^2)} + \underbrace{\frac{\partial \sigma_{\hat{x} \hat{y}}}{\partial \hat{x}}}_{\mathcal{O}({\mathcal{D}_{\hat{x} \hat{y}}}/{w})} + \underbrace{\frac{\partial \sigma_{\hat{y} \hat{y}}}{\partial \hat{y}}}_{\mathcal{O}(\mathcal{P}_c/{d})} + \underbrace{\frac{\partial \sigma_{\hat{y} \hat{z}}}{\partial \hat{z}}}_{\mathcal{O}(\epsilon\mathcal{P}_c/\ell)} &= 0,\label{cauchy-y} \\
     \underbrace{\rho_s\frac{\partial^2 u_{\hat{z}}}{\partial t^2}}_{\mathcal{O}(\rho_s \mathcal{U}_{c,z}/\mathcal{T}_f^2)} + \underbrace{\frac{\partial \sigma_{\hat{x} \hat{z}}}{\partial \hat{x}}}_{\mathcal{O}({\mathcal{D}_{\hat{x} \hat{z}}}/{w})} + \underbrace{\frac{\partial \sigma_{\hat{y} \hat{z}}}{\partial \hat{y}}}_{\mathcal{O}(\epsilon\mathcal{P}_c/{d})} + \underbrace{\frac{\partial \sigma_{\hat{z} \hat{z}}}{\partial \hat{z}}}_{\mathcal{O}({\mathcal{D}_{\hat{z} \hat{z}}}/{\ell})} &= 0.\label{cauchy-z} 
\end{align}
\end{subequations}
Here, $\sigma_{\hat{x} \hat{x}}$, $\sigma_{\hat{x} \hat{y}}$, $\sigma_{\hat{x} \hat{z}}$, $\sigma_{\hat{y} \hat{y}}$, $\sigma_{\hat{y} \hat{z}}$ and $\sigma_{\hat{z} \hat{z}}$ are the six independent components of the Cauchy stress in the solid. The order of magnitude of each term is listed underneath, based on the scales from table~\ref{tab:solid-scales}.

In table~\ref{tab:solid-scales}, $\mathcal{U}_{c,x}$, $\mathcal{U}_c$ and $\mathcal{U}_{c,z}$ are the characteristic scales for $u_{\hat{x}}$, $u_{\hat{y}}$ and $u_{\hat{z}}$, respectively. We immediately assume that  $\mathcal{U}_{c,x}\ll \mathcal{U}_c$ and $\mathcal{U}_{c,z}\ll \mathcal{U}_c$, meaning that the wall is primarily bulging upwards, as in experiments. This assumption has previously been quantitatively validated against experiments \citep{CCSS17,SC18,WC19}. Then the most prominent inertial term in the solid is in equation \eqref{cauchy-y}. Note that the time scale for equation \eqref{cauchy-eq} is still the fluid's axial advection time scale, $\mathcal{T}_f$, in order to ensure the coupling between the solid and the fluid mechanics problems. Note that this choice of time scale is different from the so-called ``viscous--elastic'' one used in related works \citep{EG14,MCSPS19}. In the latter papers, the characteristic (common) time scale $\mathcal{T}_c$ was chosen based on the kinematic boundary condition at the fluid--solid interface, \textit{i.e.,} $\partial\bar{u}_y/\partial t = v_y$, leading to a fluid time scale of  $\mathcal{T}_c=\bar{\mathcal{U}}_c/(\epsilon \mathcal{V}_c) = (\bar{\mathcal{U}}_c/h_0)(\ell/\mathcal{V}_c)= \beta \mathcal{T}_f$. However, since $\beta=\bar{\mathcal{U}}_c/h_0$ is typically at $\mathcal{O}(1)$ in our work (discussed in \S~\ref{subsec: nodim_fsi_param}), these two different choices of the fluid time scale, $\mathcal{T}_c$ and $\mathcal{T}_f$, do not differ significantly.) To further elucidate the time scales involved, the magnitude of the inertial term in equation \eqref{cauchy-y} can be written as $\rho_s\mathcal{U}_c/\mathcal{T}_f^2=\rho_s\mathcal{U}_c/\mathcal{T}_s^2 \times (\mathcal{T}_s/\mathcal{T}_f)^2$, where we have explicitly introduced the solid time scale, $\mathcal{T}_s$. {Note that when the soft solid's density is similar to the fluid's density (see example values in table~\ref{table:param} below), the elastic wave speed in the solid is  $\sqrt{E/\rho_s}\gg \mathcal{V}_c$. Therefore, the development of the solid deformation is expected to be much faster than the flow, \textit{i.e.}, $\mathcal{T}_s \ll \mathcal{T}_f$, which makes the solid's inertia a weak effect. We will quantify the solid's weak inertia in \S~\ref{subsec: nodim_fsi_param}, but for now it can be neglected to find the deformation profile (bulging of the fluid--solid interface due to the hydrodynamic pressure).}

To this end, let us consider the balance of the Cauchy stresses in equations~\eqref{cauchy-eq}. Due to the traction balance at the fluid--solid interface, it can be inferred that $\mathcal{D}_{\hat{y}\hat{y}} = \mathcal{P}_c$ and $\mathcal{D}_{\hat{y} \hat{z}} = \epsilon \mathcal{P}_c$, as tabulated in table~\ref{tab:solid-scales}. For convenience, we introduce $\gamma = d/w$ as the spanwise aspect ratio of the solid wall. Then, a balance in equation \eqref{cauchy-y} can only occur between the second and the third terms, yielding $D_{\hat{x} \hat{y}} = \mathcal{P}_c/\gamma$. At the same time, the balance of the three terms in equation \eqref{cauchy-z} gives $\mathcal{D}_{\hat{x} \hat{z}} = \epsilon \mathcal{P}_c w/d = \epsilon \mathcal{P}_c/\gamma $ and $\mathcal{D}_{\hat{z} \hat{z}} = \epsilon \mathcal{P}_c\ell/d = \delta \mathcal{P}_c/\gamma $. Finally, from equation \eqref{cauchy-x}, the only remaining possibility is that the second term balances the third term, indicating $\mathcal{D}_{\hat{x} \hat{x}}=w\mathcal{D}_{\hat{x} \hat{y}}/d = \mathcal{P}_c/\gamma^2$.

So far, we have only required that {the elastic solid is slender, \textit{i.e.},} $d\ll \ell$, which is equivalent to $\gamma\ll \delta /\epsilon$, and covers a large range of wall thicknesses {(recall that $\gamma=d/w$, $\epsilon=h_0/\ell$ and $\delta=h_0/w$)}. However, it is also expected that  $\gamma\gg \epsilon$, such that $\mathcal{D}_{\hat{x} \hat{z}}$ is a small quantity, excluding the case of an extremely thin wall. In fact, recalling that the application of linear elasticity requires that $\hat{u}_y\ll d$, any prominent deformation in a thin-walled microchannel is likely out of the scope of the linear elastic theory. 

Therefore, with $\epsilon\ll \gamma \ll \delta/\epsilon$ as well as $\epsilon\ll \delta\ll 1$, we conclude that $\sigma_{\hat{x} \hat{z}}$ and $\sigma_{\hat{y} \hat{z}}$ are negligible in comparison to the other stress components. Depending on the wall thickness, the relative magnitude among the remaining four stress components can change. For example, if $\gamma^2 \gg 1$, we can further neglect $\sigma_{\hat{x} \hat{x}}$ as in \citep{WC19}. Nevertheless, no matter how $d$ varies, the dominant balances in equations \eqref{cauchy-x} and \eqref{cauchy-y} occur in the cross-sectional $(\hat{x},\hat{y})$ plane, which reduces the original 3D elasticity problem to a 2D plane-strain problem. Since we showed in \S~\ref{sec:fluid} that $p$ is a function of $z$ only at the leading order (in $\epsilon$), the deformation of the $(\hat{x},\hat{y})$ cross-sections at different $z$-locations  (recalling $z=\hat{z}$) decouple from each other. 
At each cross-section, the deformation is then determined by the local hydrodynamic pressure $p(z,t)$. Therefore, generally, we can express the displacement of the fluid--solid interface at the leading order (in $\epsilon$) as
\begin{equation}\label{uy-O1}
    u_y(x,z,t) = u_{\hat{y}}(\hat{x}, \hat{z}, t)= \mathfrak{f}(x)p(z,t), 
\end{equation}
with $\mathfrak{f}(x)$ being the spanwise deformation profile. The separation-of-variables form of equation \eqref{uy-O1} suggests that the cross-sectional deformation profiles at different $z$-locations are, in a sense, self-similar. The displacement is fully determined by the local pressure, showing that the fluid--solid interface behaves like a Winkler foundation \citep{W67,DMKBF18}, with a variable stiffness represented by $1/\mathfrak{f}(x)$. Importantly, this Winkler-foundation-like mechanism is \emph{not} an assumption here, but rather it is a \emph{consequence} of the slenderness of the top wall. Also, note that the assumption of $\mathcal{T}_s\ll \mathcal{T}_f$ has been applied here, meaning that the solid promptly responds to pressure changes in the flow.

It is also worth mentioning that if the top wall is thin with $\epsilon\ll \gamma \lesssim 1$, the elasticity problem is usually taken to be a plane stress problem, and a 1D engineering model is usually available for the displacement out of plane (\textit{i.e.}, $u_y$ here), such as the  Kirchhoff--Love \citep{L88, TWK59} and Reissner--Mindlin \citep{R45, M51} plate theories. However, this fact does not fundamentally contradict with our plane-strain reduction because the decoupling of the cross-sections remains true \citep{CCSS17, SC18, AMC20} due to the separation of scales, $w\ll \ell$. 

Moreover, the discussion above is only based on the balance of Cauchy stresses, and does not involve the boundary conditions either on the sides  (\textit{i.e.}, at $x=\pm w/2$) or at the upper surface of the wall (\textit{i.e.}, at $\hat{y}=d$ or $y=h_0 +d$). The decoupling of the cross-sectional deformation is just a consequence of the wall slenderness. However, the boundary conditions do have an important influence on the displacement field in the solid, which gives rise to different forms of $\mathfrak{f}(x)$ in equation \eqref{uy-O1} \citep{WC21}.

\subsection{Reduction: introducing the width-averaged (effective) height}
\label{subsec:eff_height}

Since we seek a 1D model dependent on $z$ only, the $x$-dependence can be eliminated by averaging the displacement of the fluid--solid interface over $x$:
\begin{equation}\label{uyavg-O1}
    \bar{u}_y(z,t) = \frac{1}{w}\int_{-w/2}^{+w/2} u_y(x,z,t)\, \rd x = \underbrace{\left[ \frac{1}{w}\int_{-w/2}^{+w/2} \mathfrak{f}(x)\, \rd x\right]}_{1/k} p(z,t),
\end{equation}
having used equation \eqref{uy-O1}. Then, the width-averaged height $\bar{h}$ of the channel is
\begin{equation}\label{havg-O1}
    \bar{h}(z,t) = \frac{1}{w}\int_{-w/2}^{+w/2} h_0 + u_y(x,z,t)\, \rd x = h_0 + \frac{1}{k}p(z,t).
\end{equation}
An important quantity in the above equations is the proportionality constant, $k$, which represents the \emph{effective stiffness} of the interface and further highlights the Winkler-foundation-like mechanism of deformation of the fluid--solid interface. 

Equation \eqref{havg-O1} simplifies the FSI problem in two aspects. First, the deformation of the interface is further reduced from 2D to 1D. Second, the flow in the channel is reduced from 3D to 2D, and this reduction does not change any key statements made before {because we can still appeal to the lubrication approximation for 2D flows due to $h_0\ll \ell$}. In other words, if we start from the 2D incompressible Navier--Stokes equations (simply neglecting $v_x$ and $x$-dependent terms), we still find the lubrication approximation is applicable up to $\hat{Re} = \mathcal{O}(1)$, as in \S~\ref{subsec:fluid-reduction}. 

In the following discussion, we will take $\bar{h}$ as the effective height of the deformed channel and consider the reduced system sketched in figure~\ref{subfig:gen1d-geo-2d}. {Note that the 2D configuration in figure~\ref{subfig:gen1d-geo-2d} is not  \emph{a priori} assumed but, rather, it is derived from the 3D configuration in figure~\ref{subfig:gen1d-geo-3d} based on averaging the deformation across $x$, via equation~\eqref{uyavg-O1} (or equation~\eqref{havg-O1}). This approach is in contrast to the earlier work of, \textit{e.g.}, \cite{SM04}, who assumed a 2D configuration of the elastic solid in the $(y,z)$ plane from the outset.} Taking $\bar{h}$ as the effective deformed height was first suggested by \citet{GEGJ06}. As shown by \citet{WC21}, the error introduced by this averaging approach is controllable.

\subsection{Extension: introducing weak deformation effects}
\label{subsec:weak_effects}

So far, the discussion in the previous  subsections was based on the leading-order (in $\epsilon$) theory, which does not take into account the possible restrictions imposed at the inlet and the outlet (\textit{i.e.}, at $z=0$ and $z=\ell$). As shown in figure~\ref{subfig:gen1d-geo-2d}, the movement of the fluid--solid interface at both ends is often physically restricted. In this sense, we can think of the solid mechanics problem as being essentially a boundary layer problem. While the Winkler-foundation-like mechanism is dominant outside any ``boundary layers,'' with equation \eqref{uyavg-O1} being the (outer) solution there, another mechanism plays a role within thin (boundary) layers near $z=0,\ell$, each potentially admitting inner solutions that regularize the problem and allow the enforcement of end constraints. 

Since the bulging of the top wall unavoidably introduces stretching along $z$ in the solid, a simple extension of equation \eqref{uyavg-O1} can be achieved by introducing weak constant tension into the formulation \citep{WC21}. First, for the Winkler-foundation-like mechanism to be dominant, the tension needs to be ``weak.'' Second, the tension force can be taken to be constant because its variation along $z$ is proportional to the fluid's tangential traction at the interface, which is a small quantity per the lubrication approximation, as shown in \S~\ref{sec:fluid} \citep[see also][]{HBDB14}. The ``regularized'' governing equation for the {width-averaged fluid--solid interface's displacement} $\bar{u}_y$ is then written as
\begin{equation}\label{dimension-1d-solid-eq}
    \underbrace{\rho_s b^{\star} \frac{\partial^2 \bar{u}_y}{\partial t^2}}_{\text{inertia}} + \underbrace{k\bar{u}_y}_{\text{stiffness}} - \underbrace{\chi_t\frac{\partial^2 \bar{u}_y}{\partial z^2}}_{\text{tension}} = \underbrace{p(z,t)}_{\text{load}},
\end{equation}
where $\rho_s$ denotes the solid density, $b^{\star}$ represents the effective thickness of the interface (discussed in \S~\ref{subsec:eff_thick}), {which is introduced so that the first term can represent the bulk inertial effects of the solid. Recalling that $k$ is the effective stiffness introduced from equation \eqref{uyavg-O1}, the second term represents the dominant Winkler-foundation-like effect, while $\chi_t$ is introduced to model the tension per unit width (discussed in \S~\ref{subsec:tension_coeff}). In the case of $\chi_t$ being constant, the tension can be written in terms of the transverse displacement $\bar{u}_y$ \citep[see \textit{e.g.}][\S~4.3]{HKO09}.} The weakness of the solid inertia and the tension may not be obvious from equation~\eqref{dimension-1d-solid-eq}, although the scaling analysis in \S~\ref{subsec:solid-O1} anticipates it.

Equation \eqref{dimension-1d-solid-eq} is essentially the equation of motion of a Kramer-type surface, which has been used extensively in the study of high-$Re$ (boundary layer) flows over compliant coatings. The goal of the latter studies is to understand how to delay the laminar--turbulence transition \citep{G02}. However, microchannel flows cannot reach such high $Re$ values. More surprisingly, the application of equation \eqref{dimension-1d-solid-eq} in modeling soft microchannels {leads to different conclusions from the compliant coating studies. Compliance of the channel wall can actually promote (instead of delay) the laminar--turbulence transition in pressure-driven flow thanks} to the FSI-induced instabilities. This effect can be successfully exploited for micromixing.


\subsection{Nondimensionalization: introducing the FSI parameter}
\label{subsec: nodim_fsi_param}

We can now make the 1D interface motion equation \eqref{dimension-1d-solid-eq} dimensionless to better illustrate the relative magnitude of the different terms (effects). Capital letters are used to denote the dimensionless variables. 

The first step is to determine the characteristic scale, $\bar{\mathcal{U}}_c$, for the fluid--solid interface displacement. The dominant deformation effect in equation \eqref{uyavg-O1}, suggests that we should take $\bar{\mathcal{U}}_c = \mathcal{P}_c/k$, recalling that the scale for $p$ is $\mathcal{P}_c$. Then, the dimensionless version of equation \eqref{uyavg-O1} is simply
\begin{equation}\label{UYavg-O1}
    \bar{U}_Y(Z,T) = P(Z,T).
\end{equation}
Still using $h_0$ to scale $\bar{h}$, the dimensionless effective channel height (equation \eqref{havg-O1}) becomes
\begin{equation}\label{Havg}
    \bar{H}(Z,T) = 1 + \frac{\bar{\mathcal{U}}_c}{h_0}\bar{U}_Y(Z,T) = 1 + \beta \bar{U}_Y(Z,T).
\end{equation}
Here, we have introduced another dimensionless parameter, $\beta=\bar{\mathcal{U}}_c/h_0=\mathcal{P}_c/(kh_0)$. It is clear from equation \eqref{Havg} that $\beta$ translates the interface displacement into the deformation of the fluid domain, capturing the ``strength'' of the fluid--solid coupling. Thus, $\beta$ is the ``FSI parameter'' of our model.

The dependence of $\beta$ on the system properties comes through $\mathcal{P}_c$ and $k$. While $\mathcal{P}_c$ is determined by the flow conditions (\textit{i.e.}, the viscosity of the fluid, the flow rate, and the geometry of the undeformed channel), $k$ is determined by the material properties, the geometry, and the boundary conditions on the compliant wall. To explicitly show this, we write
\begin{equation}\label{1_to_k}
    \frac{1}{k} =  \frac{1}{w}\int_{-w/2}^{+w/2} \mathfrak{f}(x)\, \rd x= \int_{-1/2}^{+1/2} \mathfrak{f}(wX)\, \rd X = \frac{\xi}{\bar{E}}\underbrace{\int_{-1/2}^{+1/2} F(X)\, \rd X}_{\mathcal{I}_1}= \frac{\xi\mathcal{I}_1}{\bar{E}}.
\end{equation}
The definition of $k$ from equation \eqref{uyavg-O1} is used in the first step. The second step is making the integral dimensionless. In the third step, the assumption of a linearly elastic solid has been invoked with $k\propto \bar{E}$ and $\mathfrak{f}(x)\propto 1/\bar{E}$. {Note that $\bar{E} = E/(1-\nu_s^2)$ here, which means that $k$ is well defined as $\nu_s\to1/2^-$, because of the plane-strain reduction from 3D to 2D}, with $E$ being the Young's modulus and $\nu_s$ being the Poisson's ratio, respectively. Then, $F(X)$ is introduced as the dimensionless self-similar deformation profile, and $\xi$ is the resulting pre-factor after $x$ is scaled by $w$. In the last step, $\mathcal{I}_1=\int_{-1/2}^{+1/2} F(X)\, \rd X$ was introduced to simplify the expression. While the effect of the material properties of the solid wall are captured by $\bar{E}$, the influence of the wall geometry and the boundary conditions are taken into account by both $\xi$ and $\mathcal{I}_1$. As mentioned in \S~\ref{subsec:solid-O1}, $\mathfrak{f}(x)$ takes different forms in different situations, thus giving different expressions for $\xi$ and $\mathcal{I}_1$. For example, for the thick-walled microchannel considered by \citet{WC19}, $\xi=w$ and $\mathcal{I}_1\approx0.542754$. Meanwhile, for the microchannels with thick-plate-like top walls considered by \citet{SC18}, $\xi=w^4/(2d^3)$ and $\mathcal{I}_1={1}/{30}+{(d/w)^2}/[{3\kappa(1-\nu_s)}]$, with $\kappa$  being a ``shear correction factor'' (typically, $\kappa=1$). Nevertheless, the key point is that both $\xi$ and $\mathcal{I}_1$ can be obtained \textit{a priori}, by solving the corresponding elasticity problem, as analytical expressions.

Using the scales from tables~\ref{tab:fluid-scales} and \ref{tab:solid-scales}, the dimensionless version of equation \eqref{dimension-1d-solid-eq} is
\begin{equation}\label{1d-solid-eq}
    \theta_I \frac{\partial^2 \bar{U}_Y}{\partial T^2} + \bar{U}_Y -\theta_t\frac{\partial^2 \bar{U}_Y}{\partial Z^2} = P,
\end{equation}
where $\theta_I = \rho_s b^{\star}\mathcal{\bar{U}}_c/(\mathcal{T}_f^2 \mathcal{P}_c)$ and $\theta_t = \chi_t \bar{\mathcal{U}}_c/(\ell^2\mathcal{P}_c) $ are introduced above as the inertial coefficient and the tension coefficient, respectively. We expect that $\theta_I\ll1$ and $\theta_t\ll 1$ because both the solid inertia and the tension effect are weak. Then, the leading-order solution (as $\theta_I,\theta_t\to0$) of equation \eqref{1d-solid-eq} is equation \eqref{UYavg-O1}, as required by the above discussion. {Even though $\theta_I\ll1$ and $\theta_t\ll 1$, it is important to properly model these weak effects because they have significant influence on the global instability of the system (as we will see in \S~\ref{subsec:eig}).}

First, let us justify $\theta_I\ll 1$. Recalling that $\mathcal{P}_c=\mu\mathcal{V}_c/(\epsilon h_0)$, $\hat{Re}=\epsilon \rho_f\mathcal{V}_c h_0/\mu$, and $\beta = \bar{\mathcal{U}}_c/h_0$, $\theta_I$ can be written as
\begin{equation}\label{thetaI1}
    \theta_I = \frac{\rho_s b^{\star}\mathcal{\bar{U}}_c}{\mathcal{T}_f^2 \mathcal{P}_c}= \epsilon\frac{\rho_f\mathcal{V}_ch_0}{\mu}\frac{\mathcal{\bar{U}}_c}{{h_0}}\frac{b^{\star}h_0}{\ell^2}\frac{\rho_s}{\rho_f} = \epsilon \hat{Re}\beta\frac{b^{\star} }{\ell }\frac{\rho_s}{\rho_f}.
\end{equation}
Since $\epsilon\ll 1$, $b^{\star} \leq d\ll \ell$, $\hat{Re}= \mathcal{O}(1)$, $\beta$ is typically $\mathcal{O}(1)$ and $\rho_s\simeq \rho_f$ in the microchannel setting (because PDMS has a similar density to water), we have justified $\theta_I\ll 1$. Note time in equation \eqref{dimension-1d-solid-eq} is scaled by $\mathcal{T}_f$, as before, to ensure the  fluid--solid coupling. Thus, $\theta_I\ll 1$ corresponds to $\mathcal{T}_s^2\ll \mathcal{T}_f^2$, meaning that the solid responds to pressure changes in the flow promptly, as discussed in \S~\ref{subsec:solid-O1}. It is also helpful to note that, using equation \eqref{1_to_k}, we can also write
\begin{equation}\label{thetaI2}
    \theta_I = \mathcal{I}_1\frac{\rho_s b^\star \xi}{\mathcal{T}_f^2 \bar{E}},
\end{equation}
which indicates that, for a more rigid solid (\textit{i.e.,} with the increase of $\bar{E}$), the solid deformation develops much faster than the flow. 

Next, using equation \eqref{1_to_k} again, the dimensionless tension coefficient can be written as
\begin{equation}\label{tension_coeff1}
    \theta_t = \frac{\chi_t \bar{\mathcal{U}}_c}{\ell^2\mathcal{P}_c} =\mathcal{I}_1\frac{\chi_t  \xi}{\bar{E}\ell^2}.
\end{equation}
If $\chi_t$ is deformation-induced (thus time-dependent), substituting equation \eqref{ft-eq} into equation \eqref{tension_coeff1}, we obtain
\begin{equation}\label{tension_coeff2}
\begin{aligned}
    \theta_t(T) &=\mathcal{I}_1\frac{\xi}{\bar{E}\ell^2} \frac{E b^\star}{\ell} \int_0^{\ell} \frac{1}{2}\left(\frac{\partial \bar{u}_y}{\partial z}\right)^2\, \rd z  \\
    &= (1-\nu_s^2)\mathcal{I}_1 \frac{\xi b^{\star}\bar{\mathcal{U}}_c^2 }{\ell^4} \int_0^1 \frac{1}{2} \left(\frac{\partial \bar{U}_Y}{\partial Z}\right)^2 \, \rd Z \\
    &= \Tilde{\theta}_t \int_0^1 \frac{1}{2} \left(\frac{\partial \bar{H}}{\partial Z}\right)^2 \, \rd Z,
\end{aligned}
\end{equation}
where 
\begin{equation}\label{theta_tt}
    \Tilde{\theta}_t = \frac{1}{\beta^2}\times (1-\nu_s^2)\mathcal{I}_1 \frac{\xi b^{\star}\bar{\mathcal{U}}_c^2 }{\ell^4} =(1-\nu_s^2)\mathcal{I}_1 \epsilon^2\frac{\xi b^\star}{\ell^2}.
\end{equation}
Note that we have used equation \eqref{Havg} in the last step of the manipulations in equation \eqref{tension_coeff2}. Also, note that $\tilde{\theta}_t$ is not related to the flow conditions, and $\Tilde{\theta}_t \ll 1$ for microchannels. Within linear elasticity, the integral in equation \eqref{tension_coeff2} is  $\mathcal{O}(1)$, thus $\theta_t\ll 1$ as well. 

Finally, the key dimensional and dimensionless parameters are summarized in table~\ref{table:param}. Typical values for microfluidic systems are given to justify the bigness/smallness assumptions made.

\begin{table}[t]
    \centering
    \begin{tabular}{l@{\quad} l@{\quad} l@{\quad} l}
    \hline\hline
    Quantity & Notation & Typical value & Units \\
    \hline
     channel's length & $\ell$ & $1.0$ & \si{\centi\meter} \\
    channel's undeformed height & $h_{0}$ & $30$ & \si{\micro\meter} \\
    channel's width & $ w $ & $500$ & \si{\micro\meter}\\
    top wall's thickness & $t$ & $2.0$ & \si{\milli\meter}\\
    solid's Young's modulus & $E$ & See table~\ref{table:case-param} & \si{\mega\pascal}  \\
    solid's Poisson's ratio & $\nu$ & $0.5$ & -- \\
    solid's density & $\rho_s$ & $1.0 \times 10^3$ & \si{\kilo\gram\per\meter\tothe{3}} \\
    fluid's density & $\rho_f$ &$1.0\times10^{3}$ & \si{\kilo\gram\per\meter\tothe{3}} \\
    fluid's dynamic viscosity & $\mu$ & $1.0\times10^{-3}$ & \si{\pascal\second} \\
    inlet flow rate & $q$ & See table~\ref{table:case-param} & \si{\micro\litre\per\minute} \\
    effective interface thickness & $b^{\star}$ & $0.6141w=307.05$ & \si{\micro\meter} \\ 
    \hline
    channel's height-to-length aspect ratio & $\epsilon={h_0}/{\ell}$ & $0.003$ &  --\\
    channel's height-to-width aspect ratio & $\delta = {h_0}/{w} $ & $0.06$ & --\\
    reduced Reynolds number & $\hat{Re}={\epsilon\rho_f q}/{(w\mu)}$  & See table~\ref{table:case-param} &  --\\
    FSI parameter & $\beta=\mathcal{I}_1 \mathcal{P}_c \xi/(\bar{E} h_0) $ &  See table~\ref{table:case-param}  & -- \\
    solid's inertia coefficient & $\theta_I=\epsilon \hat{Re} \beta b^{\star}\rho_s/(\ell \rho_f)$ & See table~\ref{table:case-param}  & -- \\
    deformation-induced tension coefficient & $\theta_t (T) = \Tilde{\theta}_t\displaystyle\int_0^1 \frac{1}{2}\left(\frac{\partial \bar{H}}{\partial Z}\right)^2 \, \rd Z $ & Variable & -- \\[-2mm]
    &&($\Tilde{\theta}_t$ in table~\ref{table:case-param}) & -- \\
    \hline\hline
    \end{tabular}
    \caption{The dimensional and dimensionless parameters of the 1D FSI model.}
    \label{table:param}
\end{table}

\section{Fluid--solid coupling: A new 1D FSI model}
\label{sec:coupling}

The final step in deriving the 1D FSI model involves coupling the fluid mechanics problem to the solid mechanics problem. This coupling is achieved by depth averaging the Navier--Stokes equations at the leading order in $\epsilon$. Recall that, even though we averaged across $X$ in \S~\ref{subsec:eff_height}, the flow field is 2D at this stage, \textit{i.e.}, $V_Z=V_Z^{2D}(Y,Z,T)$. Vertically integrating the axial velocity, $V_Z^{2D}$, introduces the flow rate, $Q(Z,T) =  \int_{0}^{\bar{H}(Z,T)}V_Z^{2D}(Y,Z,T)\,\rd Y $, into the formulation. Invoking a von K\'arm\'an--Pohlhausen approximation \citep{Pohl21} (see also, \textit{e.g.}, \cite[\S~4-6.5]{W06_book} or \cite[p.~541]{panton}), we assume a parabolic velocity profile across any deformed cross-section of the channel:
\begin{equation}\label{vel-profile}
    V_Z^{2D}(Y,Z,T) = \frac{6QY[\bar{H}(Z,T)-Y]}{\bar{H}^3(Z,T)}.
\end{equation}
The kinematic boundary conditions at the top wall requires that
\begin{equation}\label{kinematic-bc}
    V_Y^{2D}|_{Y=\bar{H}} = \frac{\partial \bar{H}}{\partial T}.
\end{equation}
Making equations \eqref{COM} and \eqref{COLM-z} dimensionless, neglecting $x$-dependent terms and small terms, and integrating over $Y$, we obtain
\begin{subequations}\label{gen1d-flow-eq}
\begin{align}
   \frac{\partial Q}{\partial Z}+ \frac{\partial \bar{H}}{\partial T}&=0,\label{gen1d-cont-eq}\\
    \hat{Re} \frac{\partial Q}{\partial T} + {\hat{Re}} \frac{6}{5} \frac{\partial }{\partial Z} \left(\frac{Q^2}{\bar{H}}\right) &= - \bar{H}\frac{\partial P}{\partial Z} - \frac{12Q}{\bar{H}^2},\label{gen1d-mom-eq}
\end{align}
\end{subequations}
where the no-slip boundary condition has been applied at $Y=0$ and $Y=\bar{H}$. Note equations \eqref{gen1d-flow-eq} were also previously derived by \citet{SWJ09} and \citet{IWC20}. 

Equations \eqref{gen1d-cont-eq}, \eqref{gen1d-mom-eq}, \eqref{1d-solid-eq} and \eqref{Havg} define a 1D FSI model. In this work, we consider the case in which the flow rate at the inlet is fixed, while the pressure at the outlet is set to gauge, \textit{i.e.},
\begin{equation}\label{bc1}
    Q(0,T) = 1, \qquad P(1,T)=0.
\end{equation}
Also, there are no displacements at the inlet and the outlet of the channel:
\begin{equation}\label{bc2}
   \bar{U}_Y(0,T)=\bar{U}_Y(1,T)=0 \quad \Rightarrow \quad \bar{H}_Y(0,T)=\bar{H}_Y(1,T)=1.
\end{equation}
Initially, we assume the wall is undeformed and the flow is uniform through the channel, \textit{i.e.,}
\begin{equation}\label{ic}
     Q(Z, 0) = 1,\qquad \bar{U}_Y(Z,0)=0 \quad \Rightarrow \quad \bar{H}_Y(Z,0)=1.
\end{equation}

\subsection{Exemplar cases and preview of results}

In the remainder of this paper, we address the steady-state features, the dynamic response, and also the linear stability of the non-flat steady state of the proposed 1D FSI model. To explore these issues, we have chosen exemplar cases with typical dimensional and dimensionless values given in table~\ref{table:param} and table~\ref{table:case-param}. The values for the geometrical and material properties are taken and/or modified from \citep{GEGJ06}. The long and shallow microchannels are assumed fabricated via soft lithography, with a thick top wall. The leading-order steady response of such microchannels has been solved by \citet{WC19}, according to which $\mathcal{I}_1 = 0.542754$ and $\xi=w$ for calculating $\beta$ in table~\ref{table:case-param}. 

\begin{table}[ht]
    \centering
    \begin{tabular}{c | c c c c c c }
    \hline\hline
    Case & $E$ & $q$ & $\hat{Re}$ & $\beta$  & $\theta_I$ & $\Tilde{\theta}_t$ \\
         & (\si{\mega\pascal}) & (\si{\micro\litre\per\minute}) & (--) & (--) &  (--) & (--) \\
    \hline
    C1  & $1$ & $1500$ & $0.15$  & $0.1256$  & $1.7360\times 10^{-6}$ & $5.6245\times 10^{-9}$\\
    C2  & $1$ & $6000$ & $0.60$  & $0.5026$  & $2.7775\times 10^{-5}$ & $5.6245\times 10^{-9}$\\
    C3  & $1$ & $9000$ & $0.90$  & $0.7538$  & $6.2495\times 10^{-5}$ & $5.6245\times 10^{-9}$\\
    C4  & $2$ & $1500$ & $0.15$  & $0.0628$  & $8.6798\times 10^{-7}$ & $5.6245\times 10^{-9}$ \\
    C5  & $2$ & $6000$ & $0.60$  & $0.2513$  & $1.3888\times 10^{-5}$ & $5.6245\times 10^{-9}$ \\
    C6  & $2$ & $9000$ & $0.90$  & $0.3769$  & $3.1247\times 10^{-5}$ & $5.6245\times 10^{-9}$ \\
    \hline\hline
    \end{tabular}
    \caption{The dimensional and dimensionless parameters for the exemplar cases considered.}
    \label{table:case-param}
\end{table}

Similar to the experiments of \citet{VK13}, cases C1 to C3 and C4 to C6 summarized in table~\ref{table:case-param} are each based on a single microchannel, operated under different flow conditions. As catalogued in table~\ref{tab:comparison}, the steady and dynamic responses of the new 1D FSI model match several experimental observations qualitatively, which indicates that the proposed model can provide unique   insights into  this unstable FSI problem. However, we cannot perform direct quantitative comparisons between our 1D model and the experiments of \citet{VK13}, for the following reasons. First, the experiments' soft wall was compressed upon a rigid outer surface, unlike our model wherein a soft wall bulges outwards in an unconstrained manner (being stress-free on its outer surface). Further, in the experiments, the wall thickness was comparable to the channel's width, with two side walls made rigid. Consequently, the deformation field within the compliant wall in the microchannels fabricated by \citet{VK13} is described by a different leading-order theory of the flow-induced deformation than the theories considered herein (recall \S~\ref{sec:solid}). At this time, it is not clear whether an exact solution (along the lines of \cite{WC19}) could be obtained for the deformation in the configuration fabricated by \citet{VK13}. The main difference would be in the definition of $\beta$. Nevertheless, since the experiments did consider long and shallow microchannels with a slender deformable wall, the assumptions made in \S\S~\ref{sec:fluid} and \ref{sec:solid} apply. Therefore, the FSI \emph{physics} in these experiments are expected to be captured by the theoretical framework proposed herein. Indeed, as discussed by \citet{VK13}, the FSI-induced instabilities are generic, thus are not expected to be an ``accidental'' phenomenon occurring only in some specific experimental devices. It follows that our qualitative comparisons below are meaningful and useful for validating the proposed 1D FSI model.

\begin{table}[ht!]
\centering
\begin{tabular}{c|p{0.4\textwidth}|p{0.4\textwidth}}
\hline
\hline
   & Experimental observation  & Proposed 1D FSI model behavior \\
   \hline
   Steady & \begin{itemize}[leftmargin=*]
       \item Wall deformation is nonuniform along the streamwise direction.
       \item There is a sharp diverging section after  the channel entrance, followed by a longer converging section tapering towards the outlet.
   \end{itemize} &
   \begin{itemize}[leftmargin=*]
       \item The steady-state pressure and deformation profiles vary along the streamwise direction.
       \item For weak axial tension, the channel expands sharply near the inlet, reaching a  maximum deformation. Then, the deformation tapers out towards the outlet. This effect is due to the governing equation of the fluid--solid interface exhibiting a boundary-layer-like behavior for $\theta_t\ll 1$.
   \end{itemize}\\
   \hline
   Dynamic & \begin{itemize}[leftmargin=*]
       \item Dye injected in the flow oscillates/breaks-up at $Re\simeq 100$ ($\hat{Re}\simeq 1$). The dye is observed to breakup first in the converging section near the channel outlet.
       \item In the mixing experiments, vigorous mixing is observed downstream in the converging section at $Re\simeq 100$ ($\hat{Re}\simeq 1$).
       \item Under the same flow rate, the mixing in the more compliant channel is observed to be more complete.
       \item The wall oscillates as the dye breakup (instability) is observed. 
   \end{itemize}
   & \begin{itemize}[leftmargin=*]
       \item The base (steady) solution becomes linearly unstable to infinitesimal perturbations at $Re\simeq 100$ ($\hat{Re}\simeq 1$). 
       \item The global unstable modes are found to be highly oscillatory, with frequencies close to the natural frequency of the wall. The corresponding eigenfunctions are highly oscillatory in space, with the shape changing more dramatically near the outlet than that near the inlet.
       \item Under the same flow rate, the more compliant channel's most unstable mode has a larger growth rate.
       \item Self-sustained wall oscillations can be triggered in the linearly unstable cases. The wall oscillations have a peak frequency close to the natural frequency of the wall, and are found to be more violent near the channel outlet.
   \end{itemize}\\
\hline
\hline
\end{tabular}
\caption{Qualitative comparison between the experimental observations of \cite{VK13} and the predictions of the proposed global 1D FSI model.}
\label{tab:comparison}
\end{table}

\begin{figure}[t]
    \centering
    \includegraphics[width=0.99\textwidth]{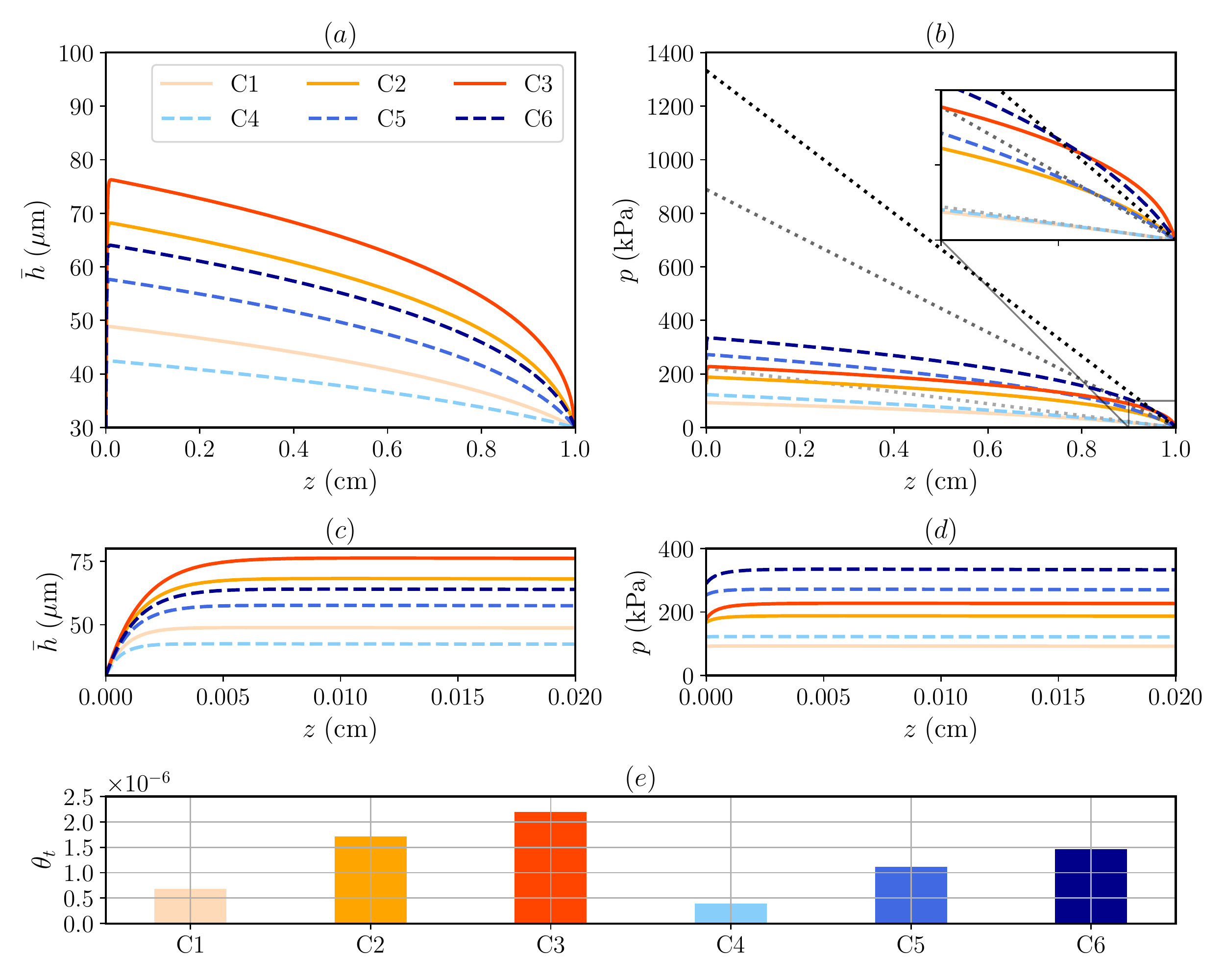}
    \caption{The steady-state response for the exemplar cases from  table~\ref{table:case-param}, for flow rates of $q = 1500$, $6000$, $9000$ \si{\micro\litre\per\minute} (higher $q$ corresponds to darker curves). (\textit{a}) The variation of the deformed channel height along $z$. (\textit{b}) The pressure distribution along $z$, with the inset window showing a zoom-in view near the $z=\ell$. The dotted lines show the Hagen--Poiseuille law for a rigid channel (linearly variation of $p$ along $z$). Panels (\textit{c}) and (\textit{d}) show zoomed-in views for $\bar{h}$ and $p$ near the inlet, $z=0$, respectively. (\textit{e}) The computed $\theta_t$ of the exemplar cases from table~\ref{table:case-param}.}
    \label{fig:steady1}
\end{figure}

\section{Base state: features of the inflated microchannel at steady state}
\label{sec:steady}

At steady state, all the time derivatives vanish. From equation \eqref{gen1d-cont-eq}, we have $Q\equiv 1$, upon imposing the fixed-flux upstream boundary condition from equation \eqref{bc1}. The remaining equations \eqref{1d-solid-eq}, \eqref{Havg} and \eqref{gen1d-mom-eq}, together with the unsatisfied boundary conditions from equations \eqref{bc1} and \eqref{bc2}, constitute a nonlinear two-point boundary value problem \citep{Keller76}. This nonlinear system is solved using the {\tt newton\_krylov} routine from the SciPy stack \citep{SciPy}, following the procedure described in \SM. 

Also, note that equations \eqref{1d-solid-eq}, \eqref{Havg} and \eqref{gen1d-mom-eq} can be combined to form a single equation, in terms of the steady-state deformed height $\bar{H}_0$, written as
\begin{equation}\label{steady-eq1}
    \frac{6}{5}\hat{Re}\frac{1}{\bar{H}_0^3}\frac{\rd \bar{H}_0}{\rd Z} = \frac{1}{\beta}\left(\frac{\rd\bar{H}_0}{\rd Z}-\theta_t\frac{\rd^3\bar{H}_0}{\rd Z^3}\right)+ \frac{12}{\bar{H}_0^3}.
\end{equation}
The boundary conditions for this third-order nonlinear ordinary differential equation (ODE) are
\begin{equation}\label{steady-bcs}
     \bar{H}_0(Z=0) = \bar{H}_0(Z=1)=1, \qquad \left.\frac{\rd^2\bar{H}_0}{\rd Z^2}\right|_{Z=1}=0,
\end{equation}
which correspond to zero displacement imposed at $Z=0,1$, along with the gauge-pressure boundary condition at the outlet.  

The steady responses of the exemplar cases from table~\ref{table:case-param} are shown in figure~\ref{fig:steady1}. The nonuniform deformation of the channel height is shown in figure~\ref{fig:steady1}(\textit{a}), along with a zoom-in view near the inlet given in  figure~\ref{fig:steady1}(\textit{c}). The channel inflates more for larger flow rates and/or for softer walls (\textit{i.e.}, with smaller $E$). For each case, there is a sharp diverging section near the channel inlet, and a much longer converging section connecting to the channel outlet, which agrees with the experimental observations of \citet{VK13}. As for the pressure distribution, figure~\ref{fig:steady1}(\textit{b}) shows that the compliance of the wall leads to a non-uniform pressure gradient so that the pressure varies nonlinearly with $z$. Furthermore, compared to the case of flow in a rigid channel, the total pressure drop is reduced significantly due to the expanded cross-sectional area resulting from the deformation of the wall. This phenomenon has been addressed and analyzed, considering different geometrical configurations and elastic response, but typically limited to the case of $\hat{Re}\to 0$ \citep[see, \textit{e.g.},][]{C21}. However, our proposed 1D FSI model pushes the limit to $\hat{Re} = \mathcal{O}(1)$. Figure~\ref{fig:steady1}(\textit{d}) also zooms into the neighborhood of the channel inlet. As the flow rate increases, a small region of positive pressure gradient appears. The reason for this effect is that, the sharp expansion of the channel's cross-section near the inlet makes the local velocity drop quickly (recall that $Q\equiv 1$ at steady state), and the positive pressure gradient aids in the deceleration of the flow \citep{IWC20,WC21}. Finally, the deformation-induced weak tension coefficients of the exemplar cases are shown in figure~\ref{fig:steady1}(\textit{e}). We observe that $\theta_t$ increases as the flow rate increases because higher flow rates induce larger deformations. Moreover, for the same flow conditions, it is observed that $\theta_t$ is larger when the wall is more compliant.

\begin{figure}[t]
    \centering
    \includegraphics[width=0.99\textwidth]{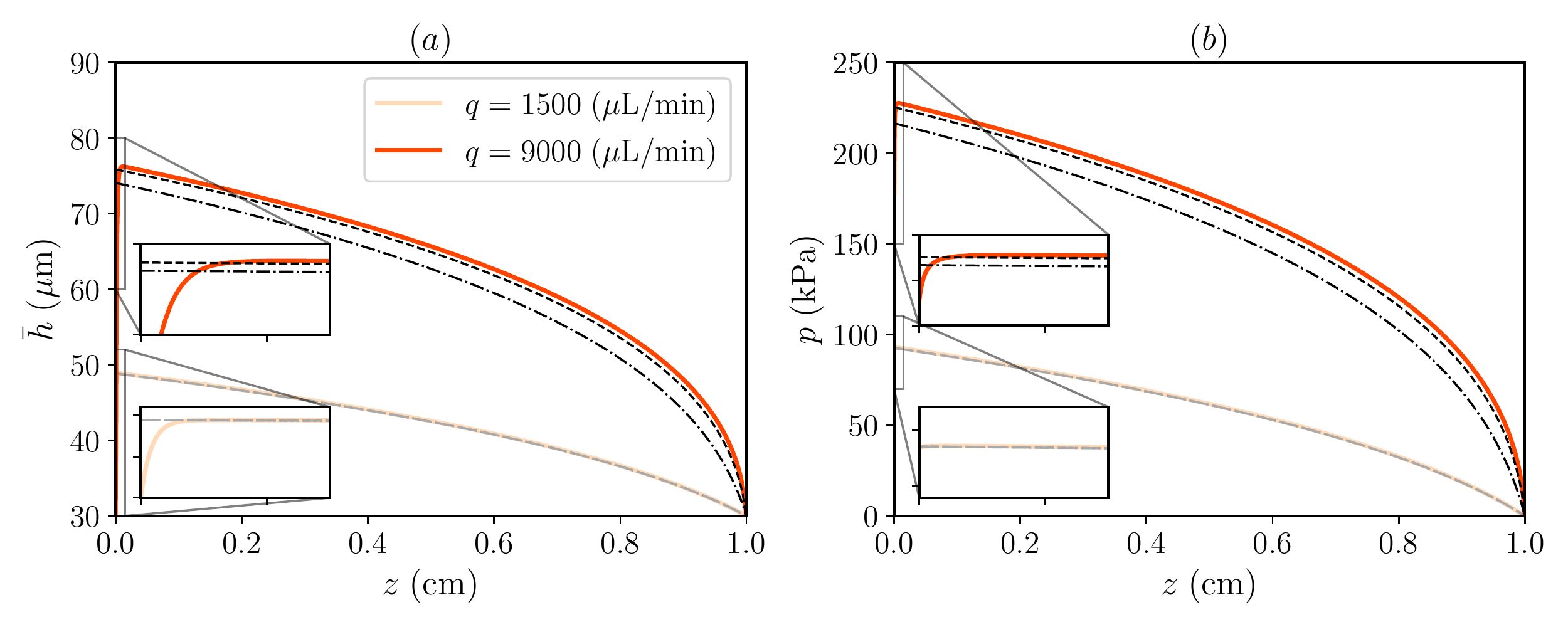}
    \caption{Comparison between the numerical solutions of the proposed 1D FSI model and previously reported analytical results for the exemplar cases of C1 and C3. (\textit{a}) The deformed channel height along $z$. (\textit{b}) The pressure distribution along $z$. The solid curves represent the numerical simulation of the 1D FSI model at steady state. The dash-dotted curves represent the results calculated by equation \eqref{steady-eq2} with $\hat{Re} = 0$ and $\theta_t = 0$. The dashed curves are calculated from equation \eqref{steady-eq3}, in which $\theta_t = 0$ but $\hat{Re}= \mathcal{O}(1)$.} 
    \label{fig:steady2}
\end{figure}

Next, to address the effect of $\hat{Re}$ and $\theta_t$, we compare the numerical results of the current 1D FSI model with previous reported analytical results. At this stage, $\theta_t$ is fixed to be the corresponding values from  figure~\ref{fig:steady1}(\textit{e}). First, taking $\hat{Re}\to 0$ and neglecting $\theta_t$, the pressure distribution and the deformation at steady state \citep{WC21} are
\begin{equation}\label{steady-eq2}
    P_0(Z)=\frac{1}{\beta}\left\{ [48\beta(1-Z)+1]^{1/4}-1 \right\}, \qquad \bar{H}_0(Z) = 1+\beta P_0(Z).
\end{equation}
Equation \eqref{steady-eq2} is the essentially the same as the model proposed by \citet{GEGJ06}. However, in the current theoretical framework, $\beta$ is obtained by solving an appropriate linear elasticity problem, instead of being calibrated by an experiment. Second, if only $\theta_t$ is neglected but $\hat{Re} = \mathcal{O}(1)$, the steady-state pressure distribution \citep{WC21} is
\begin{equation}\label{steady-eq3}
    P_0(Z)\left[ 1+\frac{3}{2}\beta P_0(Z) + \beta^2P_0^2(Z) + \frac{1}{4}\beta^3 P_0^3(Z) - \frac{6}{5}\hat{Re}\beta\right]=12(1-Z), \qquad \bar{H}_0(Z) = 1+\beta P_0(Z).
\end{equation}
Observe that equation \eqref{steady-eq3} reduces to equation \eqref{steady-eq2} for $\hat{Re}\to 0$. Finally, we mention that if both $\theta_t,\hat{Re}\ne0$, equation \eqref{steady-eq1} is essentially a singular perturbation problem, which can be solved using the method of matched asymptotic expansions. Specifically, \citet{WC21} showed that there exists a boundary layer near $Z=0$ of thickness $\mathcal{O}\big(\theta_t^{1/2}\big)$, and obtained a matched asymptotic solution, which is too lengthy to summarize here.

In figure~\ref{fig:steady2}, we show a comparison between the steady-state solution obtained by numerical simulation of the 1D FSI model and the analytical results mentioned above. For a low flow rate, with $\hat{Re}=0.15$, the fluid inertia is not important, thus the numerical results agree well with the analytical results, except near the inlet. In contrast, for a high flow rate, with $\hat{Re}=0.9$, the results neglecting the effect of $\hat{Re}$ (based on equation \eqref{steady-eq2}) tend to underestimate the channel deformation and the pressure distribution. Since equations \eqref{steady-eq2} and \eqref{steady-eq3} do not take into account the weak tension effect, the no-displacement restriction at the inlet is not satisfied. With $\theta_t$ included, whether the flow rate is low or high, a short diverging section of the channel height emerges near the inlet, indicating the feature of a boundary layer problem, as described above.

\section{Linear stability of the inflated base state}
\label{sec:linear-stability}

In this section, we address the linear stability of the base (steady-state) solutions obtained in \S~\ref{sec:steady}. We have shown, in figure~\ref{fig:steady1}, that both the deformation and the pressure gradient are nonuniform along $z$ in the inflated (non-flat) base state. This observation makes the linear stability problem nontrivial, as the linearized operators have variable coefficients and are not self-adjoint. The key question that this section will address is: is the non-flat base state linearly stable to infinitesimal perturbations? 

To answer this question, we perturb the base state with an infinitesimal disturbance as:
\begin{align}\label{perturb1}
    Q(Z,T) = Q_0+\alpha \hat{Q}(Z,T), \qquad \bar{H}(Z,T)=\bar{H}_0(Z) + \alpha \hat{H}(Z,T),
\end{align}
with $\alpha\ll 1$. Note $Q_0=1$ for fixed flux upstream. Substituting the above into the governing equations \eqref{gen1d-flow-eq} and \eqref{1d-solid-eq}, and keeping terms up to $\mathcal{O}(\alpha)$, we obtain the following linear evolution equations:
\begin{subequations}\label{linearstab-eq}
    \begin{align}
        \frac{\partial \hat{H}}{\partial T} +\frac{\partial \hat{Q}}{\partial Z} &= 0,\label{linearstab-eq1}\\
        \frac{\hat{Re}\beta}{\bar{H}_0}\frac{\partial\hat{Q}}{\partial T} + \frac{6}{5}\hat{Re}\beta\left[\left(\frac{3Q_0^2}{\bar{H}_0^4}\frac{\rd \bar{H}_0}{\rd Z}-\frac{Q_0^2}{\bar{H}_0^3}\frac{\partial}{\partial Z}\right) \hat{H} +\left(-\frac{2Q_0}{\bar{H}_0^3}\frac{\rd \bar{H}_0}{\rd Z}+\frac{2Q_0}{\bar{H}_0^2}\frac{\partial}{\partial Z}\right)\hat{Q}\right] \qquad\qquad \nonumber\\
        +~\theta_I\frac{\partial^3 \hat{H}}{\partial Z\partial T^2} +\frac{\partial \hat{H}}{\partial Z}-\theta_t\frac{\partial^3 \hat{H}}{\partial Z^3} + 12\beta\left(\frac{\hat{Q}}{\bar{H}_0^3}-\frac{3Q_0}{\bar{H}_0^4}\hat{H}\right) &= 0.\label{linearstab-eq2}
    \end{align}
\end{subequations}
We further note that $\theta_t$ in the above equation is fixed to be the steady-state value rather than computed from equation \eqref{tension_coeff2}, \textit{i.e.}, we have neglected any modifications of $\theta_t$ introduced by the initial perturbations. It can be shown (see \S~\ref{subsec:dynamic1}) that this effect is negligible. 

In this work, we only consider the asymptotic behavior of infinitesimal initial perturbations, \textit{i.e.}, the modal analysis. To this end, we write
\begin{align}\label{perturb2}
    \hat{Q}(Z,T) =\widetilde{Q}(Z)e^{-\ri \omega_G T}, \qquad \hat{H}(Z,T)=\widetilde{H}(Z)e^{-\ri \omega_G T}.
\end{align}
Since the base state is non-flat, the eigenfunctions $\widetilde{Q}$ and $\widetilde{H}$ are not homogeneous in $Z$. Then, $\omega_G\in\mathbb{C}$ denotes the ``global'' growth/decay rate of the eigenmode \citep{HM90}.

For computational convenience, equation \eqref{linearstab-eq} is rewritten in matrix form as
\begin{equation}\label{linearstab-mat}
\underbrace{ \begin{pmatrix}
    0 & \frac{\rd}{\rd Z} \\[6pt] \mathscr{L}_H & \mathscr{L}_Q
    \end{pmatrix}}_{\boldsymbol{A}} \underbrace{\begin{pmatrix} \widetilde{H} \\[6pt] \widetilde{Q}\end{pmatrix}}_{\boldsymbol{\psi}}
    = \ri \omega_G \underbrace{ \begin{pmatrix} 1 & 0\\[6pt] 0 &  \frac{\hat{Re}\beta}{\bar{H}_0}-\theta_I\frac{\rd^2}{\rd Z^2}
\end{pmatrix}}_{\boldsymbol{B}} \underbrace{\begin{pmatrix} \widetilde{H} \\[6pt] \widetilde{Q}\end{pmatrix}}_{\boldsymbol{\psi}},
\end{equation}
with the operators $\mathscr{L}_H$ and $\mathscr{L}_Q$ defined as
\begin{subequations}
\begin{align}
\mathscr{L}_H &= \frac{6}{5}\hat{Re}\beta\left(\frac{3Q_0^2}{\bar{H}_0^4}\frac{\rd \bar{H}_0}{\rd Z}-\frac{Q_0^2}{\bar{H}_0^3}\frac{\rd}{\rd Z}\right)+\frac{\rd}{\rd Z}-\theta_t\frac{\rd^3 }{\rd Z^3} -\frac{36\beta Q_0}{\bar{H}_0^4}
,\\
\mathscr{L}_Q &= \frac{6}{5}\hat{Re}\beta\left(-\frac{2Q_0}{\bar{H}_0^3}\frac{\rd \bar{H}_0}{\rd Z}+\frac{2Q_0}{\bar{H}_0^2}\frac{\rd}{\rd Z}\right) + \frac{12\beta}{\bar{H}_0^3}.
\end{align}
\end{subequations}
Note that $\mathscr{L}_H$ and $\mathscr{L}_Q$ are linear operators with \emph{non-constant} coefficients, as a consequence of the non-flat base state. Also note in $\bm{B}$, $\theta_I \rd^2\widetilde{Q}/\rd Z^2$ originates from  $\theta_I\partial^3 \hat{H}/\partial Z\partial T^2 = \theta_I \partial^3 \hat{Q}/\partial Z^2\partial T$, using equation \eqref{linearstab-eq1}. 

Since the base state has satisfied all the boundary conditions from equations \eqref{bc1} and \eqref{bc2}. The boundary conditions for the infinitesimal perturbations are homogeneous:
\begin{subequations}\label{linearstab-bc}
\begin{align}
    \widetilde{Q}|_{Z=0}=\left.\frac{\rd \widetilde{Q}}{\rd Z}\right|_{Z=0}=\left.\frac{\rd \widetilde{Q}}{\rd Z}\right|_{Z=1}=0,\\
    \widetilde{H}|_{Z=0}=\widetilde{H}|_{Z=1}=\left.\frac{\rd^2 \widetilde{H}}{\rd Z^2}\right|_{Z=1}=0.
\end{align}
\end{subequations}                    
The first boundary condition is deduced from the fixed flux upstream boundary condition, while the boundary conditions in terms of $\widetilde{H}$ correspond to the no-displacement restrictions at both ends and the outlet pressure set to gauge. The remaining two boundary conditions on $\widetilde{Q}$ are derived from the equation \eqref{gen1d-cont-eq} by imposing zero displacement at the channel inlet and outlet. 

Equation \eqref{linearstab-mat} subject to equation \eqref{linearstab-bc} gives rise to a generalized eigenvalue problem, which can be solved numerically using the Chebyshev pseudospectral method (see \SM\ for details). For all the eigenmodes resolved, if the corresponding $\mathrm{Im}(\omega_G)>0$, we say the non-flat base state of the 1D FSI system is linearly unstable to infinitesimal disturbances.

\subsection{Eigenspectra of the exemplar cases}
\label{subsec:eig}

Here, it is illustrative to plot \emph{dimensional} quantities to show how a corresponding physical system would behave. To this end, we write the dimensional frequency as
\begin{equation}\label{fg}
    f_g = \frac{\omega_G}{2\pi \mathcal{T}_f} = \frac{\omega_G q}{2\pi \ell w h_0}.
\end{equation}
Note that $f_g\in\mathbb{C}$, such that $\mathrm{Re}(f_g)$ is the oscillatory frequency of the corresponding eigenmode, \textit{i.e.}, $[\widetilde{H}, \widetilde{Q}]^T$ in our formulation, while $\mathrm{Im}(f_g)$ is the eigenmode's growth/decay time constant.

On the other hand, since equation \eqref{1d-solid-eq} essentially represents a mechanical oscillator, the dimensionless and dimensional natural frequency of the oscillator, denoted as $F_N$ and $f_n$ respectively, can be approximated as 
\begin{equation}\label{fn}
    F_N \approx \frac{1}{2\pi \sqrt{\theta_I}}, \qquad f_n \approx \frac{F_N}{\mathcal{T}_f} = \frac{F_N q}{\ell w h_0}.
\end{equation}
Unlike $f_g$, both $F_N,f_n\in\mathbb{R}$. Also note that, equation \eqref{fn} does not take the tension effect into account, but since tension is weak, we believe equation \eqref{fn} provides a good approximation to the natural frequency of the system.

\begin{figure}[ht!]
    \centering
    \includegraphics[width=\textwidth]{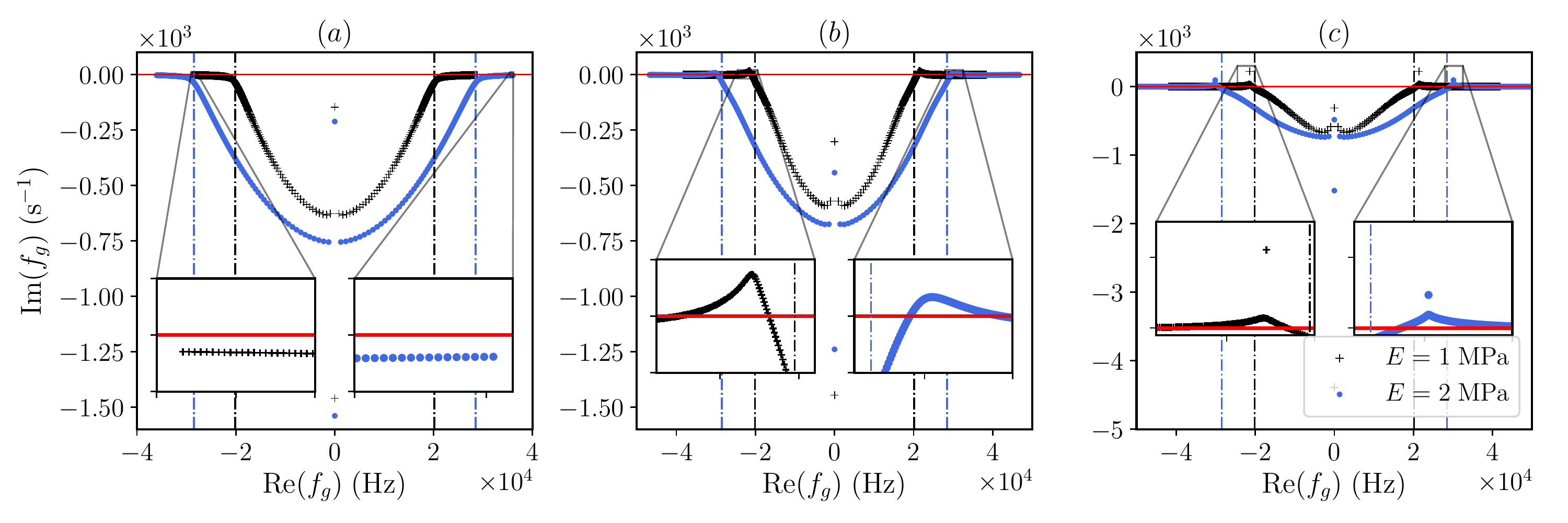}
    \caption{Eigenspectra of the exemplar cases from table~\ref{table:case-param}: (\textit{a}) $q=1500$ \si{\micro\litre\per\minute} {{(C1 and C4)}}; (\textit{b}) $q=6000$ \si{\micro\litre\per\minute} {{(C2 and C5)}}; (\textit{c}) $q=9000$ \si{\micro\litre\per\minute} {{(C3 and C6)}}. The symbols represent the discrete eigenvalues for $E=1$ \si{\mega\pascal} and $E=2$ \si{\mega\pascal}, respectively. The red horizontal line mark the position of real axis, \textit{i.e.}, $\mathrm{Im}(f_g)=0$. The dash-dotted lines mark the natural frequencies calculated by equation \eqref{fn}. The insets in each panel have the same vertical axes.}
    \label{fig:eig}
\end{figure}

\begin{figure}[ht!]
    \centering
    \begin{subfigure}[b]{0.49\textwidth}
        \caption{}
        \includegraphics[width=\textwidth]{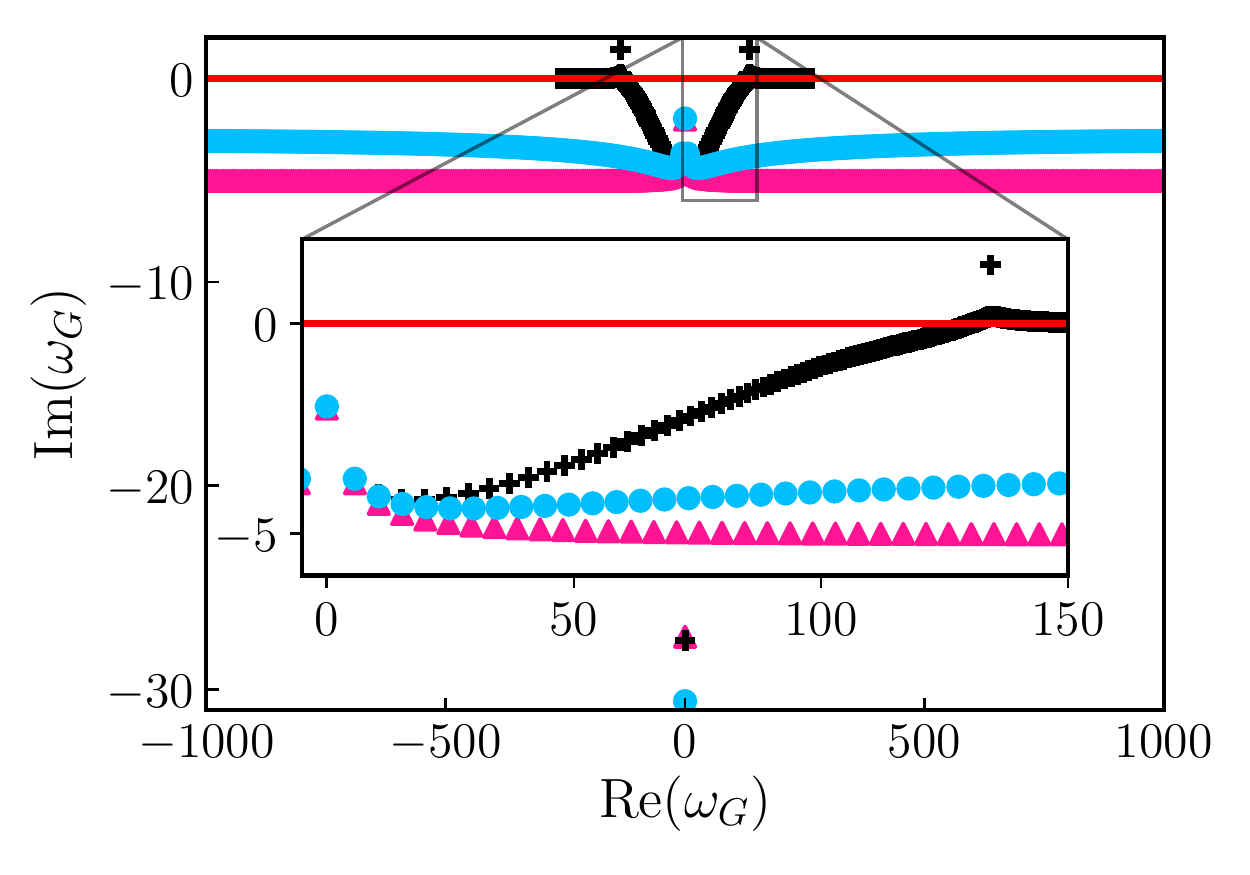}
    \end{subfigure}
    \hfill
    \begin{subfigure}[b]{0.49\textwidth}
        \caption{}
        \includegraphics[width=\textwidth]{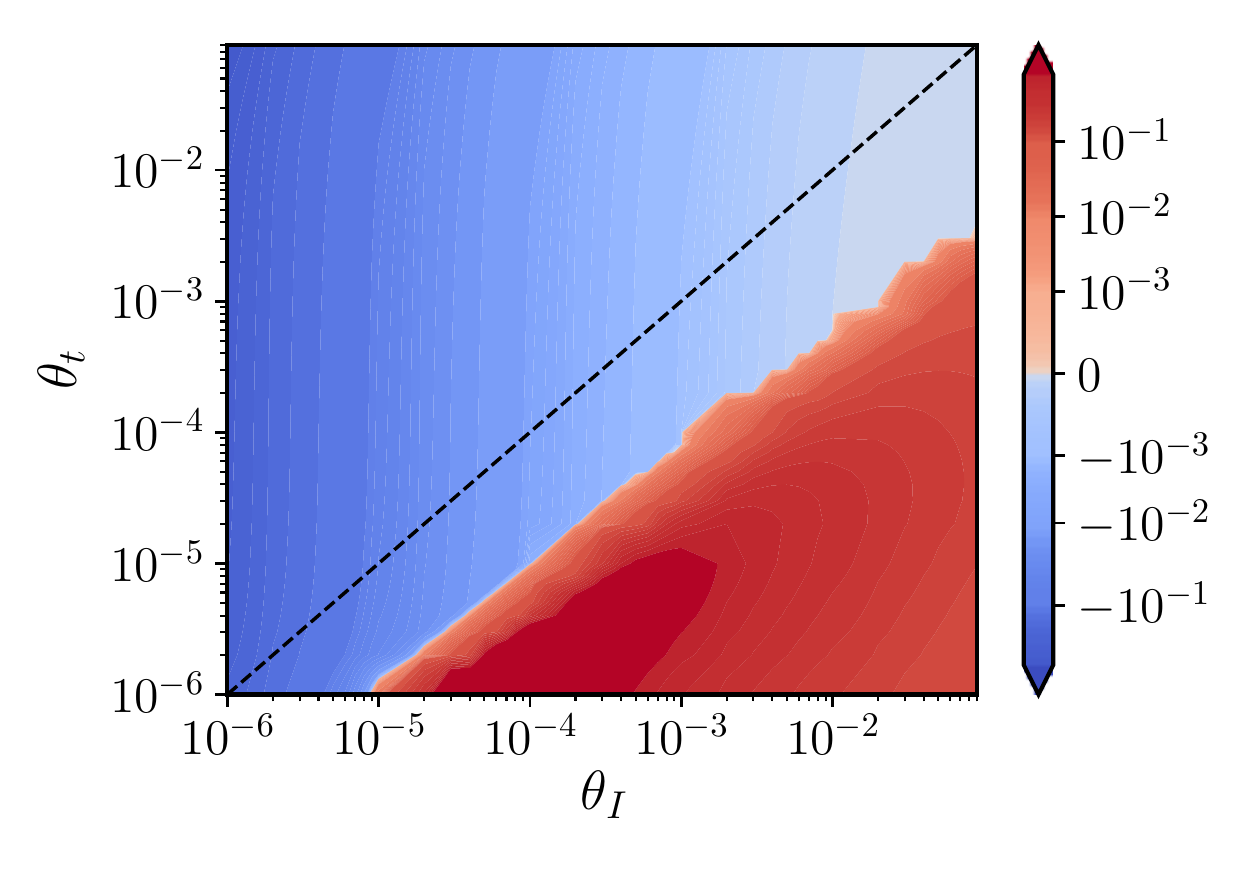}
    \end{subfigure}
    \caption{(\textit{a}) Comparison of the eigenspectra of the case with $\theta_I\ll 1$ and $\theta_t\ll 1$ (\texttt{+}), the case with $\theta_t\ll 1$ but $\theta_I=0$ (\mycirc), and the case with $\theta_I=0$ and $\theta_t=0$ (\mytriang). The dimensionless parameters are taken as per case C3 in table~\ref{table:case-param}. (\textit{b}) Contour plot of $\mathrm{Im}(\omega_G)$ of the least stable mode as a function of $\theta_I$ and $\theta_t$, with $\hat{Re} $ and $\beta$ taken from case C3 in table~\ref{table:case-param}. The dashed line marks $\theta_I = \theta_t$. }
    \label{fig:q9000-eig}
\end{figure}

In figure~\ref{fig:eig}, we show the calculated eigenspectra for the six exemplar cases from table~\ref{table:case-param}. First, observe that the eigenspectra are symmetric about the imaginary axis in the complex plane. This symmetry is a consequence of the formulation of the generalized eigenvalue problem (see equation \eqref{linearstab-mat}), as the matrix $\bm{A}$ on the left-hand side is purely real, while the matrix $\ri\bm{B}$ on the right-hand side is purely imaginary. This symmetry is also a feature of the eigenvalue analyses in \citep{IWC20, WC20}.

Second, the 1D FSI system transitions from stability (panel (\textit{a})) to instability (panels (\textit{b}) and (\textit{c})) as the flow rate is increased. The fluid inertia becomes more prominent as the flow rate is increased, indicated by the magnitude of $\hat{Re}$, which is associated with the nonlinear terms on the left-hand side of equation \eqref{gen1d-mom-eq}. For the six exemplar cases from table~\ref{table:case-param}, we observe that the linear instability typically occurs for $\hat{Re}\simeq 1$, or equivalently $Re\simeq 100$, which is close to the values reported for the microchannel experiments showing instability \citep{VK13}.

Third, it is observed that the unstable regions in figure~\ref{fig:eig}(\textit{b}) and (\textit{c}) are close to the natural frequency, indicating that the instabilities are related to the resonance of the wall. Due to the weak solid inertia, the natural frequency calculated from equation \eqref{fn}, as well as $\mathrm{Re}(f_g)$ in the unstable region, is as high as $\simeq 10^4~\si{\hertz}$. This fact also matches the local linear stability analysis of \citet{VK13}, who reported that the least stable mode oscillates at a frequency on the order of $10^4~\si{\hertz}$. Moreover, one of the common features of the eigenspectra in figure~\ref{fig:eig} is that, {away from the unstable region $\mathrm{Im}(f_g)\to 0^{-}$ as $|\mathrm{Re}(f_g)|$ becomes large.} This observation indicates that, apart from those modes that grow for the given flow conditions, the stable modes that oscillate with higher frequency decay slowly, which highlights the \emph{computational} stiffness of the 1D FSI system.

Lastly, the compliance of the wall does influence the shape of the eigenspectra. Since the stiffer wall has a larger natural frequency, its unstable region is located at higher frequencies than its softer counterpart. Note that in figure~\ref{fig:eig}, the two zoom-in insets in each panel have the same range for the vertical axis. Then it can be shown that the more compliant wall has the larger growth rate (or the smaller decay rate in the linearly stable case) for the least stable mode, which could be related to the experimental observations that the softer microchannel is more prone to instabilities \citep{VK13}. 

Since both $\theta_I$ and $\theta_t$ are small quantities in the solid mechanics equation \eqref{1d-solid-eq}, it is informative to show a comparison of the linear stability between cases either with $\theta_I=0$ or $\theta_t=0$. We consider three different such cases: (\textit{i}) $\theta_I\ll 1$ and $\theta_t\ll 1$; (\textit{ii}) $\theta_I = 0$ and $\theta_t\ll 1$; (\textit{iii}) $\theta_I=0$ and $\theta_t=0$. The base states of both case (\textit{i}) and (\textit{ii}) are the same, governed by equation \eqref{steady-eq1}. However, in case (\textit{iii}), the system cannot satisfy $\bar{H}(0)=0$, thus the corresponding base state should be taken from equation \eqref{steady-eq3}. As shown in figure~\ref{fig:q9000-eig}(\textit{a}), we observe that even though the solid inertia and tension are weak in the system, the inclusion of these weak effects changes the eigenspectrum fundamentally. The non-flat base state from equation \eqref{steady-eq3} is shown to be linearly stable. With only $\theta_t\ne0$, case (\textit{ii}) can predict linear stability only while case (\textit{i}) with both $\theta_I,\theta_t\ne0$ is linearly unstable. Furthermore, as shown in both case (\textit{ii}) and (\textit{iii}), for $\theta_I=0$, the eigenmodes oscillate with higher frequencies. The reason is that, in this case, the natural frequency $F_N \to \infty$ as $\theta_I\to 0$.

To further investigate the effect of $\theta_I$ and $\theta_t$, we calculated the growth/decay rate of the least stable mode by taking different combination of $\theta_I$ and $\theta_t$, fixing $\hat{Re}$ and $\beta$ as the values corresponding to case C3. As shown in panel (\textit{b}) of figure~\ref{fig:q9000-eig}, for both $\theta_I$ and $\theta_t$ across five orders, linear instabilities are only observed when $\theta_I$ is at least one order larger than $\theta_t$. Since the dimensionless phase speed of the transverse waves along the fluid--solid interface is $\sqrt{\theta_t/\theta_I}$, then the linear instability occurs when the transverse waves propagate (much) slower than the flow.

\begin{table}
    \centering
    \begin{tabular}{c c c  }
    \hline\hline
    Case & Pure decay mode & Least stable mode \\
    \hline
    C1  & $-5.5501\ri$ & -- \\
    C2  & $-2.8538\ri$ & $202.3535+0.1393\ri$ \\
    C3  & $-1.9762\ri$ & $134.2670+1.3892\ri$ \\
    C4  & $-8.0068\ri $ & --  \\
    C5  & $-4.1750\ri $ & $284.8114+0.0637\ri$  \\
    C6  & $-3.0423\ri$ & $189.2292+0.5867\ri$ \\
    \hline\hline
    \end{tabular}
    \caption{The dimensionless eigenvalues, $\omega_G$, for the pure decay modes and for the least stable modes of the exemplar cases from table~\ref{table:case-param}.}
    \label{table:case-eig}
\end{table}

\subsection{Eigenfunctions of the examplar cases}
\label{subsec:eigfunc}

\begin{figure}[p]
    \centering
    \includegraphics[width=0.99\textwidth]{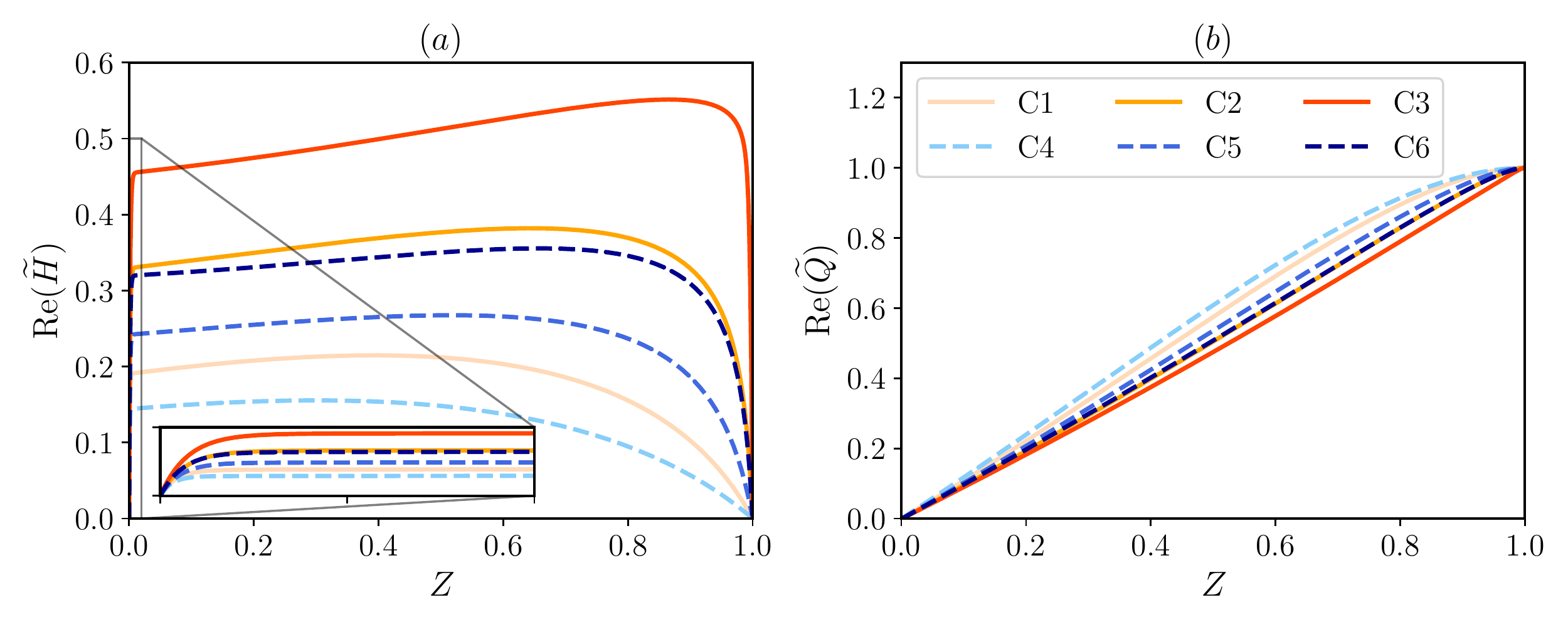}
    \caption{The eigenfunctions of the pure decay eigenmodes of the six exemplar cases from table~\ref{table:case-param}. The eigenfunctions have been scaled such that $\max|\widetilde{Q}|=1$.}
    \label{fig:eigfunc0}
\end{figure}

\begin{figure}[p]
    \centering
    \includegraphics[width=0.99\textwidth]{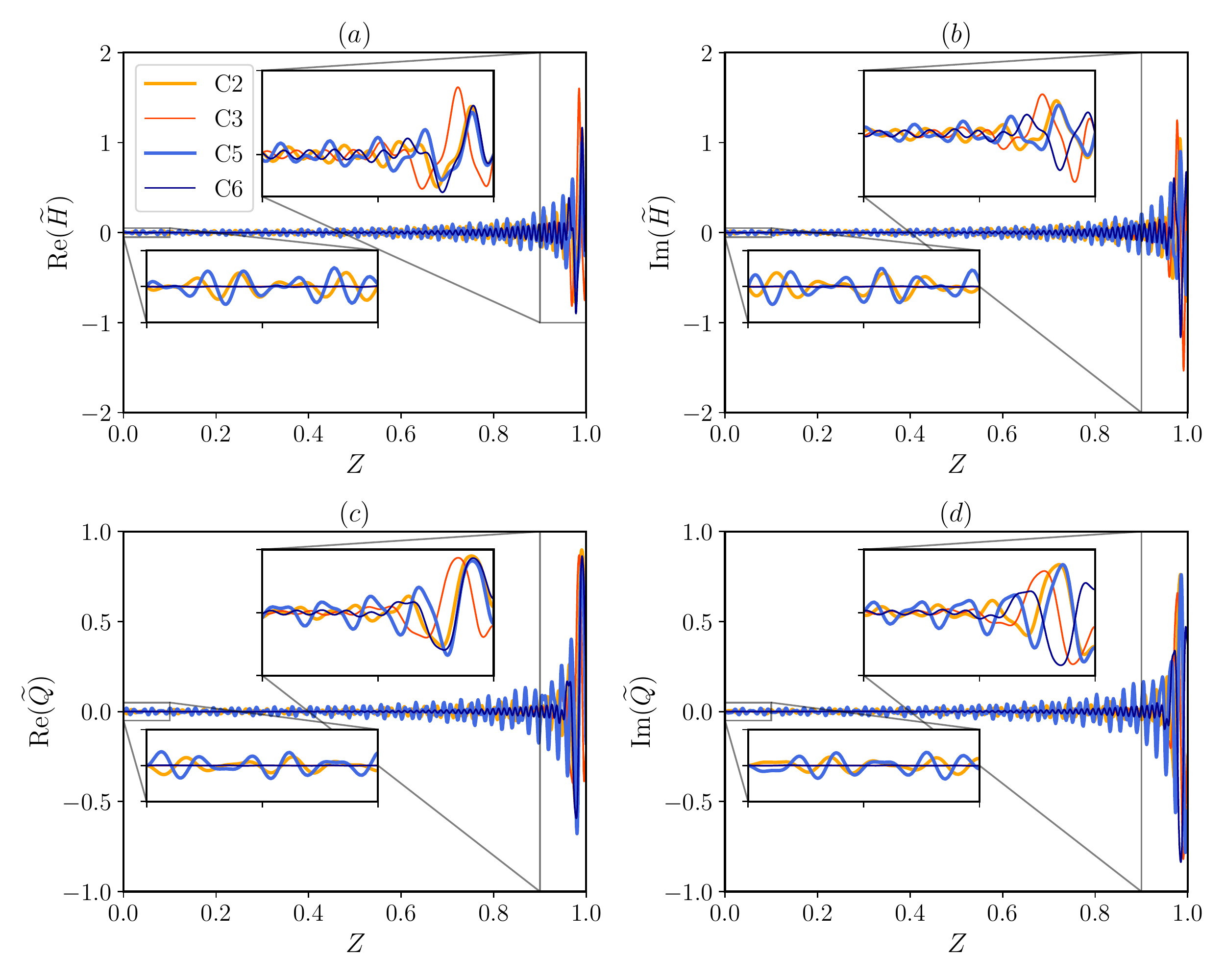}
    \caption{The eigenfunctions of the least stable modes for the linearly unstable cases, \textit{i.e.}, C2, C3, C5 and C6 in table~\ref{table:case-param}. The eigenfunctions have been scaled such that $\max|\widetilde{Q}|=1$.}
    \label{fig:eigfunc_least_stable}
\end{figure}

Each eigenvalue is associated with an eigenfunction pair, \textit{i.e.}, $[\widetilde{H}, \widetilde{Q}]^T$ via equation \eqref{linearstab-mat}. For the eigenvalues with larger $|\mathrm{Re}(\omega_G)|$, the corresponding eigenfunctions are more oscillatory in space. For example, for the purely decay mode with $\mathrm{Re}(\omega_G)=0$, the corresponding eigenfunctions are found to be purely real and non-oscillatory, as shown in figure~\ref{fig:eigfunc0}. However, for the least stable modes of the linearly unstable cases from table~\ref{table:case-param}, as shown in figure~\ref{fig:eigfunc_least_stable}, with $|\mathrm{Re}(\omega_G)|\gg 1$, the corresponding eigenfunctions are highly oscillatory in space. The corresponding eigenvalues for the eigenfunctions in figure~\ref{fig:eigfunc0} and figure~\ref{fig:eigfunc_least_stable} are tabulated in table~\ref{table:case-eig}. 

We do not tabulate the least stable modes of the linearly stable cases in table~\ref{table:case-eig} because it is hard to pick out the least stable mode due to the limitation of the numerical method. In that case, we observe that the least stable mode is always the farthest eigenvalue away from the imaginary axis, if computed with the Chebyshev pseudospectral method using different number of Gauss--Lobatto points $N$. In principle, there are infinite number of eigenvalues in the 1D FSI system, resolving all of which would require an infinite number of Gauss--Lobatto points. Unfortunately, $N$ cannot be arbitrarily large because matrices in equation \eqref{linearstab-mat} will become ill-conditioned.

Let us take a closer look at the eigenfunctions of the least stable modes in figure~\ref{fig:eigfunc_least_stable}. For all the cases, both $\widetilde{H}$ and $\widetilde{Q}$ exhibit large oscillations near the channel outlet ($Z=1$), which echoes the experimental observation that the instabilities were always first observed near the outlet in the converging section of the microchannel \citep{VK13}. Furthermore, for a larger growth rate, the difference in oscillations between the outlet and the inlet is more prominent (comparing C2 and C3, or C5 and C6).

\begin{figure}[ht]
    \centering
    \includegraphics[width=0.99\textwidth]{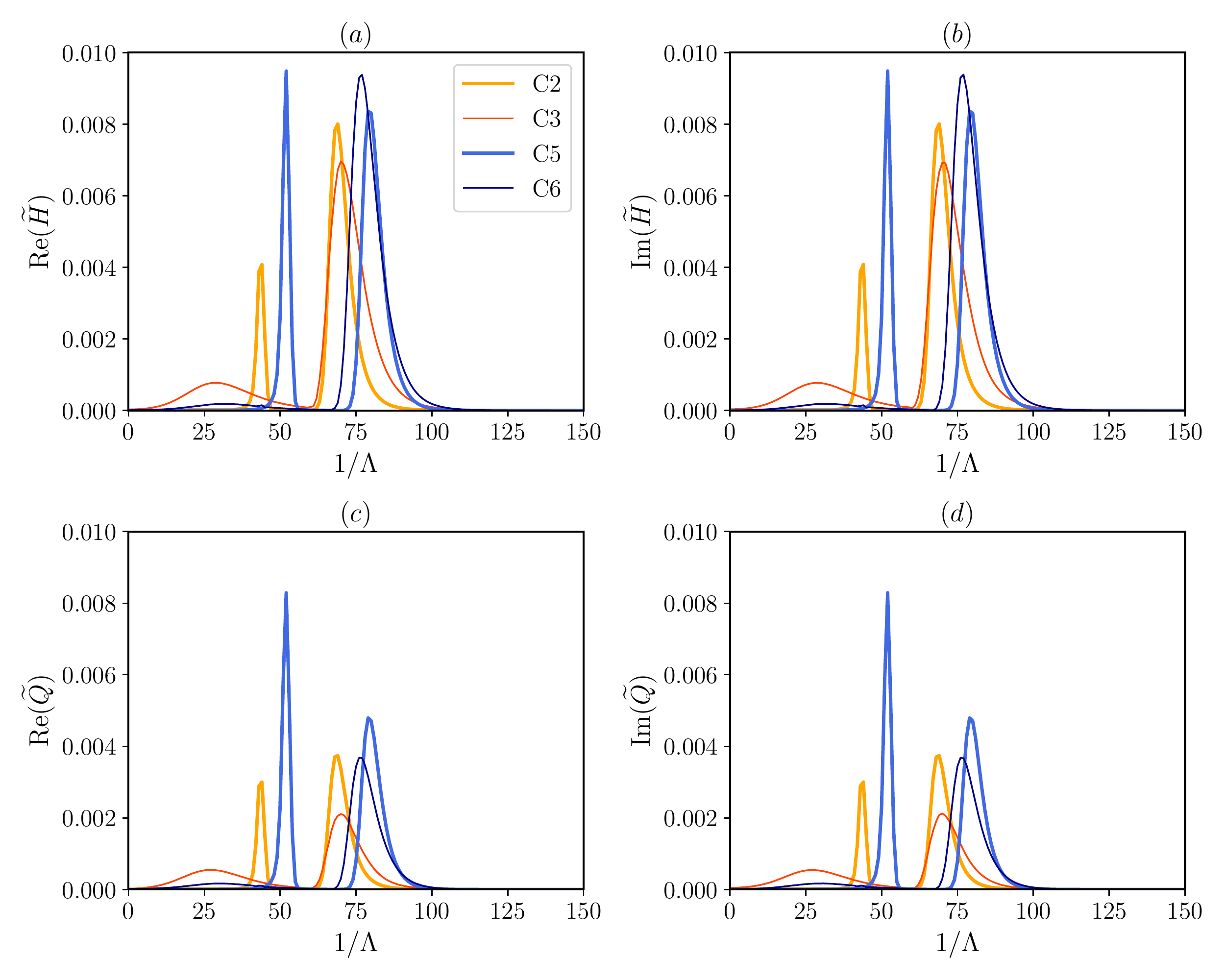}
    \caption{The spatial Fourier transform of the eigenfunctions from figure~\ref{fig:eigfunc_least_stable}. Case C3 is scaled up by a factor of $10$, while case C6 is scaled up a factor of $4$.}
    \label{fig:eigfunc_least_stable_fft}
\end{figure}

The wavy forms of the eigenfunctions in figure~\ref{fig:eigfunc_least_stable} further inspire us to conduct a Fourier transform in space for each case. We have used SciPy's {\tt fft} routine with a Blackman window. The results are summarized in figure~\ref{fig:eigfunc_least_stable_fft} and the abscissa represents the reciprocal of the dimensionless wavelength, denoted by $\Lambda=\lambda/\ell$. Here, $\lambda$ is the dimensional wavelength. Interestingly, there are always two peaks for all the cases. The major peak is located at $1/\Lambda\simeq 100$, meaning the dominant wavelength is on the order of the channel height (recall $\ell/h_0\approx 333$). This observation could be related to the results of the local linear stability analysis of \citet{VK13}, who found that the most unstable modes have a wavelength comparable to the channel height.

\section{Dynamic response of the microchannel}
\label{sec:dynamic}

In this section, we solve the 1D FSI model, by discretizing the governing equations in space and time, to investigate the dynamic responses. The spatial discretization is based on the Chebyshev pseudospectral method \citep{Boyd00, STW11}, while the ``Newmark-$\beta$'' method \citep{SD89} is used for the time integration; see \SM\ for further details about the numerical method and its benchmarking. 


\begin{figure}[t]
    \centering
    \begin{subfigure}[b]{0.49\textwidth}
        \caption{C1}
        \includegraphics[width=\textwidth]{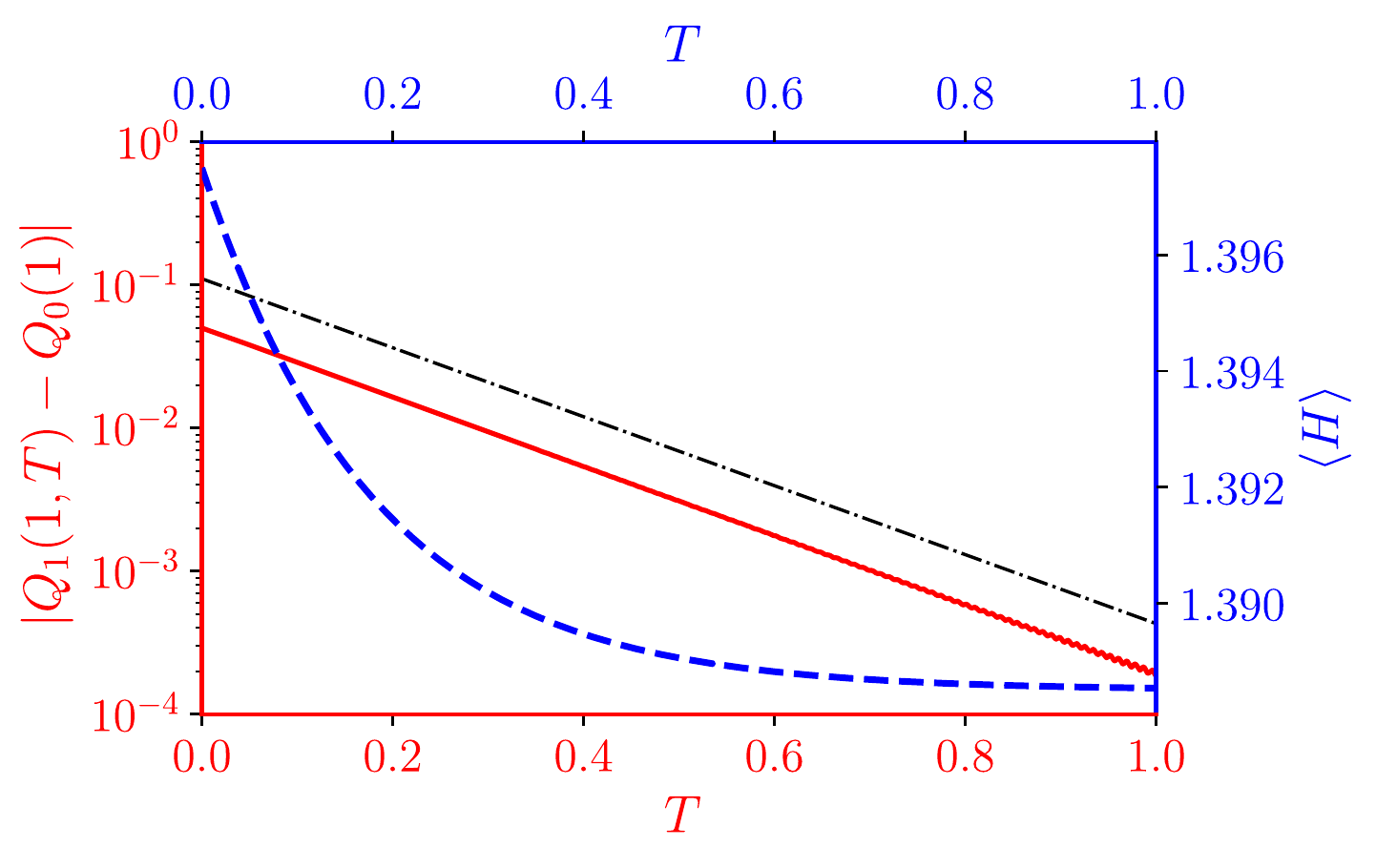}
    \end{subfigure}
    \begin{subfigure}[b]{0.49\textwidth}
        \caption{C3}
        \includegraphics[width=\textwidth]{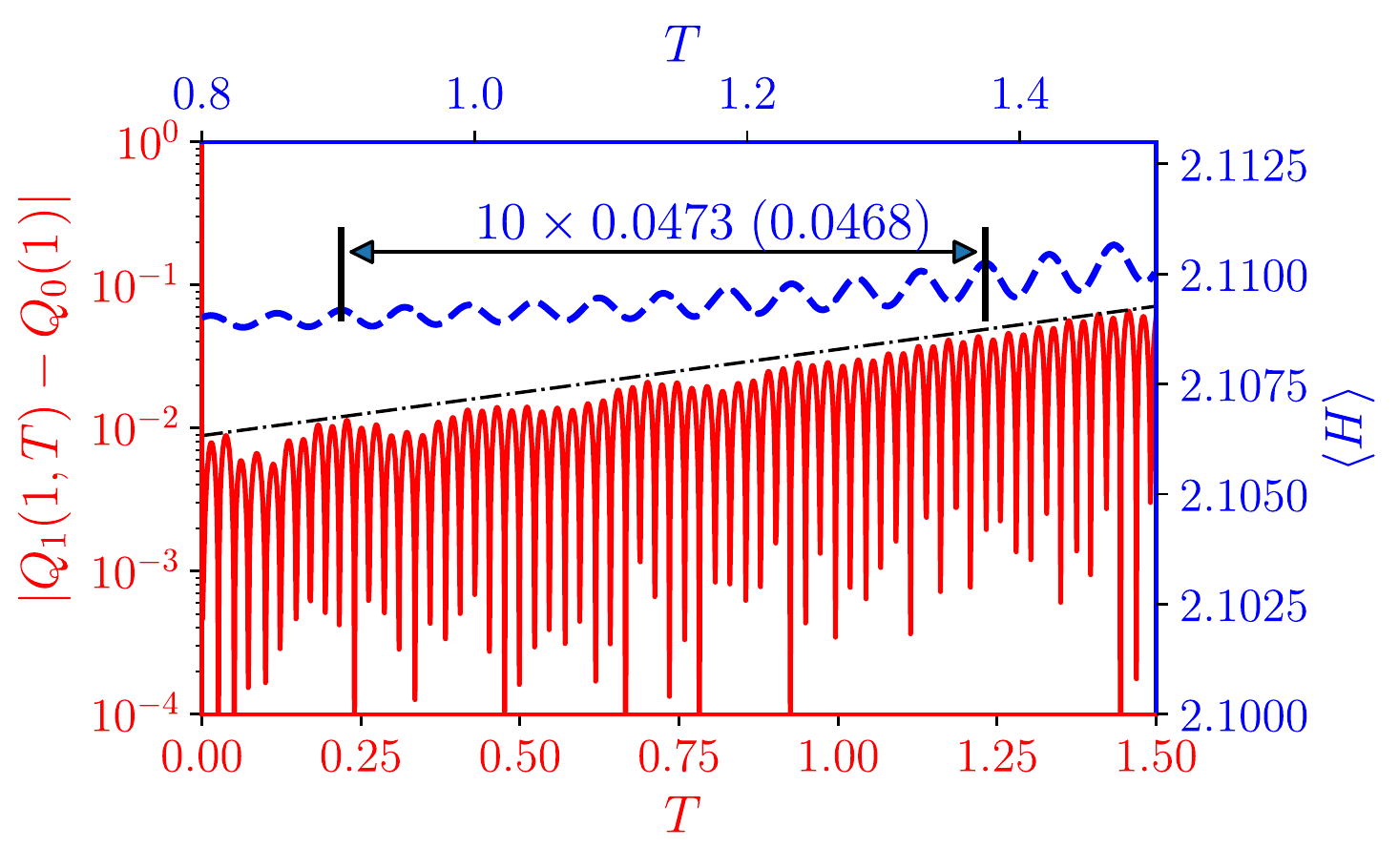}
    \end{subfigure}
    \caption{Time history of the difference between the instantaneous outlet flow rate and the steady one, \textit{i.e.}, $|Q(1,T)-Q_0(1)|$ (solid, left axes, also recall that $Q_0(Z)\equiv 1$ for the chosen boundary conditions), and the axially average deformed height, $\langle H\rangle(T) = \int_0^1 \bar{H}(Z,T)\, \rd Z$ (dashed, right axes). (\textit{a}) The steady state is perturbed using the eigenfunctions of the pure decay mode of case C1. (\textit{b}) The steady state is perturbed using the eigenfunctions of the most unstable mode of case C3. The dot-dashed trendlines represent the growth/decay of  perturbations, based on the imaginary part of the eigenvalues from table~\ref{table:case-eig}. In panel (\textit{b}), the computed period of oscillation is marked. The value in parenthesis is from linear stability analysis, \textit{i.e.}, $2\pi /\mathrm{Re}(\omega_G)$.} 
    \label{fig:simstab}
\end{figure}


\subsection{Evolution from a perturbed inflated state}
\label{subsec:dynamic1}

First, to validate the linear stability results from \S~\ref{sec:linear-stability}, at $T=0$, we perturb the steady-state solution of the 1D FSI model using the eigenfunction of a specific mode. Then, $\mathrm{Im}(\omega_G)$ indicates the growth/decay rate of the perturbation, while $2\pi /\mathrm{Re}(\omega_G)$ is the period of the perturbation's oscillations.

The first example is the linearly stable case C1 ($q=1500$ \si{\micro\litre\per\min}) perturbed from the steady state with the eigenfunctions of the pure decay mode tabulated in table \ref{table:case-eig}. The shapes of the corresponding eigenfunctions are shown in figure~\ref{fig:eigfunc0}. The decay rate of the perturbation to the steady flow agrees well with the corresponding $\mathrm{Im}(\omega_G)$, as shown in figure~\ref{fig:simstab}(\textit{a}). No oscillations are observed in the evolution  because $\mathrm{Re}(\omega_G)=0$ in this case. 

The second example corresponds to the linearly unstable case C3. At $T=0$, the steady state is perturbed by the eigenfunction of the corresponding most unstable mode. The shapes of the eigenfunctions are shown in figure~\ref{fig:eigfunc_least_stable}, while the corresponding eigenvalues are given in table~\ref{table:case-eig}. Both the simulated growth rate and the oscillation period agree well with the linear stability analysis, as shown in figure~\ref{fig:simstab}(\textit{b}). Here, note that the tension coefficient $\theta_t$ in equation \eqref{1d-solid-eq} is estimated from the instantaneous wall deformation, thus it is time dependent. Meanwhile $\theta_t$ is fixed to be the steady-state value for the purposes of the linear stability analysis. The good agreement between the numerical simulation and the linear stability analysis indicates that neglecting the time dependence of $\theta_t$ for the linear stability analysis is valid. The actual simulations for each case are conducted for a longer time window. Interestingly, although the oscillations of the system are sustained, no saturated periodic state emerges during the course of the simulations. The simulations for cases C3 and C5 are available in appendix~\ref{sec:app-a}.

\begin{figure}
    \centering
    \begin{subfigure}[b]{0.49\textwidth}
        \caption{C3}
        \includegraphics[width=\textwidth]{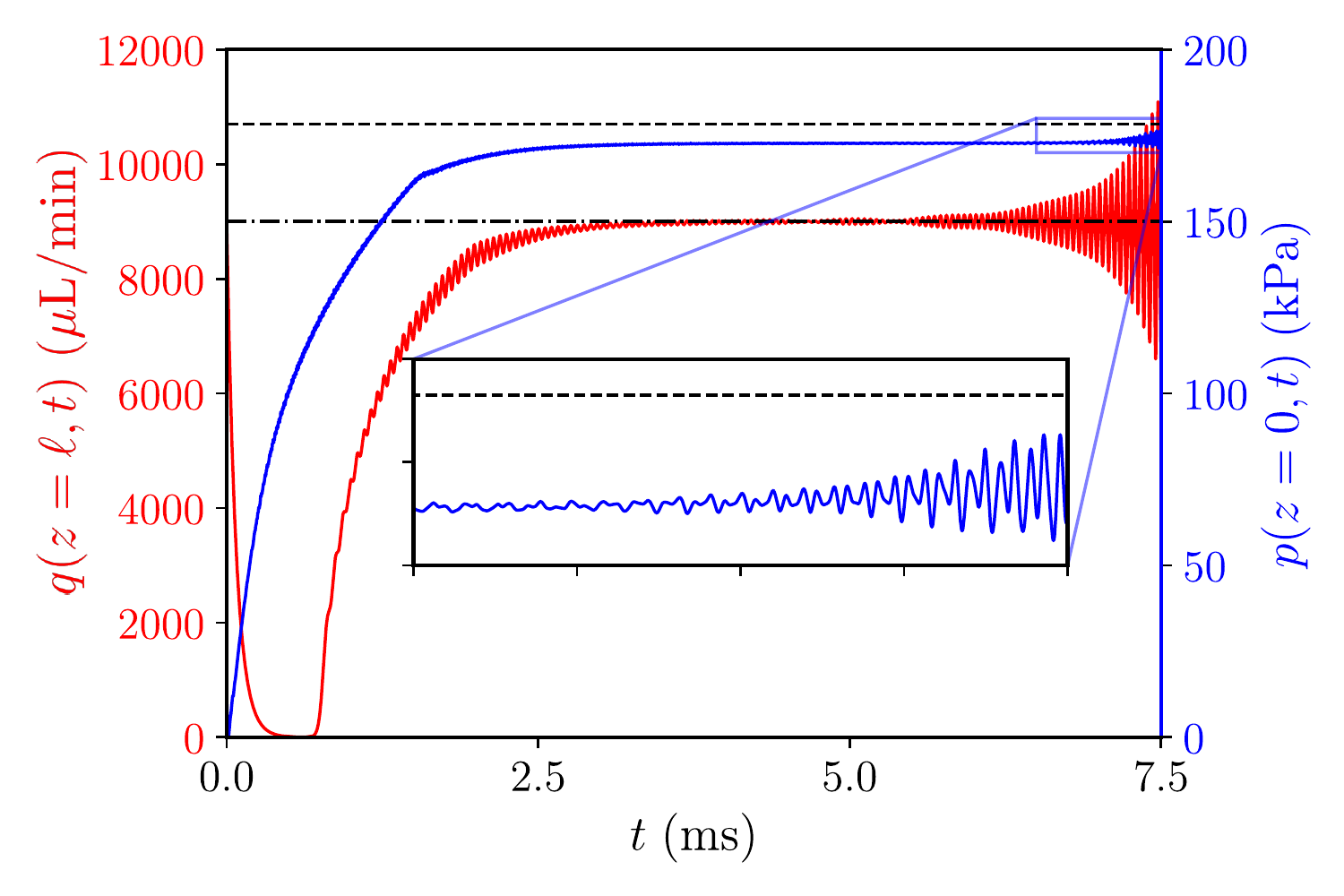}
    \end{subfigure}
    \begin{subfigure}[b]{0.49\textwidth}
        \caption{C6}
        \includegraphics[width=\textwidth]{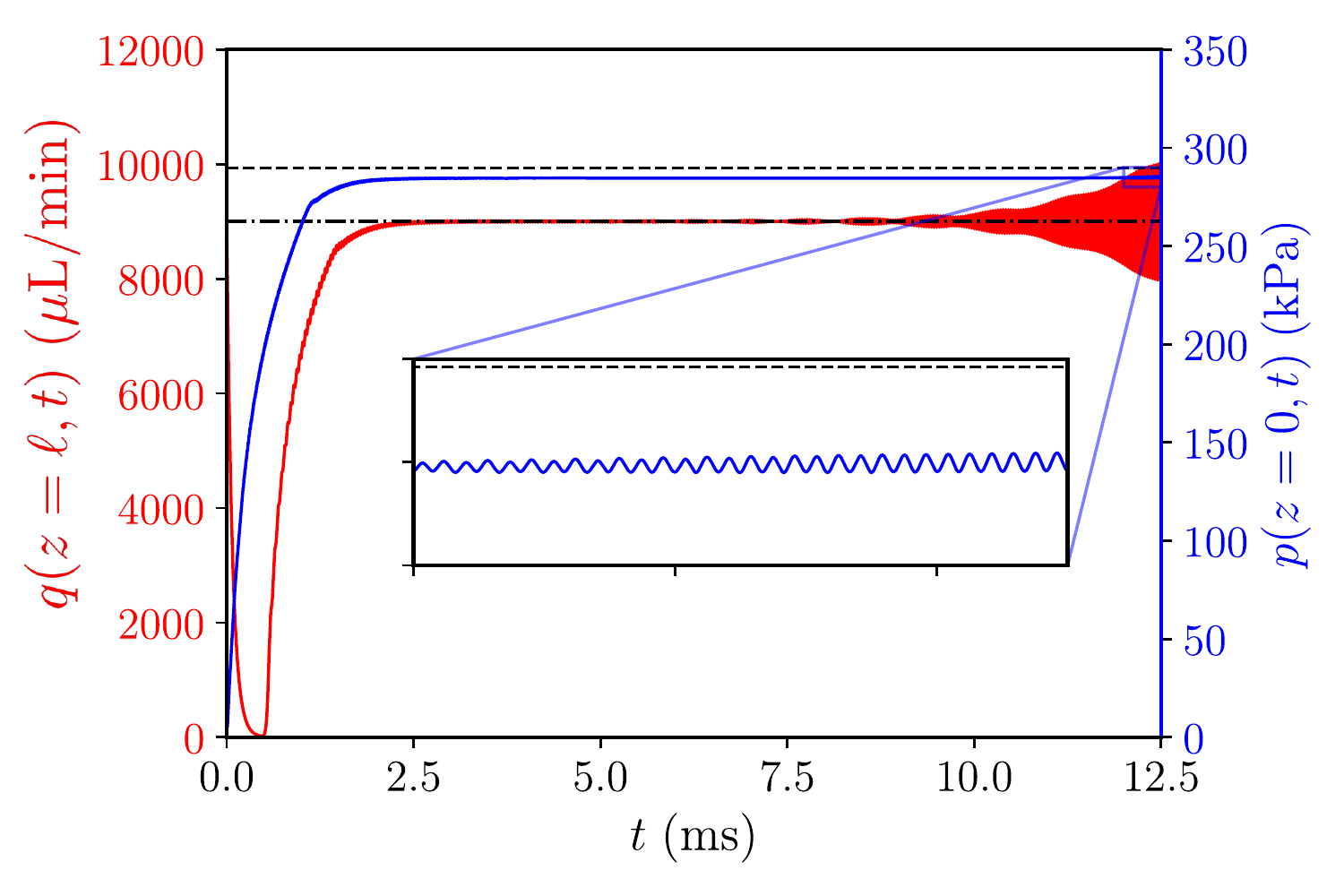}
    \end{subfigure}
    \caption{Time histories of the outlet flow rate and the inlet pressure for (\textit{a}) case C3 and (\textit{b}) case C6, respectively, with equation \eqref{ic} being the initial condition. The dot-dashed lines mark the flow rate at steady state, while the dashed lines mark the inlet pressure at the steady state.} 
    \label{fig:flatini-qp}
\end{figure}

\begin{figure}[ht!]
    \centering
    \begin{subfigure}[b]{\textwidth}
        \caption{}
        \centering
        \includegraphics[width=0.75\textwidth]{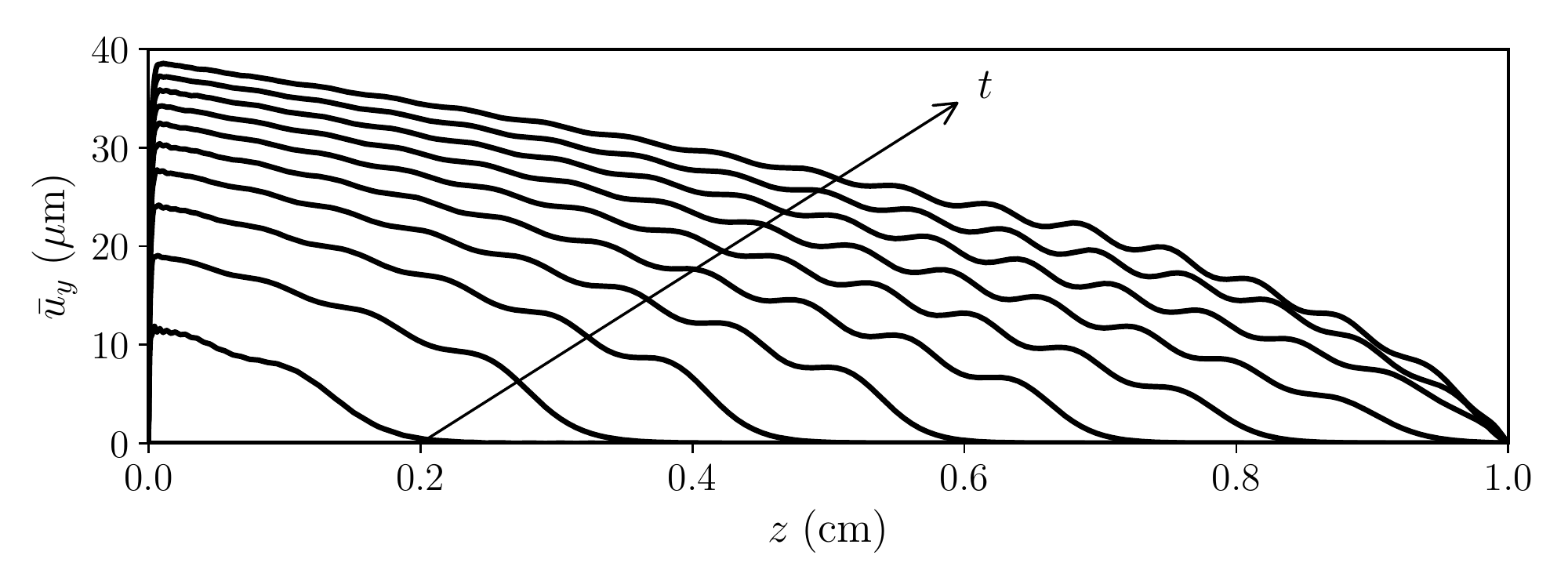}
    \end{subfigure}
    \hfill
    \begin{subfigure}[b]{\textwidth}
        \caption{}
        \centering
        \includegraphics[width=0.75\textwidth]{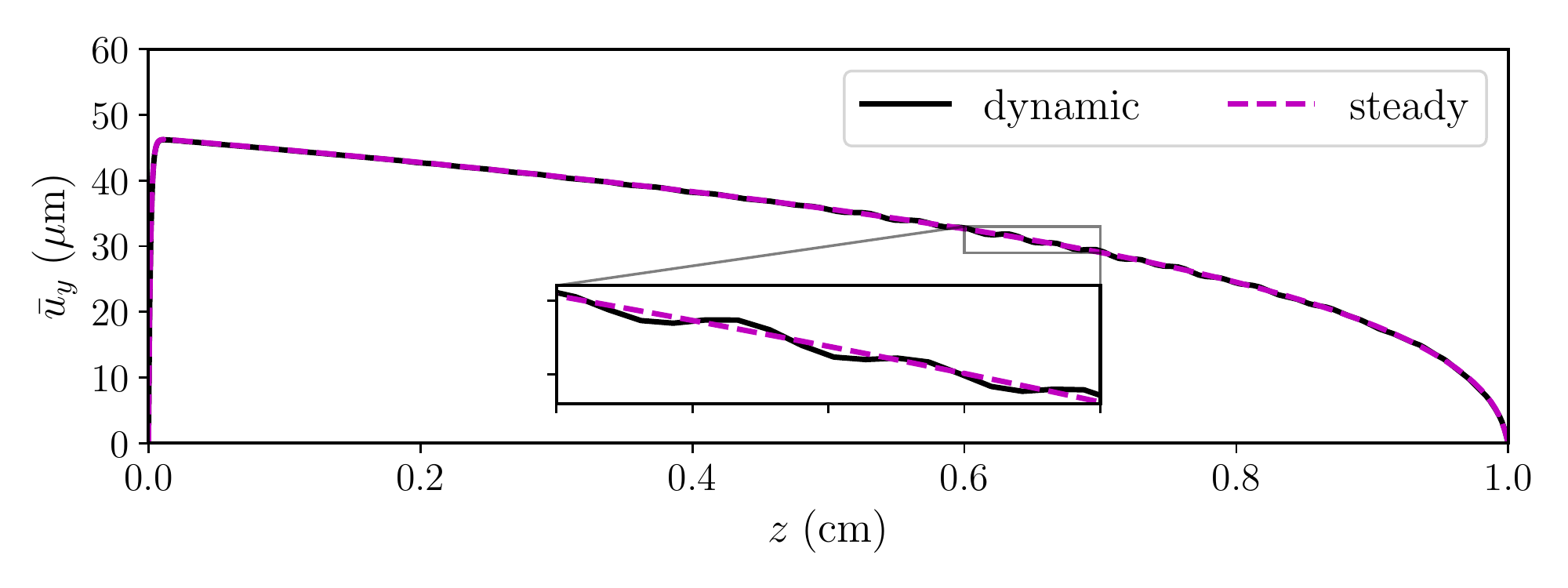}
    \end{subfigure}
    \hfill
    \begin{subfigure}[b]{\textwidth}
        \caption{}
        \centering
        \includegraphics[width=0.75\textwidth]{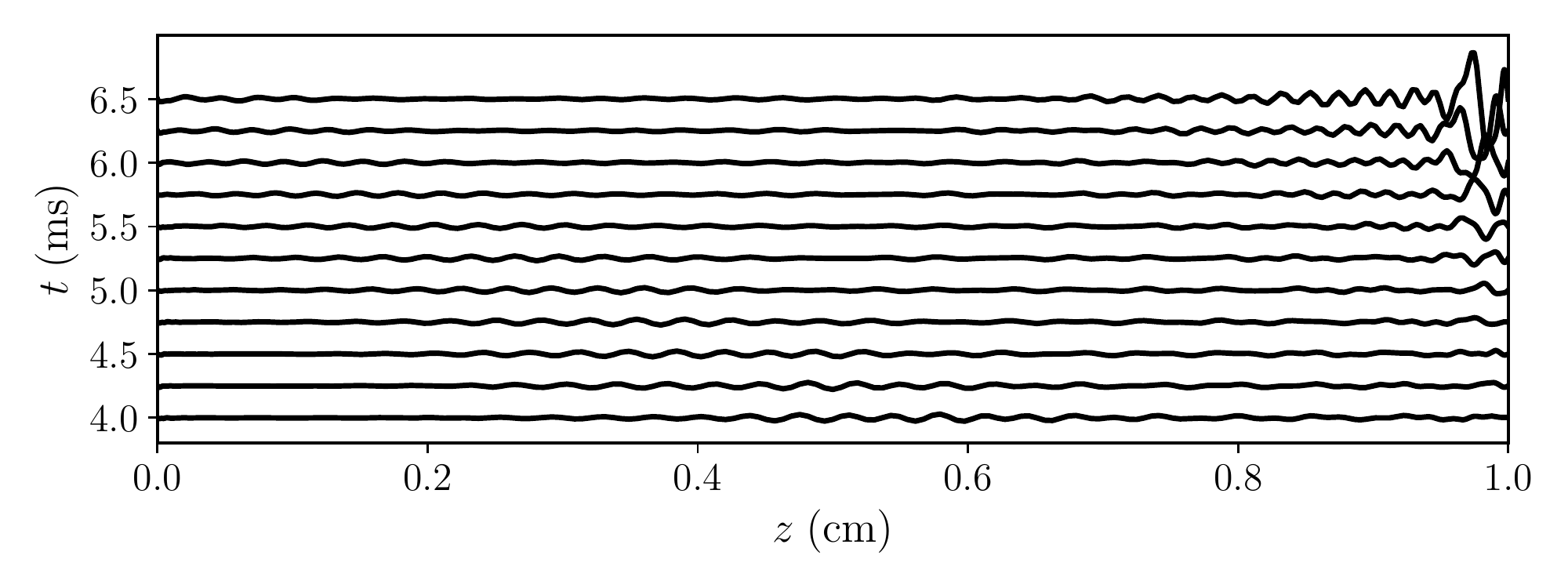}
    \end{subfigure}
    \caption{Evolution of the shape of the fluid--solid interface from the flat initial condition \eqref{ic} (movie available in \SM). (\textit{a}) Shape of the interface for $0 \leq t \leq 1$ \si{\milli\second}. (\textit{b}) Comparison of the interface shape at $t=3.5$ \si{\milli\second} with the steady state. (\textit{c}) Space-time plot of the difference between the instantaneous interface shape $\bar{u}_y$ and the steady state $\bar{u}_y^s$, \textit{i.e.}, $\bar{u}_y - \bar{u}_y^s$, for $4.0~\si{\milli\second}\leq t \leq 6.5~\si{\milli\second}$.}
    \label{fig:flatini-evolution}
\end{figure}

\subsection{Evolution from a flat initial state}
\label{subsec:dynamic2}

Starting the simulations with an undeformed channel initial condition (as in equation \eqref{ic}) would be more realistic of how a microfluidic device might be operated. With equation \eqref{ic} as the initial condition, cases C1 and C4 are linearly stable and reach  the steady state without detectable oscillations. The evolution of the representative quantities for case C4 are shown in figure~\ref{fig:grid-independence} in \SM, wherein a video of the evolution under case C1 is also provided. In this subsection, we focus on the two linearly unstable cases C3 and C6. All of the results shown below have been verified by time step refinement (see \SM\ and figure~\ref{fig:sim-diffdt} therein, for example).

First, the evolution of the outlet flow rate and the inlet pressure for cases C3 and C6 are shown in figure~\ref{fig:flatini-qp}. In both cases, the outlet flow rate first decreases and then increases, reaching a value close to the imposed flow rate at late times. Meanwhile, the inlet pressure increases to a value slightly below the steady-state inlet pressure. Small-amplitude oscillations are observed in the evolution of both quantities. More importantly, the oscillations become magnified at later times in the simulation, which suggests that these unstable cases will not reach the steady state. It can be shown (by running longer-time simulations) the that these oscillations are not unbounded. Nevertheless, similar to the cases discussed in \S~\ref{subsec:dynamic1}, no saturated periodic state appears to emerge during the time window of the simulations shown in this subsection.

It is more enlightening to contrast the two simulations shown in figure~\ref{fig:flatini-qp}. For these two cases, all system parameters are the same, except that the Young's modulus for case C3 is half of that for case C6. With a more compliant wall, the instabilities under case C3 develop more quickly. Specifically, more ``violent'' oscillations are observed in case C3 for the outlet flow rate and the inlet pressure than in case C6. These oscillation amplitudes could be, qualitatively, representative of the observations in  dye-stream experiments. In other words, dye breakup could be expected when more violent oscillations occur in softer channels, while the dye steam may just oscillate (without breaking up) if the channel is less compliant (thus, the oscillations in the flow rate and pressure are milder). Indeed, it was observed in the experiments that the dye breaks-up at lower $Re$ in softer channels \citep{VK13}.

Next, let us take a closer look at the evolution of the fluid--solid interface. The example in figure~\ref{fig:flatini-evolution} corresponds to case C3 in figure~\ref{fig:flatini-qp} (\textit{a}), while the evolution under case C6 is qualitatively similar. Initially, for $0\leq t\leq 1~\si{\milli\second}$ as shown in  figure~\ref{fig:flatini-evolution}(\textit{a}), the interface bulges near the channel inlet first because the pressure is relatively high there. At the same time, transverse waves are shed and propagate from the inlet to the outlet until they are reflected at the downstream boundary of the domain. There is a dramatic increase in the total volume of fluid in the channel at this stage. Thereafter, the deformation of the wall stops growing, but the transverse waves still propagate back and forth along the fluid--solid interface. Compared with figure~\ref{fig:flatini-evolution}(\textit{a}), the transverse waves have smaller wavelength and amplitudes. Furthermore, the interface shape at this stage is close to the steady-state shape, as seen in figure~\ref{fig:flatini-evolution}(\textit{b}). The deviation of the interface's dynamic deformation from the steady one is plotted for $4.0~\si{\milli\second} \leq t\leq 6.5~\si{\milli\second}$ in figure~\ref{fig:flatini-evolution}(\textit{c}), where the wave propagation can be clearly observed. Interestingly, after a while, the oscillations near the channel's outlet continue to grow and become larger than the oscillations anywhere else along the channel, which explains why the variations of the outlet flow rate appear more prominent compared than those of the inlet pressure in figure~\ref{fig:flatini-qp}. This observation is also corroborated by the experimental observation that the instabilities always initiate near the channel's outlet.

\begin{figure}[t]
    \centering
    \includegraphics[width=\textwidth]{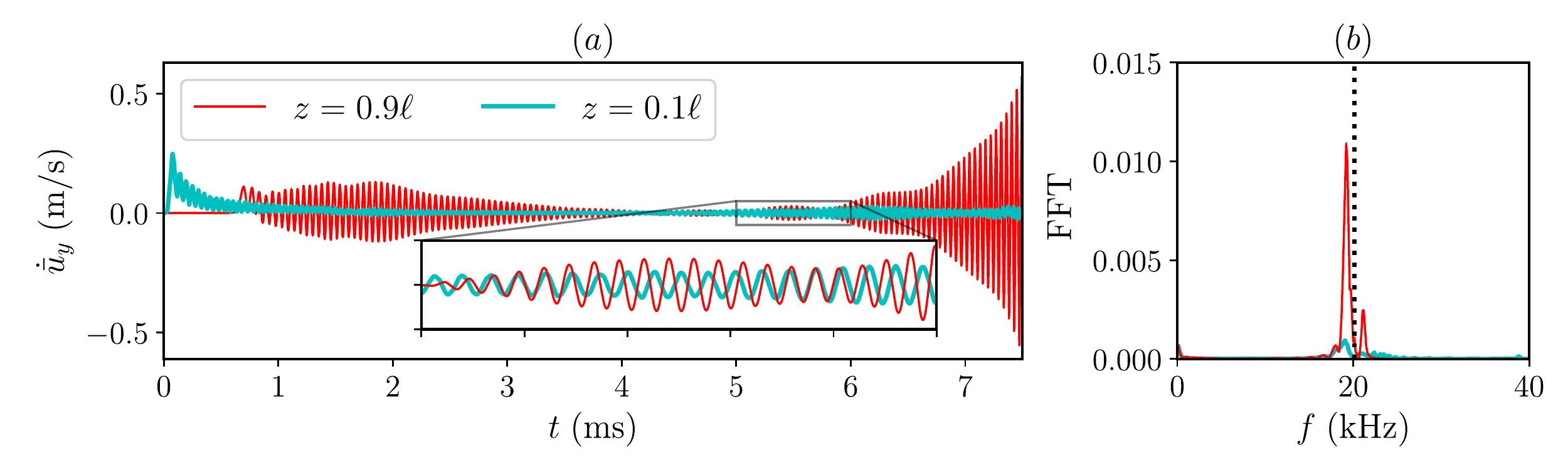}
    \caption{(\textit{a}) Time histories of the vertical velocity of the fluid--solid interface of case C3 at $z=0.9\ell$ (near the outlet) and $z=0.1\ell$ (near the inlet), respectively. (\textit{b}) Fourier transform of the corresponding time histories from (\textit{a}). The dotted line marks the natural frequency calculated from equation \eqref{fn}.}
    \label{fig:flatini_vel1}
\end{figure}

To better emphasize the difference in the wall motions near the inlet versus near the outlet, the vertical velocities of the fluid--solid interface at $z=0.9\ell$ (near the outlet) and $z=0.1\ell$ (near the inlet) of case C3 are plotted in figure~\ref{fig:flatini_vel1} (see \SM~ for the details of reconstructing the velocities). The three stages discussed in figure~\ref{fig:flatini-evolution} can also be identified from the time histories of the vertical velocities shown in figure~\ref{fig:flatini_vel1}. At early times, since the wall bulges first near the channel inlet, the vertical velocity at $z=0.1\ell$ is larger. The motion at $z=0.9\ell$ starts after the transverse waves reach the channel outlet. In the intermediate stage, during which the channel volume does not change significantly, the oscillations at both positions remain relatively small, until the motion near the outlet becomes amplified and leads to a striking difference in the oscillatory amplitudes at the two positions.

Figure~\ref{fig:flatini_vel1}(\textit{b}) shows the corresponding time histories in the frequency (Fourier) domain. We observe that the peak is near the natural frequency of the wall (predicted by equation \eqref{fn}), indicating a resonant phenomenon. Due to the FSI, which gives rise to the transverse waves along the fluid--solid interface, the pressure oscillations exhibit a frequency close to the natural frequency of the wall as well, causing a feedback. Note that, the resonances are \emph{self-excited} as no oscillatory components are introduced in the initial condition \eqref{ic}. Further, the oscillations are self-sustained as they do not die out during the entire simulation time window. Consequently, the demonstrated FSI-induced instabilities could be an effective and inexpensive way of enhancing mixing at the microscale.

\section{Conclusion}
\label{sec:discussion}

We derived a new 1D (reduced) FSI model for the physics underlying FSI-induced instabilities of flows conveyed in long and shallow microchannels with a deformable top wall.
The key advance in our 1D FSI model, compared to previous work, lies in the accurate modeling of the wall deformation due to two-way FSI. For collapsible tubes, a constant large tension is always included, though bending was also considered in some computational studies \citep{LCLP08, LLC12, WLS21}. Similarly, the 1D FSI model of \citet{IWC20} considered the top wall as a beam and took large-deformation-induced tension and bending into account. However, in a typical long and shallow rectangular microchannel, previous studies \citep{CCSS17, SC18, WC19} have demonstrated that, under linear elasticity, at the leading order, the soft wall deforms more like a Winkler foundation with a variable stiffness {(without assuming the Winkler model \textit{a priori} or suffering from its limitations)}. In other words, the deformation of the channel's cross-section at different streamwise locations is fully determined by the local pressure. In contrast to previous 1D models, the 1D FSI model proposed herein maintains the dominance of the Winkler-foundation-like behavior of the soft wall by modeling weak tension as a ``boundary effect'' near the inlet  of the channel. Moreover, the inertia of the solid is also modeled consistently, just like the inertia of the fluid was taken into account by lubrication theory at low, but finite, Reynolds number, extending the steady-state model in  \citep{WC21}. Our proposed 1D model establishes how the unsteady flow rate, the pressure and the channel deformation evolve together in a tightly coupled manner. 

Importantly, we found that the predictions of the proposed 1D FSI model agree qualitatively  with key experiments \citep{VK13} (summarized in table~\ref{tab:comparison}). Consequently, we believe that the present analysis leads to significant, novel insight into the experimentally observed low-$Re$ FSI-induced instabilities in compliant microchannels. In short, the physical insight provided by our new model is that FSI causes wall resonances, giving rise to self-sustained oscillations of the fluid--solid interface. {These resonances are triggered thanks to the combined effect of weak axial tension and finite solid inertia, which leads to fluctuations in the local pressure at frequencies close to the natural frequency of the wall.} Further, the experimentally observed dye breakup (and ``ultrafast'' mixing) are explained by the global instability of the non-flat (deformed) base state of our model, which was not accurately accounted for in previous work. Our proposed 1D FSI model allows for the identification (computationally) of the critical conditions for instability of this coupled system. The predicted critical Reynolds number is in agreement with the value suggested by experiments.

To the experimentalist, our proposed 1D FSI model provides a tool through which different microchannel designs can be rapidly prototyped and evaluated. Beyond that, our model provides a convenient way to evaluate operating conditions that might lead to instability and mixing. Extending the present results, the pressure drop could be prescribed across the channel (instead of fixing the flux at the inlet), similarly to the works of \citet{SWJ09, SHWJ10}. Further, the proposed 1D modeling framework can be easily used to analyze soft conduits of different cross-sectional geometries and other boundary conditions, as long as the basic assumptions on the separation of scales (and weak versus dominant effects in the solid) are not violated.

The current work was motivated by microfluidic experiments and aimed to provide new qualitative and quantitative physical insights into these phenomena. Nevertheless, further work is needed to understand the full range of dynamic behaviors possible under the proposed 1D FSI model. For example, in the linearly unstable case, the numerical simulations of the model using different time step sizes begin to diverge after a certain (long) integration time. This observation reminds us of the similarly chaotic behavior observed in a 1D FSI model derived by \citet{J92} in the somewhat different context of collapsible tubes. Similar to our 1D FSI model, Jensen's model also exhibits multiple unstable modes, and its dynamics may be sensitive to initial conditions (due to the interactions of multiple unstable modes). Therefore, understanding the nonlinear dynamics of the proposed 1D FSI model could be a fruitful avenue for future work. Further, since the observed oscillations are low-amplitude and high-frequency, asymptotic analysis could yield  the stability boundaries \citep{JH03}. On the other hand, an Orr--Sommerfeld-type local stability analysis (similarly to the Kumaran family of studies, recall \S~\ref{sec:intro}) could once again be conducted on the proposed model to complement to the global stability analysis. Investigating the connections between the local and global instabilities could provide insight regarding what types of excitations trigger unstable global modes \citep{SWJ09, SHWJ10}, opening the door towards more controllable ``ultrafast mixing'' due to FSIs in compliant microchannels.


\paragraph{Acknowledgements.}
We thank E.\ Boyko for feedback on an earlier draft.

\paragraph{Funding.}
This work originates from research partially supported by the US National Science Foundation (NSF) under grant No.\ CBET-1705637. The manuscript was completed with partial support from NSF grant No.\ CMMI-2029540.

\paragraph{Declaration of interests.} 
The authors report no conflict of interest.

\paragraph{Data availability statement.}
The data that support the findings of this study are openly available in the Purdue University Research Repository (PURR)  at \url{http://doi.org/10.4231/SX6D-1370}.

\paragraph{Author ORCID.} 
X.\ Wang, \url{https://orcid.org/0000-0002-3487-1600}; \\
I.\ C.\ Christov, \url{https://orcid.org/0000-0001-8531-0531}

\paragraph{Author contributions.}
X.W.\ led the analysis of the problem and the derivation of the mathematical model, to which also I.C.C.\ contributed. X.W.\ wrote the Python scripts and conducted all the case studies, numerical simulations, and data analysis. Both authors contributed to the analysis of the data and conclusions drawn thereof. X.W.\ and I.C.C.\ jointly drafted and revised the manuscript for publication.


\begin{appendices}


\section{Modeling of weak deformation effects}

\subsection{Weak inertia and the effective interface thickness}
\label{subsec:eff_thick}

\begin{figure}[ht]
    \centering
    \includegraphics[width=0.95\textwidth]{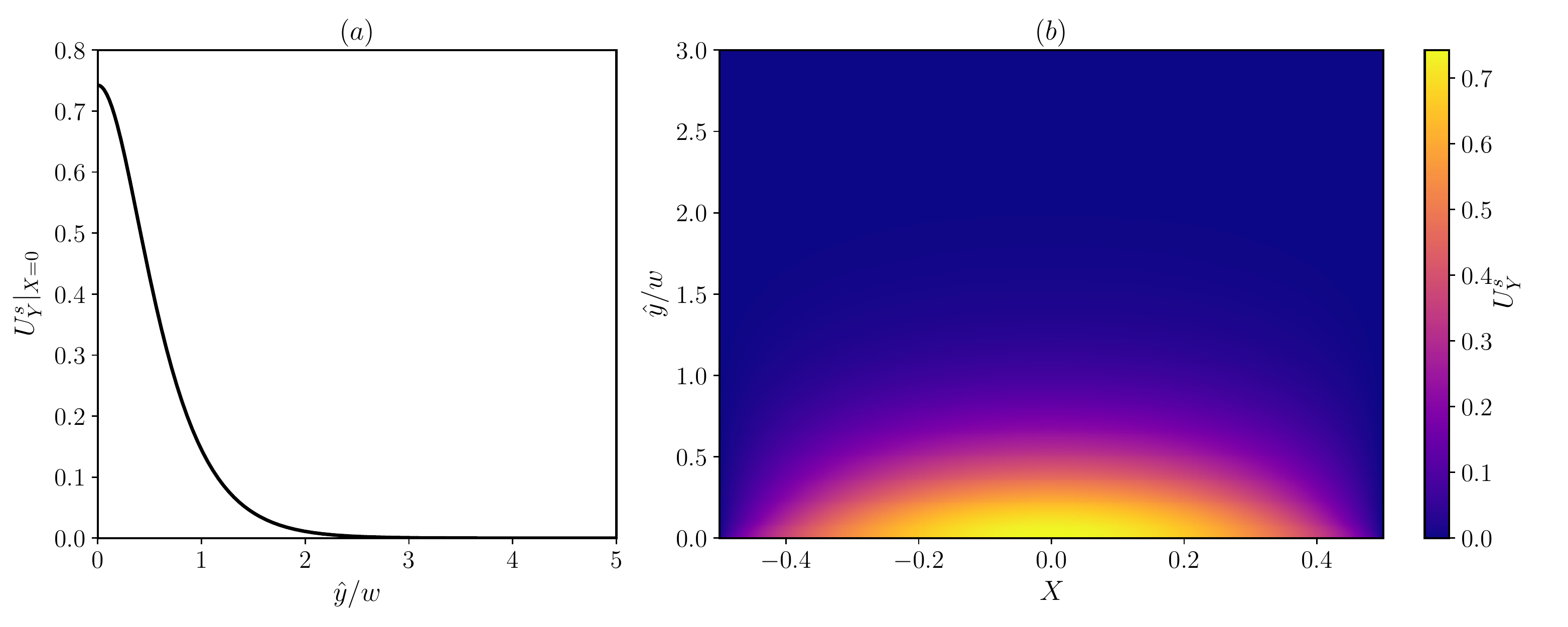}
    \caption{Illustration of the displacement field in a thick wall predicted by equation (3.14) from \cite{WC19}. (\textit{a}) Centerline displacement at $x/w=X=0$ versus dimensionless vertical distance from the fluid--solid interface, $\hat{y}/w$. 
    (\textit{b}) Contour plot of the displacement field. Here, $\hat{y} = y+h_0$ and $\mathcal{U}_c = w\mathcal{P}_c/(h_0 \bar{E})$.  }
    \label{fig:disp-thick}
\end{figure}

The goal of previous studies was to find a solution for the fluid--solid interface displacement, $\bar{u}_y$, from which to determine the cross-sectional area of the deformed fluidic channel. To illustrate this point, consider the microchannel studied by \citet{WC19}, which has a similar configuration to  figure~\ref{subfig:gen1d-geo-3d}. In this case, the theory of the flow-induced deformation predicts that the vertical displacement of the solid, $U_Y^s= u_y^s(x,\hat{y},z)/\mathcal{U}_c$, varies with the vertical distance from the fluid--solid interface ($\hat{y}=0)$), as shown in figure~\ref{fig:disp-thick}. 
For the unsteady problem in this work, however, we must properly connect the motion of the fluid--solid interface to the non-uniform motion of the entire top wall. Since this variation is rapidly decaying, it is reasonable to expect that a suitable \emph{effective thickness} of the fluid--solid interface, $b^{\star}$, can be introduced and used in equation~\eqref{dimension-1d-solid-eq}. In doing so, the unsteady motion of the \emph{whole} solid (of nonuniform vertical displacements) will be captured by the vertical motion of an \emph{interface} of ``virtual'' thickness $b^{\star}$.

In analogy to the definition of the hydrodynamic boundary layer thickness \citep{panton,W06_book}, we define $b^{\star}$ by requiring that the momentum of the solid wall's motion is equivalent to the momentum of the reduced interface's motion, \textit{i.e.},
\begin{equation}\label{eq-bstar1}
    \rho_s \int_0^d \int_{-w/2}^{+w/2} \dot{u}^s_y(x,\hat{y},z) \,\rd x\,\rd\hat{y} = \rho_s w b^\star \dot{\bar{u}}_y(z),
\end{equation}
where the overdot denotes a time derivative. 
Since we have assumed that the solid deformation is governed by linear elasticity, the domain of integration is unchanged after deformation. Then, we can take the time derivative out of the integral, and consider an instantaneous balance. Substituting the definition of $\bar{u}_y$, we obtain 
\begin{equation}\label{eq-bstar2}
    b^\star = \frac{ \int_0^d \int_{-w/2}^{+w/2} {u}_y^s(x,\hat{y},z) \,\rd x\,\rd\hat{y}}{w \bar{u}_y} = \frac{  \int_0^d \int_{-w/2}^{+w/2} {u}_y^s(x,\hat{y},z) \,\rd x\,\rd\hat{y}}{\int_{-w/2}^{+w/2} {u}_y^s(x,0,z) \,\rd x}.
\end{equation}

Substituting equation (3.14) from \cite{WC19} into equation \eqref{eq-bstar2}, we find that  $b^{\star}\approx 0.6141w$ for thick-walled microchannel ($d^2/w^2\gg1$). However, if the top wall is thin ($d\simeq w$), plate theory can be invoked, and the mid-plane displacement can represent the bulk motion of the interface; in this case, $b^\star = d$. 

\subsection{Weak deformation-induced tension}
\label{subsec:tension_coeff}

Having introduced $b^{\star}$, we are ready to give an expression for the weak tension, $\chi_t$. One possible scenario is that $\chi_t$ arises from the bulging of the wall. In principle, the deformation-induced tension is nonuniform along $z$. However, as mentioned, the variation of tension in $z$ is balanced with the shear stress in the flow, and thus can be neglected. Then, assuming the in-plane displacement (along $z$) is negligible, $\chi_t$ can be estimated by the average stretch of the wall \citep{HBDB14}, written as 
\begin{equation}\label{ft-eq}
    \chi_t = \frac{Eb^\star}{\ell} \int_0^{\ell} \frac{1}{2}\left(\frac{\partial \bar{u}_y}{\partial z}\right)^2 \, \rd z.
\end{equation}
Deformation-induced tension is expected to occur when the outlet of the channel is open to air, as in \citep{GEGJ06, VK13}, or the pre-tension provided by external connectors is negligible. In the unsteady case, $\chi_t$ is time-dependent.

Another possible situation is that the microchannel is pre-stretched and installed between an upstream and downstream section, as in the research on  collapsible tubes mentioned in \S~\ref{sec:intro}. With the increase of the flow rate, the bulging of the wall is more prominent, leading to larger $\chi_t$. However, beyond a certain flow rate, the deformation-induced tension will not be sufficient to hold the channel at the inlet and the outlet. In other words, the boundary conditions cannot be satisfied. The upper bound on the flow rate before the model breaks down is related to and found to increase with $\chi_t$ \citep{WC21}.
Therefore, in this case when the deformation-induced tension is not sufficient, if the system is still to operate at such a high flow rate, external pre-tension needs to be provided. Nevertheless, for the validity of equation \eqref{dimension-1d-solid-eq}, the pre-tension in this case should be much larger than the deformation-induced tension. On the other hand, the third term in equation \eqref{dimension-1d-solid-eq} needs to be small compared with the second term, to ensure the dominance of the Winkler-foundation-like mechanism.

Apart from weak tension, other elastic forces might also be relevant in other physical scenarios. For example, if the top wall is thin, bending could play a role. Another example comes from the elastic structures on top of thin fluid films, wherein (in addition to tension) bending and gravity are invoked to regularize the problem \citep[see, \textit{e.g.},][]{PL20}.

\section{Long-time simulation of the evolution from a perturbed inflated state}
\label{sec:app-a}

\begin{figure}[ht!]
    \centering    \includegraphics[width=0.85\textwidth]{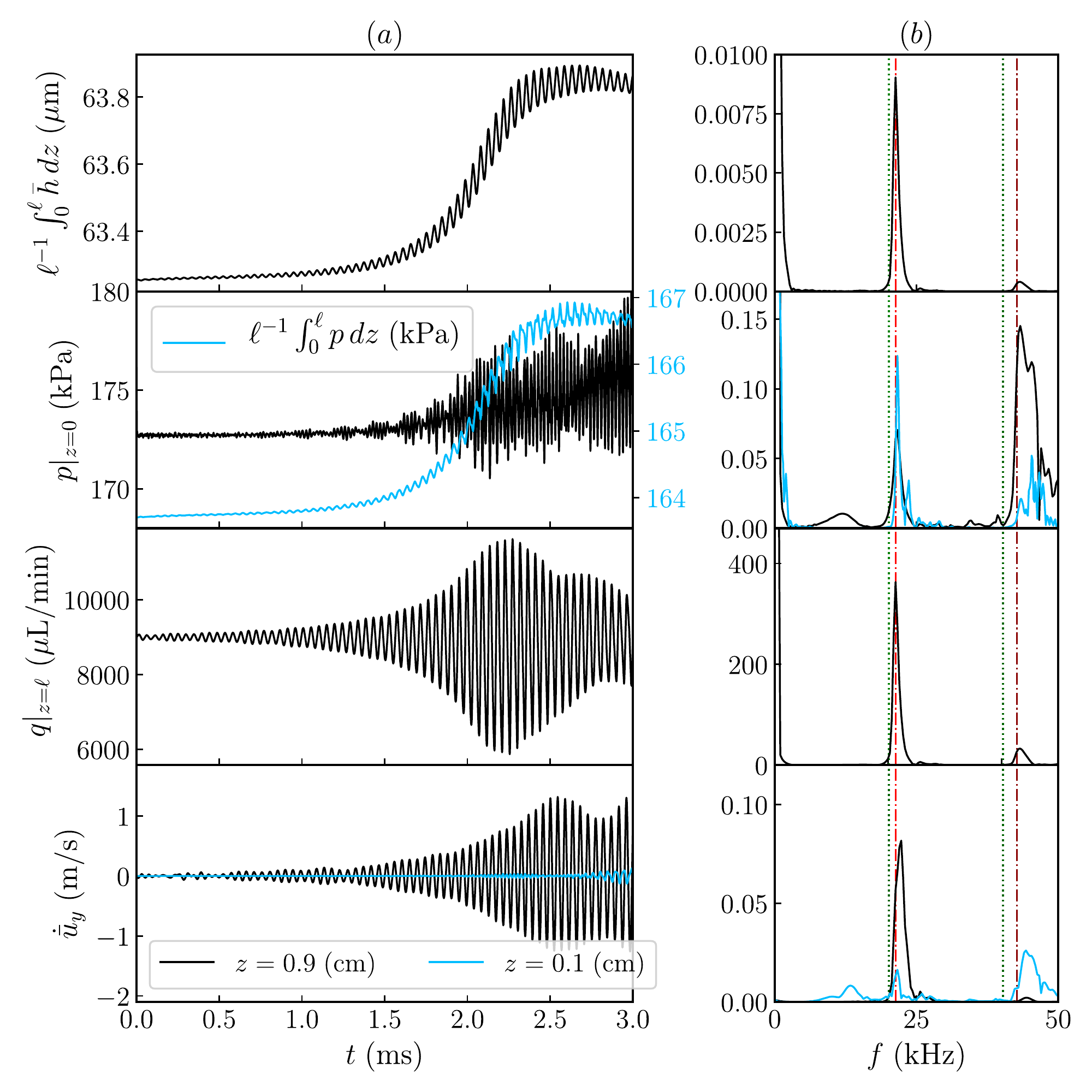}
    \caption{Dynamic simulations of case C3, by perturbing the steady state with the eigenfunctions of the most unstable mode from table~\ref{table:case-eig}. (\textit{a}) Time histories of representative quantities: the axially averaged deformed height, the inlet pressure and the axially averaged pressure, the outlet flow rate, and the vertical velocity of the interface at $z=0.9$ \si{\centi\meter} and $z=0.1$ \si{\centi\meter}, respectively, from top to below. (\textit{b}) Fourier transform of the time signals from (\textit{a}). Note that in the second and fourth rows, the vertical axis has been rescaled for a clearer view. The dot-dashed lines mark $f_g$ and $2f_g$ (see equation \eqref{fg}), while the dotted lines mark $f_n$ and $2f_n$ (see equation \eqref{fn}).}
    \label{fig:simstab-time-1}
\end{figure}

\begin{figure}[ht!]
    \centering
    \includegraphics[width=0.85\textwidth]{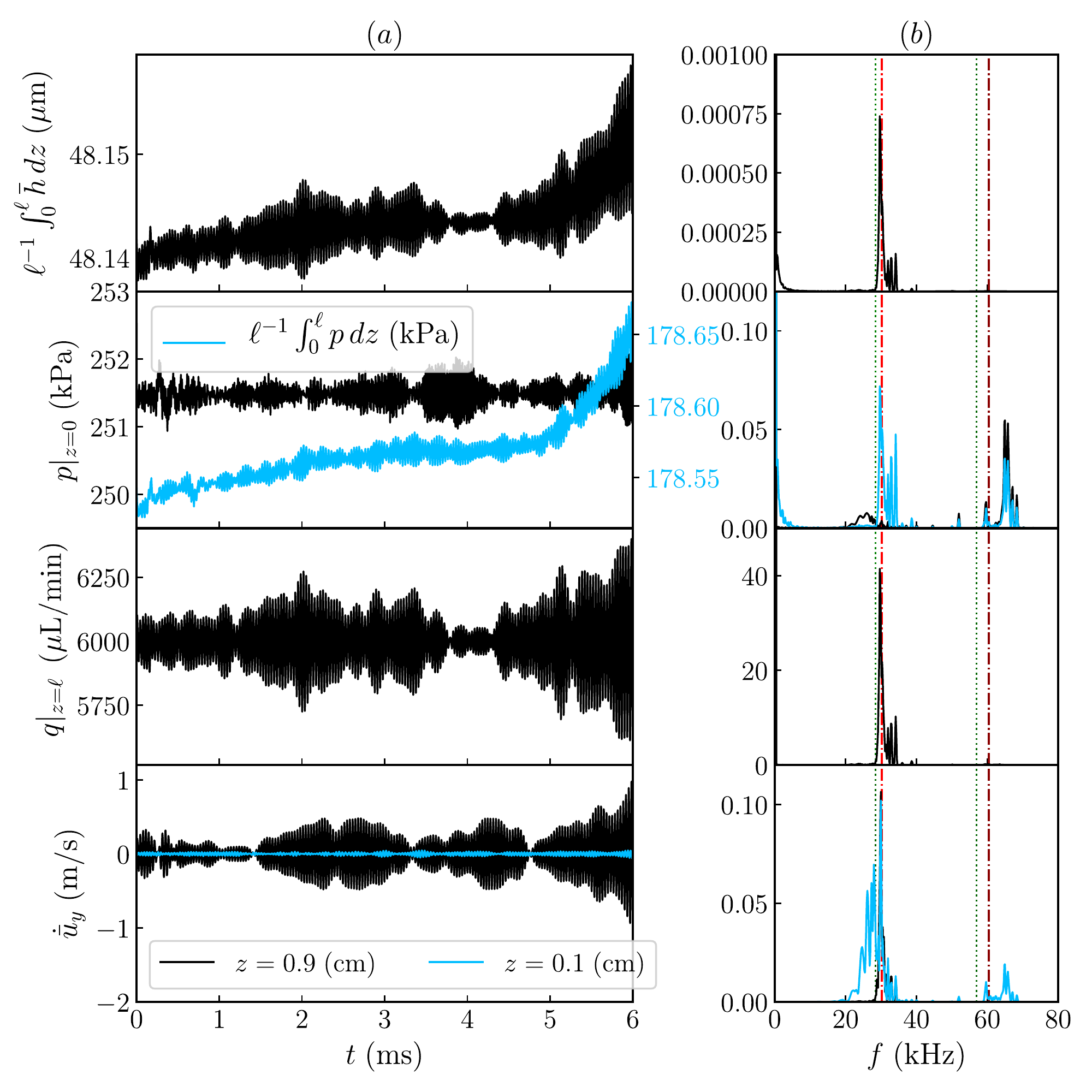}
    \caption{Dynamic simulations of case C5, by perturbing the steady state with the eigenfunctions of the most unstable mode from table~\ref{table:case-eig}. (\textit{a}) Time histories of the representative quantities: the axially averaged deformed height, the inlet pressure and the axially averaged pressure, the outlet flow rate, and the vertical velocity of the interface at $z=0.9$\ \si{\centi\meter} and $z=0.1$\ \si{\centi\meter}, respectively, from top to below. (\textit{b}) Fourier transform of the time signals from (\textit{a}). Note that in the second and fourth rows, the vertical axis has been rescaled to highlight the smaller-scale details. The dot-dashed lines mark $f_g$ and $2f_g$ (see equation \eqref{fg}), while the dotted lines mark $f_n$ and $2f_n$ (see equation \eqref{fn}).}
    \label{fig:simstab-time-2}
\end{figure}

The linear stability analysis only predicts the evolution of the perturbation in the vicinity of the steady state, \textit{i.e.}, for early times. The simulations in \S~\ref{subsec:dynamic1} are actually conducted for a longer time window. The results for cases C3 and C5 are shown in figures~\ref{fig:simstab-time-1} and \ref{fig:simstab-time-2}, respectively. For the time window shown, the numerical results are verified by time step refinement. All representative quantities shown in figures~\ref{fig:simstab-time-1} and \ref{fig:simstab-time-2} experience high-frequency oscillations with nonlinear variations in their amplitudes. No saturated (periodic) state is found in any of the cases. Actually, beyond the given time window, the simulation results diverge when using different time step sizes, which suggests that this nonlinear dynamical system may exhibit chaotic behavior. However, pursing this possibility is beyond the scope of the present work. The most important conclusion from these simulations is that the perturbed steady state is unstable, and the system undergoes self-sustained oscillations, instead of returning to the inflated steady state.

In figure~\ref{fig:simstab-time-1} and figure~\ref{fig:simstab-time-2}, the axially averaged deformed height (first rows), the inlet pressure and the axially averaged pressure (second rows) are observed to vary in a small range ($<1$ \si{\micro\meter} for the axially averaged deformation and $<10$ \si{\kilo\pascal} for the pressure), which is consistent with the fact that no dramatic changes in the channel volume and the inlet pressure were reported in the experiments \citep{VK13}. On the contrary, the outlet flow rate (third rows) experiences more violent oscillations, which are prominent  near the channel outlet, as shown in figure~\ref{fig:eigfunc_least_stable}. In the fourth rows, the vertical velocity of the fluid--solid interface is shown. It is obtained via equation \eqref{vel-profile} and conservation of mass (see \SM\ for further details). Like the flow rate, the vertical velocity of the interface is observed to experience larger oscillatory amplitude near the channel outlet than that near the channel inlet, which again matches the experimental observation that the instabilities initiate near the channel's outlet.

Figures~\ref{fig:simstab-time-1}(\textit{b}) and \ref{fig:simstab-time-2}(\textit{b}) show the Fourier transforms of the time histories of the corresponding representative quantity. It is observed that there is a peak near $f_g$ (see equation \eqref{fg}), showing a good agreement with the linear stability analysis. Also observe that $f_g$ is close to the natural frequency of the wall, $f_n$ (see equation \eqref{fn}), which indicates that the wall oscillations are a resonance phenomenon. Further, nonlinearity generates higher harmonics. In figure~\ref{fig:simstab-time-1}, there is another peak at $\approx 2f_g$, while in figure~\ref{fig:simstab-time-2}, the second peak is at a frequency higher than $2f_g$. The higher-frequency oscillations are more prominent near the channel inlet. For example, the second peak of the vertical velocity of the interface at $z=0.1$ \si{\centi\meter} is taller than the first peak. Furthermore, the higher-frequency oscillations in the inlet pressure are more prominent than the lower-frequency oscillations as shown in both figures~\ref{fig:simstab-time-1} and \ref{fig:simstab-time-2}.

\end{appendices}

\phantomsection
\addcontentsline{toc}{section}{References}

\bibliography{references2.bib}

\begin{thebibliography}{71}
\expandafter\ifx\csname natexlab\endcsname\relax\def\natexlab#1{#1}\fi
\providecommand{\url}[1]{\texttt{#1}}
\providecommand{\href}[2]{#2}
\providecommand{\path}[1]{#1}
\providecommand{\DOIprefix}{doi:}
\providecommand{\ArXivprefix}{arXiv:}
\providecommand{\URLprefix}{URL: }
\providecommand{\Pubmedprefix}{pmid:}
\providecommand{\doi}[1]{\href{http://dx.doi.org/#1}{\path{#1}}}
\providecommand{\Pubmed}[1]{\href{pmid:#1}{\path{#1}}}
\providecommand{\bibinfo}[2]{#2}
\ifx\xfnm\relax \def\xfnm[#1]{\unskip,\space#1}\fi
\bibitem[{Anand et~al.(2020)Anand, Muchandimath and Christov}]{AMC20}
\bibinfo{author}{Anand, V.}, \bibinfo{author}{Muchandimath, S.C.},
  \bibinfo{author}{Christov, I.C.}, \bibinfo{year}{2020}.
\newblock \bibinfo{title}{{Hydrodynamic Bulge Testing: Materials
  Characterization Without Measuring Deformation}}.
\newblock \bibinfo{journal}{ASME J. Appl. Mech.} \bibinfo{volume}{87},
  \bibinfo{pages}{051012}.
\newblock \DOIprefix\doi{10.1115/1.4046297}.
\bibitem[{Boyd(2000)}]{Boyd00}
\bibinfo{author}{Boyd, J.P.}, \bibinfo{year}{2000}.
\newblock \bibinfo{title}{{Chebyshev and Fourier Spectral Methods}}.
\newblock \bibinfo{edition}{2} ed., \bibinfo{publisher}{Dover Publications},
  \bibinfo{address}{Mineola, NY}.
\bibitem[{Christov(2022)}]{C21}
\bibinfo{author}{Christov, I.C.}, \bibinfo{year}{2022}.
\newblock \bibinfo{title}{{Soft hydraulics: from Newtonian to complex fluid
  flows through compliant conduits}}.
\newblock \bibinfo{journal}{J. Phys.: Condens. Matter} \bibinfo{volume}{34},
  \bibinfo{pages}{063001}.
\newblock \DOIprefix\doi{10.1088/1361-648X/ac327d}.
\bibitem[{Christov et~al.(2018)Christov, Cognet, Shidhore and Stone}]{CCSS17}
\bibinfo{author}{Christov, I.C.}, \bibinfo{author}{Cognet, V.},
  \bibinfo{author}{Shidhore, T.C.}, \bibinfo{author}{Stone, H.A.},
  \bibinfo{year}{2018}.
\newblock \bibinfo{title}{{Flow rate--pressure drop relation for deformable
  shallow microfluidic channels}}.
\newblock \bibinfo{journal}{J. Fluid Mech.} \bibinfo{volume}{814},
  \bibinfo{pages}{267--286}.
\newblock \DOIprefix\doi{10.1017/jfm.2018.30}.
\bibitem[{Dhong et~al.(2018)Dhong, Edmunds, Ram{\'{i}}rez, Kayser, Chen,
  Jokerst and Lipomi}]{Dhong18}
\bibinfo{author}{Dhong, C.}, \bibinfo{author}{Edmunds, S.J.},
  \bibinfo{author}{Ram{\'{i}}rez, J.}, \bibinfo{author}{Kayser, L.V.},
  \bibinfo{author}{Chen, F.}, \bibinfo{author}{Jokerst, J.V.},
  \bibinfo{author}{Lipomi, D.J.}, \bibinfo{year}{2018}.
\newblock \bibinfo{title}{{Optics-free, non-contact measurements of fluids,
  bubbles, and particles in microchannels using metallic nano-islands on
  graphene}}.
\newblock \bibinfo{journal}{Nano Lett.} \bibinfo{volume}{18},
  \bibinfo{pages}{5306--5311}.
\newblock \DOIprefix\doi{10.1021/acs.nanolett.8b02292}.
\bibitem[{Di~Carlo et~al.(2007)Di~Carlo, Irimia, Tompkins and Toner}]{DITT07}
\bibinfo{author}{Di~Carlo, D.}, \bibinfo{author}{Irimia, D.},
  \bibinfo{author}{Tompkins, R.G.}, \bibinfo{author}{Toner, M.},
  \bibinfo{year}{2007}.
\newblock \bibinfo{title}{{Continuous inertial focusing, ordering, and
  separation of particles in microchannels}}.
\newblock \bibinfo{journal}{Proc. Natl Acad. Sci. USA} \bibinfo{volume}{104},
  \bibinfo{pages}{18892--18897}.
\newblock \DOIprefix\doi{10.1073/pnas.0704958104}.
\bibitem[{Dillard et~al.(2018)Dillard, Mukherjee, Karnal, Batra and
  Frechette}]{DMKBF18}
\bibinfo{author}{Dillard, D.A.}, \bibinfo{author}{Mukherjee, B.},
  \bibinfo{author}{Karnal, P.}, \bibinfo{author}{Batra, R.C.},
  \bibinfo{author}{Frechette, J.}, \bibinfo{year}{2018}.
\newblock \bibinfo{title}{{A review of Winkler's foundation and its profound
  influence on adhesion and soft matter applications}}.
\newblock \bibinfo{journal}{Soft Matter} \bibinfo{volume}{14},
  \bibinfo{pages}{3669--3683}.
\newblock \DOIprefix\doi{10.1039/c7sm02062g}.
\bibitem[{Elbaz and Gat(2014)}]{EG14}
\bibinfo{author}{Elbaz, S.B.}, \bibinfo{author}{Gat, A.D.},
  \bibinfo{year}{2014}.
\newblock \bibinfo{title}{{Dynamics of viscous liquid within a closed elastic
  cylinder subject to external forces with application to soft robotics}}.
\newblock \bibinfo{journal}{J. Fluid Mech.} \bibinfo{volume}{758},
  \bibinfo{pages}{221--237}.
\newblock \DOIprefix\doi{10.1017/jfm.2014.527}.
\bibitem[{{Gaurav} and Shankar(2009)}]{GS09}
\bibinfo{author}{{Gaurav}}, \bibinfo{author}{Shankar, V.},
  \bibinfo{year}{2009}.
\newblock \bibinfo{title}{{Stability of fluid flow through deformable
  neo-Hookean tubes}}.
\newblock \bibinfo{journal}{J. Fluid Mech.} \bibinfo{volume}{627},
  \bibinfo{pages}{291--322}.
\newblock \DOIprefix\doi{10.1017/S0022112009005928}.
\bibitem[{Gervais et~al.(2006)Gervais, El-Ali, G{\"{u}}nther and
  Jensen}]{GEGJ06}
\bibinfo{author}{Gervais, T.}, \bibinfo{author}{El-Ali, J.},
  \bibinfo{author}{G{\"{u}}nther, A.}, \bibinfo{author}{Jensen, K.F.},
  \bibinfo{year}{2006}.
\newblock \bibinfo{title}{{Flow-induced deformation of shallow microfluidic
  channels}}.
\newblock \bibinfo{journal}{Lab Chip} \bibinfo{volume}{6},
  \bibinfo{pages}{500--507}.
\newblock \DOIprefix\doi{10.1039/b513524a}.
\bibitem[{Gkanis and Kumar(2005)}]{GK05}
\bibinfo{author}{Gkanis, V.}, \bibinfo{author}{Kumar, S.},
  \bibinfo{year}{2005}.
\newblock \bibinfo{title}{{Stability of pressure-driven creeping flows in
  channels lined with a nonlinear elastic solid}}.
\newblock \bibinfo{journal}{J. Fluid Mech.} \bibinfo{volume}{524},
  \bibinfo{pages}{357--375}.
\newblock \DOIprefix\doi{10.1017/S0022112004002472}.
\bibitem[{Gad-el Hak(2002)}]{G02}
\bibinfo{author}{Gad-el Hak, M.}, \bibinfo{year}{2002}.
\newblock \bibinfo{title}{{Compliant coatings for drag reduction}}.
\newblock \bibinfo{journal}{Prog. Aerospace Sci.} \bibinfo{volume}{38},
  \bibinfo{pages}{77--99}.
\newblock \DOIprefix\doi{10.1016/S0376-0421(01)00020-3}.
\bibitem[{Heil and Boyle(2010)}]{HB10}
\bibinfo{author}{Heil, M.}, \bibinfo{author}{Boyle, J.}, \bibinfo{year}{2010}.
\newblock \bibinfo{title}{{Self-excited oscillations in three-dimensional
  collapsible tubes: simulating their onset and large-amplitude oscillations}}.
\newblock \bibinfo{journal}{J. Fluid Mech.} \bibinfo{volume}{652},
  \bibinfo{pages}{405--426}.
\newblock \DOIprefix\doi{10.1017/S0022112010000157}.
\bibitem[{Hewitt et~al.(2015)Hewitt, Balmforth and De~Bruyn}]{HBDB14}
\bibinfo{author}{Hewitt, I.J.}, \bibinfo{author}{Balmforth, N.J.},
  \bibinfo{author}{De~Bruyn, J.R.}, \bibinfo{year}{2015}.
\newblock \bibinfo{title}{{Elastic-plated gravity currents}}.
\newblock \bibinfo{journal}{Eur. J. Appl. Math.} \bibinfo{volume}{26},
  \bibinfo{pages}{1--31}.
\newblock \DOIprefix\doi{10.1017/S0956792514000291}.
\bibitem[{Hosokawa et~al.(2002)Hosokawa, Hanada and Maeda}]{HHM02}
\bibinfo{author}{Hosokawa, K.}, \bibinfo{author}{Hanada, K.},
  \bibinfo{author}{Maeda, R.}, \bibinfo{year}{2002}.
\newblock \bibinfo{title}{{A polydimethylsiloxane (PDMS) deformable diffraction
  grating for monitoring of local pressure in microfluidic devices}}.
\newblock \bibinfo{journal}{J. Microelectromechan. Syst.} \bibinfo{volume}{12},
  \bibinfo{pages}{1--6}.
\newblock \DOIprefix\doi{10.1088/0960-1317/12/1/301}.
\bibitem[{Howell et~al.(2009)Howell, Kozyreff and Ockendon}]{HKO09}
\bibinfo{author}{Howell, P.}, \bibinfo{author}{Kozyreff, G.},
  \bibinfo{author}{Ockendon, J.}, \bibinfo{year}{2009}.
\newblock \bibinfo{title}{{Applied Solid Mechanics}}.
\newblock \bibinfo{publisher}{Cambridge University Press},
  \bibinfo{address}{Cambridge, UK}.
\newblock \DOIprefix\doi{10.1017/CBO9780511611605}.
\bibitem[{Huerre and Monkewitz(1990)}]{HM90}
\bibinfo{author}{Huerre, P.}, \bibinfo{author}{Monkewitz, P.A.},
  \bibinfo{year}{1990}.
\newblock \bibinfo{title}{{Local and Global Instabilities in Spatially
  Developing Flows}}.
\newblock \bibinfo{journal}{Annu. Rev. Fluid Mech.} \bibinfo{volume}{22},
  \bibinfo{pages}{473--537}.
\newblock \DOIprefix\doi{10.1146/annurev.fl.22.010190.002353}.
\bibitem[{Inamdar et~al.(2020)Inamdar, Wang and Christov}]{IWC20}
\bibinfo{author}{Inamdar, T.C.}, \bibinfo{author}{Wang, X.},
  \bibinfo{author}{Christov, I.C.}, \bibinfo{year}{2020}.
\newblock \bibinfo{title}{{Unsteady fluid-structure interactions in a
  soft-walled microchannel: A one-dimensional lubrication model for finite
  Reynolds number}}.
\newblock \bibinfo{journal}{Phys. Rev. Fluids} \bibinfo{volume}{5},
  \bibinfo{pages}{064101}.
\newblock \DOIprefix\doi{10.1103/PhysRevFluids.5.064101}.
\bibitem[{Jensen(1990)}]{J90}
\bibinfo{author}{Jensen, O.E.}, \bibinfo{year}{1990}.
\newblock \bibinfo{title}{{Instabilities of flow in a collapsed tube}}.
\newblock \bibinfo{journal}{J. Fluid Mech.} \bibinfo{volume}{220},
  \bibinfo{pages}{623--659}.
\newblock \DOIprefix\doi{10.1017/S0022112090003408}.
\bibitem[{Jensen(1992)}]{J92}
\bibinfo{author}{Jensen, O.E.}, \bibinfo{year}{1992}.
\newblock \bibinfo{title}{{Chaotic Oscillations in a Simple Collapsible-Tube
  Model}}.
\newblock \bibinfo{journal}{ASME J. Biomech. Eng.} \bibinfo{volume}{114},
  \bibinfo{pages}{55--59}.
\newblock \DOIprefix\doi{10.1115/1.2895450}.
\bibitem[{Jensen and Heil(2003)}]{JH03}
\bibinfo{author}{Jensen, O.E.}, \bibinfo{author}{Heil, M.},
  \bibinfo{year}{2003}.
\newblock \bibinfo{title}{{High-frequency self-excited oscillations in a
  collapsible-channel flow}}.
\newblock \bibinfo{journal}{J. Fluid Mech.} \bibinfo{volume}{481},
  \bibinfo{pages}{235--268}.
\newblock \DOIprefix\doi{10.1017/S002211200300394X}.
\bibitem[{Jensen and Pedley(1989)}]{JP89}
\bibinfo{author}{Jensen, O.E.}, \bibinfo{author}{Pedley, T.J.},
  \bibinfo{year}{1989}.
\newblock \bibinfo{title}{{The existence of steady flow in a collapsed tube}}.
\newblock \bibinfo{journal}{J. Fluid Mech.} \bibinfo{volume}{206},
  \bibinfo{pages}{339--374}.
\newblock \DOIprefix\doi{10.1017/S0022112089002326}.
\bibitem[{Karan et~al.(2021)Karan, Chakraborty, Chakraborty, Wereley and
  Christov}]{KCCWC21}
\bibinfo{author}{Karan, P.}, \bibinfo{author}{Chakraborty, J.},
  \bibinfo{author}{Chakraborty, S.}, \bibinfo{author}{Wereley, S.T.},
  \bibinfo{author}{Christov, I.C.}, \bibinfo{year}{2021}.
\newblock \bibinfo{title}{{Profiling a soft solid layer to passively control
  the conduit shape in a compliant microchannel during flow}}.
\newblock \bibinfo{journal}{Phys. Rev. E} \bibinfo{volume}{104},
  \bibinfo{pages}{015108}.
\newblock \DOIprefix\doi{10.1103/PhysRevE.104.015108}.
\bibitem[{Karnik(2013)}]{K13}
\bibinfo{author}{Karnik, R.}, \bibinfo{year}{2013}.
\newblock \bibinfo{title}{{Microfluidic Mixing}}, in: \bibinfo{editor}{Li, D.}
  (Ed.), \bibinfo{booktitle}{Encyclopedia of Microfluidics and Nanofluidics}.
  \bibinfo{publisher}{Springer}, \bibinfo{address}{Boston, MA}, pp.
  \bibinfo{pages}{1969--1979}.
\newblock \DOIprefix\doi{10.1007/978-3-642-27758-0{\_}939-2}.
\bibitem[{Keller(1976)}]{Keller76}
\bibinfo{author}{Keller, H.B.}, \bibinfo{year}{1976}.
\newblock \bibinfo{title}{{Numerical Solution of Two Point Boundary Value
  Problems}}. volume~\bibinfo{volume}{24} of \textit{\bibinfo{series}{CBMS-NSF
  Regional Conference Series in Applied Mathematics}}.
\newblock \bibinfo{publisher}{SIAM}, \bibinfo{address}{Philadelphia, PA}.
\newblock \DOIprefix\doi{10.1137/1.9781611970449}.
\bibitem[{Krindel and Silberberg(1979)}]{KS79}
\bibinfo{author}{Krindel, P.}, \bibinfo{author}{Silberberg, A.},
  \bibinfo{year}{1979}.
\newblock \bibinfo{title}{{Flow through gel-walled tubes}}.
\newblock \bibinfo{journal}{J. Colloid Interface Sci.} \bibinfo{volume}{71},
  \bibinfo{pages}{39--50}.
\newblock \DOIprefix\doi{10.1016/0021-9797(79)90219-4}.
\bibitem[{Kumaran(1995)}]{K95}
\bibinfo{author}{Kumaran, V.}, \bibinfo{year}{1995}.
\newblock \bibinfo{title}{{Stability of the viscous flow of a fluid through a
  flexible tube}}.
\newblock \bibinfo{journal}{J. Fluid Mech.} \bibinfo{volume}{294},
  \bibinfo{pages}{259--281}.
\newblock \DOIprefix\doi{10.1017/S0022112095002886}.
\bibitem[{Kumaran(2021)}]{K21}
\bibinfo{author}{Kumaran, V.}, \bibinfo{year}{2021}.
\newblock \bibinfo{title}{{Stability and the transition to turbulence in the
  flow through conduits with compliant walls}}.
\newblock \bibinfo{journal}{J. Fluid Mech.} \bibinfo{volume}{924},
  \bibinfo{pages}{P1}.
\newblock \DOIprefix\doi{10.1017/jfm.2021.602}.
\bibitem[{Kumaran and Bandaru(2016)}]{KB16}
\bibinfo{author}{Kumaran, V.}, \bibinfo{author}{Bandaru, P.},
  \bibinfo{year}{2016}.
\newblock \bibinfo{title}{{Ultra-fast microfluidic mixing by soft-wall
  turbulence}}.
\newblock \bibinfo{journal}{Chem. Eng. Sci.} \bibinfo{volume}{149},
  \bibinfo{pages}{156--168}.
\newblock \DOIprefix\doi{10.1016/j.ces.2016.04.001}.
\bibitem[{Liu et~al.(2012)Liu, Luo and Cai}]{LLC12}
\bibinfo{author}{Liu, H.F.}, \bibinfo{author}{Luo, X.Y.}, \bibinfo{author}{Cai,
  Z.X.}, \bibinfo{year}{2012}.
\newblock \bibinfo{title}{{Stability and energy budget of pressure-driven
  collapsible channel flows}}.
\newblock \bibinfo{journal}{J. Fluid Mech.} \bibinfo{volume}{705},
  \bibinfo{pages}{348--370}.
\newblock \DOIprefix\doi{10.1017/jfm.2011.254}.
\bibitem[{Love(1888)}]{L88}
\bibinfo{author}{Love, A.E.H.}, \bibinfo{year}{1888}.
\newblock \bibinfo{title}{{The small free vibrations and deformation of a thin
  elastic shell}}.
\newblock \bibinfo{journal}{Phil. Trans. R. Soc. Lond. A}
  \bibinfo{volume}{179}, \bibinfo{pages}{491--546}.
\newblock \DOIprefix\doi{10.1098/rsta.1888.0016}.
\bibitem[{Luo et~al.(2008)Luo, Cai, Li and Pedley}]{LCLP08}
\bibinfo{author}{Luo, X.Y.}, \bibinfo{author}{Cai, Z.X.}, \bibinfo{author}{Li,
  W.G.}, \bibinfo{author}{Pedley, T.J.}, \bibinfo{year}{2008}.
\newblock \bibinfo{title}{{The cascade structure of linear instability in
  collapsible channel flows}}.
\newblock \bibinfo{journal}{J. Fluid Mech.} \bibinfo{volume}{600},
  \bibinfo{pages}{45--76}.
\newblock \DOIprefix\doi{10.1017/S0022112008000293}.
\bibitem[{Luo and Pedley(1996)}]{LP96}
\bibinfo{author}{Luo, X.Y.}, \bibinfo{author}{Pedley, T.J.},
  \bibinfo{year}{1996}.
\newblock \bibinfo{title}{{A numerical simulation of unsteady flow in a
  two-dimensional collapsible channel}}.
\newblock \bibinfo{journal}{J. Fluid Mech.} \bibinfo{volume}{314},
  \bibinfo{pages}{191--225}.
\newblock \DOIprefix\doi{10.1017/S0022112096000286}.
\bibitem[{Luo and Pedley(1998)}]{LP98}
\bibinfo{author}{Luo, X.Y.}, \bibinfo{author}{Pedley, T.J.},
  \bibinfo{year}{1998}.
\newblock \bibinfo{title}{{The effects of wall inertia on flow in a
  two-dimensional collapsible channel}}.
\newblock \bibinfo{journal}{J. Fluid Mech.} \bibinfo{volume}{363},
  \bibinfo{pages}{253--280}.
\newblock \DOIprefix\doi{10.1017/S0022112098001062}.
\bibitem[{Mart{\'{i}}nez-Calvo et~al.(2020)Mart{\'{i}}nez-Calvo, Sevilla, Peng
  and Stone}]{MCSPS19}
\bibinfo{author}{Mart{\'{i}}nez-Calvo, A.}, \bibinfo{author}{Sevilla, A.},
  \bibinfo{author}{Peng, G.G.}, \bibinfo{author}{Stone, H.A.},
  \bibinfo{year}{2020}.
\newblock \bibinfo{title}{{Start-up flow in shallow deformable microchannels}}.
\newblock \bibinfo{journal}{J. Fluid Mech.} \bibinfo{volume}{885},
  \bibinfo{pages}{A25}.
\newblock \DOIprefix\doi{10.1017/jfm.2019.994}.
\bibitem[{Mindlin(1951)}]{M51}
\bibinfo{author}{Mindlin, R.D.}, \bibinfo{year}{1951}.
\newblock \bibinfo{title}{{Influence of rotatory inertia and shear on flexural
  motions of isotropic, elastic plates}}.
\newblock \bibinfo{journal}{ASME J. Appl. Mech.} \bibinfo{volume}{18},
  \bibinfo{pages}{31--38}.
\bibitem[{Neelamegam and Shankar(2015)}]{NS15}
\bibinfo{author}{Neelamegam, R.}, \bibinfo{author}{Shankar, V.},
  \bibinfo{year}{2015}.
\newblock \bibinfo{title}{{Experimental study of the instability of laminar
  flow in a tube with deformable walls}}.
\newblock \bibinfo{journal}{Phys. Fluids} \bibinfo{volume}{27},
  \bibinfo{pages}{024102}.
\newblock \DOIprefix\doi{10.1063/1.4907246}.
\bibitem[{Ottino and Wiggins(2004)}]{OW04}
\bibinfo{author}{Ottino, J.M.}, \bibinfo{author}{Wiggins, S.},
  \bibinfo{year}{2004}.
\newblock \bibinfo{title}{{Introduction: mixing in microfluidics}}.
\newblock \bibinfo{journal}{Phil. Trans. R. Soc. A} \bibinfo{volume}{362},
  \bibinfo{pages}{923--935}.
\newblock \DOIprefix\doi{10.1098/rsta.2003.1355}.
\bibitem[{Ozsun et~al.(2013)Ozsun, Yakhot and Ekinci}]{OYE13}
\bibinfo{author}{Ozsun, O.}, \bibinfo{author}{Yakhot, V.},
  \bibinfo{author}{Ekinci, K.L.}, \bibinfo{year}{2013}.
\newblock \bibinfo{title}{{Non-invasive measurement of the pressure
  distribution in a deformable micro-channel}}.
\newblock \bibinfo{journal}{J. Fluid Mech.} \bibinfo{volume}{734},
  \bibinfo{pages}{R1}.
\newblock \DOIprefix\doi{10.1017/jfm.2013.474}.
\bibitem[{Panton(2013)}]{panton}
\bibinfo{author}{Panton, R.L.}, \bibinfo{year}{2013}.
\newblock \bibinfo{title}{{Incompressible Flow}}.
\newblock \bibinfo{edition}{4th} ed., \bibinfo{publisher}{John Wiley {\&}
  Sons}, \bibinfo{address}{Hoboken, NJ}.
\newblock \DOIprefix\doi{10.1002/9781118713075}.
\bibitem[{Patne and Shankar(2019)}]{PS19}
\bibinfo{author}{Patne, R.}, \bibinfo{author}{Shankar, V.},
  \bibinfo{year}{2019}.
\newblock \bibinfo{title}{{Stability of flow through deformable channels and
  tubes: implications of consistent formulation}}.
\newblock \bibinfo{journal}{J. Fluid Mech.} \bibinfo{volume}{860},
  \bibinfo{pages}{837--885}.
\newblock \DOIprefix\doi{10.1017/jfm.2018.908}.
\bibitem[{Peng and Lister(2020)}]{PL20}
\bibinfo{author}{Peng, G.G.}, \bibinfo{author}{Lister, J.R.},
  \bibinfo{year}{2020}.
\newblock \bibinfo{title}{{Viscous flow under an elastic sheet}}.
\newblock \bibinfo{journal}{J. Fluid Mech.} \bibinfo{volume}{905},
  \bibinfo{pages}{A30}.
\newblock \DOIprefix\doi{10.1017/jfm.2020.745}.
\bibitem[{Pihler-Puzovi{\'{c}} and Pedley(2014)}]{PPP14}
\bibinfo{author}{Pihler-Puzovi{\'{c}}, D.}, \bibinfo{author}{Pedley, T.J.},
  \bibinfo{year}{2014}.
\newblock \bibinfo{title}{{Flutter in a quasi-one-dimensional model of a
  collapsible channel}}.
\newblock \bibinfo{journal}{Proc. R. Soc. A} \bibinfo{volume}{470},
  \bibinfo{pages}{20140015}.
\newblock \DOIprefix\doi{10.1098/rspa.2014.0015}.
\bibitem[{Pohlhausen(1921)}]{Pohl21}
\bibinfo{author}{Pohlhausen, K.}, \bibinfo{year}{1921}.
\newblock \bibinfo{title}{{Zur n{\"{a}}herungsweisen Integration der
  Differentialgleichung der laminaren Grenzschicht}}.
\newblock \bibinfo{journal}{Z. Angew. Math. Mech. (ZAMM)} \bibinfo{volume}{1},
  \bibinfo{pages}{252--290}.
\newblock \DOIprefix\doi{10.1002/zamm.19210010402}.
\bibitem[{Reissner(1945)}]{R45}
\bibinfo{author}{Reissner, E.}, \bibinfo{year}{1945}.
\newblock \bibinfo{title}{{The effect of transverse shear deformation on the
  bending of elastic plates}}.
\newblock \bibinfo{journal}{ASME J. Appl. Mech.} \bibinfo{volume}{12},
  \bibinfo{pages}{A68--A77}.
\bibitem[{Sackmann et~al.(2014)Sackmann, Fulton and Beebe}]{SFB14}
\bibinfo{author}{Sackmann, E.K.}, \bibinfo{author}{Fulton, A.L.},
  \bibinfo{author}{Beebe, D.J.}, \bibinfo{year}{2014}.
\newblock \bibinfo{title}{{The present and future role of microfluidics in
  biomedical research}}.
\newblock \bibinfo{journal}{Nature} \bibinfo{volume}{507},
  \bibinfo{pages}{181--189}.
\newblock \DOIprefix\doi{10.1038/nature13118}.
\bibitem[{Seker et~al.(2009)Seker, Leslie, Haj-Hariri, Landers, Utz and
  Begley}]{SLHLUB09}
\bibinfo{author}{Seker, E.}, \bibinfo{author}{Leslie, D.C.},
  \bibinfo{author}{Haj-Hariri, H.}, \bibinfo{author}{Landers, J.P.},
  \bibinfo{author}{Utz, M.}, \bibinfo{author}{Begley, M.R.},
  \bibinfo{year}{2009}.
\newblock \bibinfo{title}{{Nonlinear pressure-flow relationships for passive
  microfluidic valves}}.
\newblock \bibinfo{journal}{Lab Chip} \bibinfo{volume}{9},
  \bibinfo{pages}{2691--2697}.
\newblock \DOIprefix\doi{10.1039/B903960K}.
\bibitem[{Shapiro(1977)}]{S77}
\bibinfo{author}{Shapiro, A.H.}, \bibinfo{year}{1977}.
\newblock \bibinfo{title}{{Steady flow in collapsible tubes}}.
\newblock \bibinfo{journal}{ASME J. Biomech. Eng.} \bibinfo{volume}{99},
  \bibinfo{pages}{126--147}.
\newblock \DOIprefix\doi{10.1115/1.3426281}.
\bibitem[{Shen et~al.(2011)Shen, Tang and Wang}]{STW11}
\bibinfo{author}{Shen, J.}, \bibinfo{author}{Tang, T.}, \bibinfo{author}{Wang,
  L.L.}, \bibinfo{year}{2011}.
\newblock \bibinfo{title}{{Spectral Methods: Algorithms, Analysis and
  Applications}}. volume~\bibinfo{volume}{41} of
  \textit{\bibinfo{series}{Springer Series in Computational Mathematics}}.
\newblock \bibinfo{publisher}{Springer-Verlag},
  \bibinfo{address}{Berlin/Heidelberg}.
\newblock \DOIprefix\doi{10.1007/978-3-540-71041-7}.
\bibitem[{Shiba et~al.(2021)Shiba, Li, Virot, Yoshikawa and Weitz}]{SLVYW21}
\bibinfo{author}{Shiba, K.}, \bibinfo{author}{Li, G.}, \bibinfo{author}{Virot,
  E.}, \bibinfo{author}{Yoshikawa, G.}, \bibinfo{author}{Weitz, D.A.},
  \bibinfo{year}{2021}.
\newblock \bibinfo{title}{{Microchannel measurements of viscosity for both
  gases and liquids}}.
\newblock \bibinfo{journal}{Lab Chip} \bibinfo{volume}{21},
  \bibinfo{pages}{2805--2811}.
\newblock \DOIprefix\doi{10.1039/D1LC00202C}.
\bibitem[{Shidhore and Christov(2018)}]{SC18}
\bibinfo{author}{Shidhore, T.C.}, \bibinfo{author}{Christov, I.C.},
  \bibinfo{year}{2018}.
\newblock \bibinfo{title}{{Static response of deformable microchannels: a
  comparative modelling study}}.
\newblock \bibinfo{journal}{J. Phys.: Condens. Matter} \bibinfo{volume}{30},
  \bibinfo{pages}{054002}.
\newblock \DOIprefix\doi{10.1088/1361-648X/aaa226}.
\bibitem[{Skotheim and Mahadevan(2004)}]{SM04}
\bibinfo{author}{Skotheim, J.M.}, \bibinfo{author}{Mahadevan, L.},
  \bibinfo{year}{2004}.
\newblock \bibinfo{title}{{Soft Lubrication}}.
\newblock \bibinfo{journal}{Phys. Rev. Lett.} \bibinfo{volume}{92},
  \bibinfo{pages}{245509}.
\newblock \DOIprefix\doi{10.1103/PhysRevLett.92.245509}.
\bibitem[{Srinivas and Kumaran(2015)}]{SK15}
\bibinfo{author}{Srinivas, S.S.}, \bibinfo{author}{Kumaran, V.},
  \bibinfo{year}{2015}.
\newblock \bibinfo{title}{{After transition in a soft-walled microchannel}}.
\newblock \bibinfo{journal}{J. Fluid Mech.} \bibinfo{volume}{780},
  \bibinfo{pages}{649--686}.
\newblock \DOIprefix\doi{10.1017/jfm.2015.476}.
\bibitem[{Srinivas and Kumaran(2017)}]{SK17}
\bibinfo{author}{Srinivas, S.S.}, \bibinfo{author}{Kumaran, V.},
  \bibinfo{year}{2017}.
\newblock \bibinfo{title}{{Transitions to different kinds of turbulence in a
  channel with soft walls}}.
\newblock \bibinfo{journal}{J. Fluid Mech.} \bibinfo{volume}{822},
  \bibinfo{pages}{267--306}.
\newblock \DOIprefix\doi{10.1017/jfm.2017.270}.
\bibitem[{Stewart and Foss(2019)}]{SF19}
\bibinfo{author}{Stewart, P.}, \bibinfo{author}{Foss, A.J.E.},
  \bibinfo{year}{2019}.
\newblock \bibinfo{title}{{Self-excited oscillations in a collapsible channel
  with applications to retinal venous pulsation}}.
\newblock \bibinfo{journal}{ANZIAM J.} \bibinfo{volume}{61},
  \bibinfo{pages}{320--348}.
\newblock \DOIprefix\doi{10.1017/S1446181119000117}.
\bibitem[{Stewart et~al.(2010)Stewart, Heil, Waters and Jensen}]{SHWJ10}
\bibinfo{author}{Stewart, P.}, \bibinfo{author}{Heil, M.},
  \bibinfo{author}{Waters, S.L.}, \bibinfo{author}{Jensen, O.E.},
  \bibinfo{year}{2010}.
\newblock \bibinfo{title}{{Sloshing and slamming oscillations in a collapsible
  channel flow}}.
\newblock \bibinfo{journal}{J. Fluid Mech.} \bibinfo{volume}{662},
  \bibinfo{pages}{288--319}.
\newblock \DOIprefix\doi{10.1017/S0022112010003277}.
\bibitem[{Stewart et~al.(2009)Stewart, Waters and Jensen}]{SWJ09}
\bibinfo{author}{Stewart, P.S.}, \bibinfo{author}{Waters, S.L.},
  \bibinfo{author}{Jensen, O.E.}, \bibinfo{year}{2009}.
\newblock \bibinfo{title}{{Local and global instabilities of flow in a
  flexible-walled channel}}.
\newblock \bibinfo{journal}{Eur. J. Mech. B/Fluids} \bibinfo{volume}{28},
  \bibinfo{pages}{541--557}.
\newblock \DOIprefix\doi{10.1016/j.euromechflu.2009.03.002}.
\bibitem[{Subbaraj and Dokainish(1989)}]{SD89}
\bibinfo{author}{Subbaraj, K.}, \bibinfo{author}{Dokainish, M.},
  \bibinfo{year}{1989}.
\newblock \bibinfo{title}{{A survey of direct time-integration methods in
  computational structural dynamics—II. Implicit methods}}.
\newblock \bibinfo{journal}{Comput. Struct.} \bibinfo{volume}{32},
  \bibinfo{pages}{1387--1401}.
\newblock \DOIprefix\doi{10.1016/0045-7949(89)90315-5}.
\bibitem[{Timoshenko and Woinowsky-Krieger(1959)}]{TWK59}
\bibinfo{author}{Timoshenko, S.}, \bibinfo{author}{Woinowsky-Krieger, S.},
  \bibinfo{year}{1959}.
\newblock \bibinfo{title}{{Theory of Plates and Shells}}.
\newblock \bibinfo{edition}{2nd} ed., \bibinfo{publisher}{McGraw-Hill},
  \bibinfo{address}{New York}.
\bibitem[{Verma and Kumaran(2012)}]{VK12}
\bibinfo{author}{Verma, M.K.S.}, \bibinfo{author}{Kumaran, V.},
  \bibinfo{year}{2012}.
\newblock \bibinfo{title}{{A dynamical instability due to fluid–wall coupling
  lowers the transition Reynolds number in the flow through a flexible tube}}.
\newblock \bibinfo{journal}{J. Fluid Mech.} \bibinfo{volume}{705},
  \bibinfo{pages}{322--347}.
\newblock \DOIprefix\doi{10.1017/jfm.2011.55}.
\bibitem[{Verma and Kumaran(2013)}]{VK13}
\bibinfo{author}{Verma, M.K.S.}, \bibinfo{author}{Kumaran, V.},
  \bibinfo{year}{2013}.
\newblock \bibinfo{title}{{A multifold reduction in the transition Reynolds
  number, and ultra-fast mixing, in a micro-channel due to a dynamical
  instability induced by a soft wall}}.
\newblock \bibinfo{journal}{J. Fluid Mech.} \bibinfo{volume}{727},
  \bibinfo{pages}{407--455}.
\newblock \DOIprefix\doi{10.1017/jfm.2013.264}.
\bibitem[{Verma and Kumaran(2015)}]{VK15}
\bibinfo{author}{Verma, M.K.S.}, \bibinfo{author}{Kumaran, V.},
  \bibinfo{year}{2015}.
\newblock \bibinfo{title}{{Stability of the flow in a soft tube deformed due to
  an applied pressure gradient}}.
\newblock \bibinfo{journal}{Phys. Rev. E} \bibinfo{volume}{91},
  \bibinfo{pages}{043001}.
\newblock \DOIprefix\doi{10.1103/PhysRevE.91.043001}.
\bibitem[{Virtanen et~al.(2020)Virtanen, Gommers, Oliphant, Haberland, Reddy,
  Cournapeau, Burovski, Peterson, Weckesser, Bright, van~der Walt, Brett,
  Wilson, Millman, Mayorov, Nelson, Jones, Kern, Larson, Carey, Polat, Feng,
  Moore, VanderPlas, Laxalde, Perktold, Cimrman, Henriksen, Quintero, Harris,
  Archibald, Ribeiro, Pedregosa and van Mulbregt}]{SciPy}
\bibinfo{author}{Virtanen, P.}, \bibinfo{author}{Gommers, R.},
  \bibinfo{author}{Oliphant, T.E.}, \bibinfo{author}{Haberland, M.},
  \bibinfo{author}{Reddy, T.}, \bibinfo{author}{Cournapeau, D.},
  \bibinfo{author}{Burovski, E.}, \bibinfo{author}{Peterson, P.},
  \bibinfo{author}{Weckesser, W.}, \bibinfo{author}{Bright, J.},
  \bibinfo{author}{van~der Walt, S.J.}, \bibinfo{author}{Brett, M.},
  \bibinfo{author}{Wilson, J.}, \bibinfo{author}{Millman, K.J.},
  \bibinfo{author}{Mayorov, N.}, \bibinfo{author}{Nelson, A.R.J.},
  \bibinfo{author}{Jones, E.}, \bibinfo{author}{Kern, R.},
  \bibinfo{author}{Larson, E.}, \bibinfo{author}{Carey, C.J.},
  \bibinfo{author}{Polat, I.}, \bibinfo{author}{Feng, Y.},
  \bibinfo{author}{Moore, E.W.}, \bibinfo{author}{VanderPlas, J.},
  \bibinfo{author}{Laxalde, D.}, \bibinfo{author}{Perktold, J.},
  \bibinfo{author}{Cimrman, R.}, \bibinfo{author}{Henriksen, I.},
  \bibinfo{author}{Quintero, E.A.}, \bibinfo{author}{Harris, C.R.},
  \bibinfo{author}{Archibald, A.M.}, \bibinfo{author}{Ribeiro, A.H.},
  \bibinfo{author}{Pedregosa, F.}, \bibinfo{author}{van Mulbregt, P.},
  \bibinfo{year}{2020}.
\newblock \bibinfo{title}{{SciPy 1.0: fundamental algorithms for scientific
  computing in Python}}.
\newblock \bibinfo{journal}{Nature Methods} \bibinfo{volume}{17},
  \bibinfo{pages}{261--272}.
\newblock \DOIprefix\doi{10.1038/s41592-019-0686-2}.
\bibitem[{Wang et~al.(2021)Wang, Luo and Stewart}]{WLS21}
\bibinfo{author}{Wang, D.}, \bibinfo{author}{Luo, X.},
  \bibinfo{author}{Stewart, P.}, \bibinfo{year}{2021}.
\newblock \bibinfo{title}{{Energetics of collapsible channel flow with a
  nonlinear fluid-beam model}}.
\newblock \bibinfo{journal}{J. Fluid Mech.} \bibinfo{volume}{926},
  \bibinfo{pages}{A2}.
\newblock \DOIprefix\doi{10.1017/jfm.2021.642}.
\bibitem[{Wang and Christov(2019)}]{WC19}
\bibinfo{author}{Wang, X.}, \bibinfo{author}{Christov, I.C.},
  \bibinfo{year}{2019}.
\newblock \bibinfo{title}{{Theory of the flow-induced deformation of shallow
  compliant microchannels with thick walls}}.
\newblock \bibinfo{journal}{Proc. R. Soc. A} \bibinfo{volume}{475},
  \bibinfo{pages}{20190513}.
\newblock \DOIprefix\doi{10.1098/rspa.2019.0513}.
\bibitem[{Wang and Christov(2020)}]{WC20}
\bibinfo{author}{Wang, X.}, \bibinfo{author}{Christov, I.C.},
  \bibinfo{year}{2020}.
\newblock \bibinfo{title}{{Soft hydraulics in channels with thick walls: The
  finite-Reynolds-number base state and its stability}}, in:
  \bibinfo{editor}{Todorov, M.D.} (Ed.), \bibinfo{booktitle}{Proceedings of the
  12th International On-line Conference for Promoting the Application of
  Mathematics in Technical and Natural Sciences - AMiTaNS’20}.
  \bibinfo{publisher}{AIP}. volume \bibinfo{volume}{2302} of
  \textit{\bibinfo{series}{AIP Conference Proceedings}}, p.
  \bibinfo{pages}{020002}.
\newblock \DOIprefix\doi{10.1063/5.0033517}.
\bibitem[{Wang and Christov(2021)}]{WC21}
\bibinfo{author}{Wang, X.}, \bibinfo{author}{Christov, I.C.},
  \bibinfo{year}{2021}.
\newblock \bibinfo{title}{{Reduced models of unidirectional flows in compliant
  rectangular ducts at finite Reynolds number}}.
\newblock \bibinfo{journal}{Phys. Fluids} \bibinfo{volume}{33},
  \bibinfo{pages}{102004}.
\newblock \DOIprefix\doi{10.1063/5.0062252}.
\bibitem[{White(2006)}]{W06_book}
\bibinfo{author}{White, F.M.}, \bibinfo{year}{2006}.
\newblock \bibinfo{title}{{Viscous Fluid Flow}}.
\newblock \bibinfo{edition}{3} ed., \bibinfo{publisher}{McGraw-Hill Higher
  Education}, \bibinfo{address}{New York, NY}.
\bibitem[{Winkler(1867)}]{W67}
\bibinfo{author}{Winkler, E.}, \bibinfo{year}{1867}.
\newblock \bibinfo{title}{{Die Lehre von der Elastizit{\"{a}}t und Festigkeit
  mit besonderer R{\"{u}}cksicht auf ihre Anwendung in der Technik}}.
\newblock \bibinfo{publisher}{Verlag von H. Dominicus},
  \bibinfo{address}{Prag}.
\newblock \URLprefix \url{https://www.google.com/books/edition/_/25E5AAAAcAAJ}.
\bibitem[{Xu et~al.(2013)Xu, Billingham and Jensen}]{XBJ13}
\bibinfo{author}{Xu, F.}, \bibinfo{author}{Billingham, J.},
  \bibinfo{author}{Jensen, O.E.}, \bibinfo{year}{2013}.
\newblock \bibinfo{title}{{Divergence-driven oscillations in a flexible-channel
  flow with fixed upstream flux}}.
\newblock \bibinfo{journal}{J. Fluid Mech.} \bibinfo{volume}{723},
  \bibinfo{pages}{706--733}.
\newblock \DOIprefix\doi{10.1017/jfm.2013.97}.
\bibitem[{Xu et~al.(2014)Xu, Billingham and Jensen}]{XBJ14}
\bibinfo{author}{Xu, F.}, \bibinfo{author}{Billingham, J.},
  \bibinfo{author}{Jensen, O.E.}, \bibinfo{year}{2014}.
\newblock \bibinfo{title}{{Resonance-driven oscillations in a flexible-channel
  flow with fixed upstream flux and a long downstream rigid segment}}.
\newblock \bibinfo{journal}{J. Fluid Mech.} \bibinfo{volume}{746},
  \bibinfo{pages}{368--404}.
\newblock \DOIprefix\doi{10.1017/jfm.2014.136}.

\end{thebibliography}


\clearpage

\addtocontents{toc}{\protect\setcounter{tocdepth}{0}}

\setcounter{section}{0}
\renewcommand*{\thesection}{SM.\arabic{section}}
\setcounter{equation}{0}
\setcounter{figure}{0}
\renewcommand*{\thefigure}{SM.\arabic{figure}}

\pretitle{\begin{center}\LARGE\textbf{Supplementary Materials}\\[5mm]}

\date{}

\maketitle

\section{Numerical scheme for the 1D FSI model}
\label{sec:app-b}

Here, we introduce a numerical scheme used for solving the coupled problem of deformation-induced tension, wall deformation, and flow. This scheme is also applied to the simpler case of constant tension with given $\theta_t$. The spatial domain is discretized using the pseudospectral method \citep{Boyd00}, with which the governing equations are satisfied at preassigned Gauss-type-quadrature nodes. In our case, the Gauss--Lobatto points are chosen. Therefore, the method is also referred to as ``Chebyshev pseudospectral method'' or ``Chebyshev collocation method.'' Note that in some literature \citep{STW11}, the pseudospectral method is specifically referred to as a Galerkin-type method with the numerical integration using a Gauss-type quadrature, which is not the case for the present method.

The Gauss--Lobatto points 
\begin{equation}
    \Tilde{Z}_j = -\cos\left(\frac{j\pi}{N} \right), \quad j =0,1,\dots,N
\end{equation}
are defined on the domain $[-1,1]$. (Note that $\Tilde{Z}_0=-1$ and $\Tilde{Z}_N=1$.) Consequently, 
we use a change of variables, $\Tilde{Z}=2Z-1$, to transform the computational domain from $\{Z\,|\, Z\in [0,1]\}$ to $\{\Tilde{Z}\,|\, \Tilde{Z}\in [-1,1] \}$. Then, $\rd^m/\rd Z^m = 2^m \rd^m/\rd \Tilde{Z}^m$. 
There are two major advantages of choosing the Gauss--Lobatto points as the collocation points. First, the Gauss--Lobatto points are nonuniformly distributed and clustered near the endpoints $\Tilde{Z}=\pm 1$, with the spacing scaling as $\mathcal{O}(N^{-2})$, which helps resolve the deformation boundary layers near the channel ends. Second, it is convenient to compute derivatives at the Gauss--Lobatto points. Essentially, the Chebyshev pseudospectral method finds a high-order polynomial-based, valid in the whole domain, to approximate the actual solution. As long as the functional values are known at the $N+1$ collocated points, an $N$-order polynomial can be uniquely determined. 

The Lagrange basis, $l_j(Z)\ (j=0,1,\hdots, N)$, is a convenient choice for the interpolating polynomial since the coefficients are just the functional values. Here, $l_j$ denotes the Lagrange polynomial which takes the value of $1$ at $\Tilde{Z}=Z_j$ while being $0$ for $Z_k$ with $k\neq j$.  Importantly, the derivatives of the Lagrange basis at the Gauss--Lobatto points are known analytically. For example, taking $Q_k \approx Q|_{\Tilde{Z}=\Tilde{Z}_k},\ k= 0,1,\hdots,N$, then $\rd Q/\rd \Tilde{Z}|_{ \Tilde{Z}=\Tilde{Z}_k } \approx \sum_{j=0}^N D_{kj}^{(1)} Q_j$. Here, the components of the first-order differentiation matrix $\bm{D}^{(1)}$ are:
\begin{equation}
    D_{kj}^{(1)}=\left.\frac{\rd l_j}{\rd \Tilde{Z}}\right|_{\Tilde{Z}=\Tilde{Z}_k}=\begin{cases}
    \displaystyle-\frac{2N^2+1}{6},&\qquad k=j=0,\\[3mm]
    \displaystyle\frac{\widetilde{c}_k}{\widetilde{c}_j}\frac{(-1)^{k+j}}{\Tilde{Z}_k-\Tilde{Z}_j},&\qquad k\ne j,\;\;0\le k,j\le N,\\[4mm]
    \displaystyle-\frac{\Tilde{Z}_k}{2(1-\Tilde{Z}_k^2)},&\qquad k= j,\;\;1\le k,j\le N-1,\\[5mm]
    \displaystyle\frac{2N^2+1}{6},&\qquad k=j=N,
    \end{cases}
\end{equation}
where $\widetilde{c}_0 = \widetilde{c}_N =2$ and $\widetilde{c}_j=1$ for $1\leq j \leq N-1$. The higher-order differentiation matrix is just the matrix multiplication of the lower ones, \textit{i.e.,} $\rd^m/\rd \Tilde{Z}^m|_{\Tilde{Z}=\Tilde{Z}_k} \approx \bm{D}^{(m)} = [\bm{D}^{(1)}]^m = \bm{D}^{(1)}\times \bm{D}^{(1)}\times  \cdots\times  \bm{D}^{(1)}$.

As for the time integration, a second-order backward-difference formula is used for the flow equations. Let the time step be $\Delta T$, and a subscript denote the functional value at the corresponding grid point, while a superscript indicates the time step. Then, equations \eqref{gen1d-cont-eq} and \eqref{gen1d-mom-eq} are discretized as:
\begin{align}\label{disc-gen1d-cont-eq}
    Q_j^{n+1} &= 1 - \int_{-1}^{\Tilde{Z}_j} \frac{1}{2}\beta \left(\dot{\bar{U}}_Y\right)^{n+1} \, \rd \Tilde{Z},\\
    \label{disc-gen1d-mom-eq}
    P_j^{n+1} &= \int_1^{\Tilde{Z}_j} \frac{1}{2} \left\{ -\frac{\hat{Re}}{\bar{H}^{n+1} } \frac{3Q^{n+1} - 4Q^n + Q^{n-1}}{2\Delta T} - \frac{6\hat{Re}}{5 \bar{H}^{n+1} } 2 D^{(1)}\frac{(Q^{n+1})^2}{\bar{H}^{n+1}} -\frac{12 Q^{n+1}}{(\bar{H}^{n+1})^3}\right\} \, \rd\Tilde{Z}.
\end{align}
Both of these equations have been integrated in space, with $Q_0 \equiv Q|_{Z=0} = 1$ imposed in equation \eqref{disc-gen1d-cont-eq} and $P_N \equiv P|_{Z=1} = 0$ imposed in equation \eqref{disc-gen1d-mom-eq}. The integral is to be evaluated numerically using the trapezoidal rule. Note $\rd \Tilde{Z} = 2\rd Z$ due to the change of variables introduced above. Also, in equation \eqref{disc-gen1d-cont-eq}, $\dot{\bar{U}}_Y$ denotes the velocity of the interface, which can be obtained from equation \eqref{newmark-vel} below. 

The so-called Newmark--$\beta$ method is applied to the governing solid equation \eqref{1d-solid-eq}. Then, the spatially discretized equation \eqref{1d-solid-eq} is written as
\begin{equation}\label{disc1-solid-eq}
    \boldsymbol{M} \ddot{\bar{U}}_Y + \boldsymbol{K}\bar{U}_Y = P,
\end{equation}
with
\begin{equation}
    \boldsymbol{M}=\theta_I \boldsymbol{I}, \quad \boldsymbol{K}= \boldsymbol{I} -\theta_t \boldsymbol{D}^{(2)}.
\end{equation}
Here, $\theta_t$ is evaluated from equation \eqref{tension_coeff2}. To ensure the accuracy of the numerical integration required to evaluate $\theta_t$, the kernel, $(\rd \bar{H}/\rd Z)^2 \approx (2 \boldsymbol{D}^{(1)} \bar{H})^2 $, is interpolated on the finer grid of $N=1024$ using the {\tt barycentric\_interpolate} subroutine in SciPy. Then, a Gauss--Lobatto quadrature is applied on the finer grids to calculate the integral in equation \eqref{tension_coeff2}. 

With the coefficients in equation \eqref{disc1-solid-eq} determined, the acceleration, velocity and displacement of the interface are calculated as
\begin{subequations}\label{disc2-solid-eq}
\begin{align}
    & \ddot{\bar{U}}_Y^{n+1} = \left(\boldsymbol{M} + \phi_2\Delta T^2 \boldsymbol{K} \right)^{-1} \left\{ P^{n+1} - \boldsymbol{K} \left[ \bar{U}_Y^n + \Delta T \dot{\bar{U}}_Y^n + \left(\frac{1}{2}-\phi_2\right)\Delta T^2 \ddot{\bar{U}}_Y^n \right]\right\}, \label{newmark-acc}\\
    & \dot{\bar{U} }_Y^{n+1} = \dot{\bar{U}}_Y^n + (1-\phi_1)\Delta T \ddot{\bar{U}}_Y^n + \phi_1 \Delta T \ddot{\bar{U}}_Y^{n+1}, \label{newmark-vel}\\
    & \bar{U}_Y^{n+1} = \bar{U}_Y^n + \Delta T \dot{\bar{U}}_Y^n + \left(\frac{1}{2} - \phi_2\right)\Delta T^2\ddot{\bar{U}}_Y^n + \phi_2\Delta T^2 \ddot{\bar{U}}_Y^{n+1},\label{newmark-disp}
\end{align}
\end{subequations}
where $\phi_1$ and $\phi_2$ are two adjustable parameters. The Newmark--$\beta$ scheme is unconditionally stable and second-order accurate if $\phi_1 = 1/2$ and $\phi_2 = 1/4$. However, to damp out numerically-induced high-frequency oscillations, $\phi_1> 1/2$ is usually needed \citep{SD89}. In our simulations, we use $\phi_1=1.0$  and $\phi_2=0.5625$. 

Finally, the discretized interface equation \eqref{Havg} is simply
\begin{equation}\label{disc-interface-eq}
    \bar{H}_j^{n+1} = 1 + \beta (\bar{U}_Y)_{j}^{n+1}.
\end{equation}

\begin{figure}[t]
    \centering
    \includegraphics[width=\textwidth]{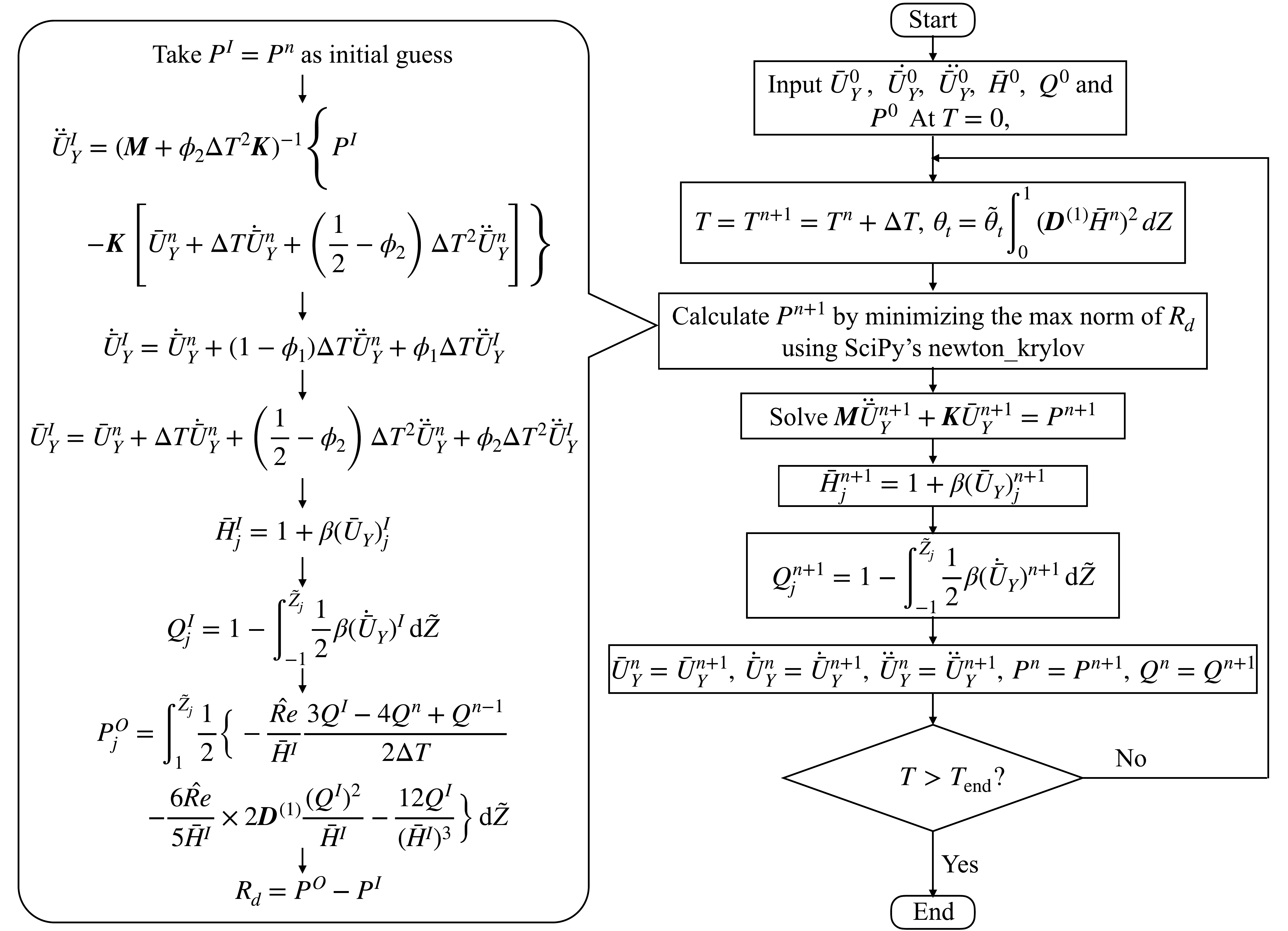}
    \caption{Flow chart of the numerical scheme for the dynamic simulations. To the left of the cell where SciPy's {\tt newton\_krylov} solver is called, the details of the construction of the residual, $R_d$, is shown. The superscript $I$ indicates that the quantity is calculated based on $P^I$.}
    \label{fig:scheme-flow-chart}
\end{figure}

\subsection{Steady-state simulation}
\label{subsec:app-b-steady}

As mentioned in the main text, the case with given constant tension can be easily solved using ScPpy's {\tt solve\_bvp}. However, in the case of the deformation-dependent tension, since $\theta_t$ is unknown, {\tt solve\_bvp} is not as robust and usually has difficulty reaching convergence. Instead, SciPy's {\tt newton\_krylov} method is applied to resolve the steady-state solution.

At steady state, all of the terms involving $\Delta T$ can be neglected. Further, we drop the subscripts on the spatial discretizations for convenience. Then, equations \eqref{disc-gen1d-cont-eq},  \eqref{disc-gen1d-mom-eq},  \eqref{disc1-solid-eq} and  \eqref{disc-interface-eq} comprise a nonlinear algebraic problem. Then, given $\bar{U}_Y$, equations \eqref{disc-interface-eq} and  \eqref{disc-gen1d-mom-eq} allow us to evaluate the pressure, denoted as $P^F$. At the same time, equation \eqref{disc1-solid-eq} can also be used to evaluate the pressure, denote as $P^S$.
Now, we define a residual as
\begin{equation}\label{Rs}
    R_s = P^F - P^S.
\end{equation}
SciPy's {\tt newton\_krylov} solver is used to minimize the max-norm of $R_s$, which yields an approximate evaluation for $\bar{U}_Y$ at steady state. The tolerance used was $6\times 10^{-6}$. 

With $\bar{U}_Y$ obtained, $\theta_t$ is calculated from equation \eqref{tension_coeff2} using the Gauss--Lobatto quadrature, as discussed before. The steady-state solution is then validated with {\tt solve\_bvp} by solving equation \eqref{steady-eq1}, where both the initial guess and $\theta_t$ are based on the outputs of {\tt newton\_krylov}. 

\begin{figure}[ht!]
    \centering
    \includegraphics[width=\textwidth]{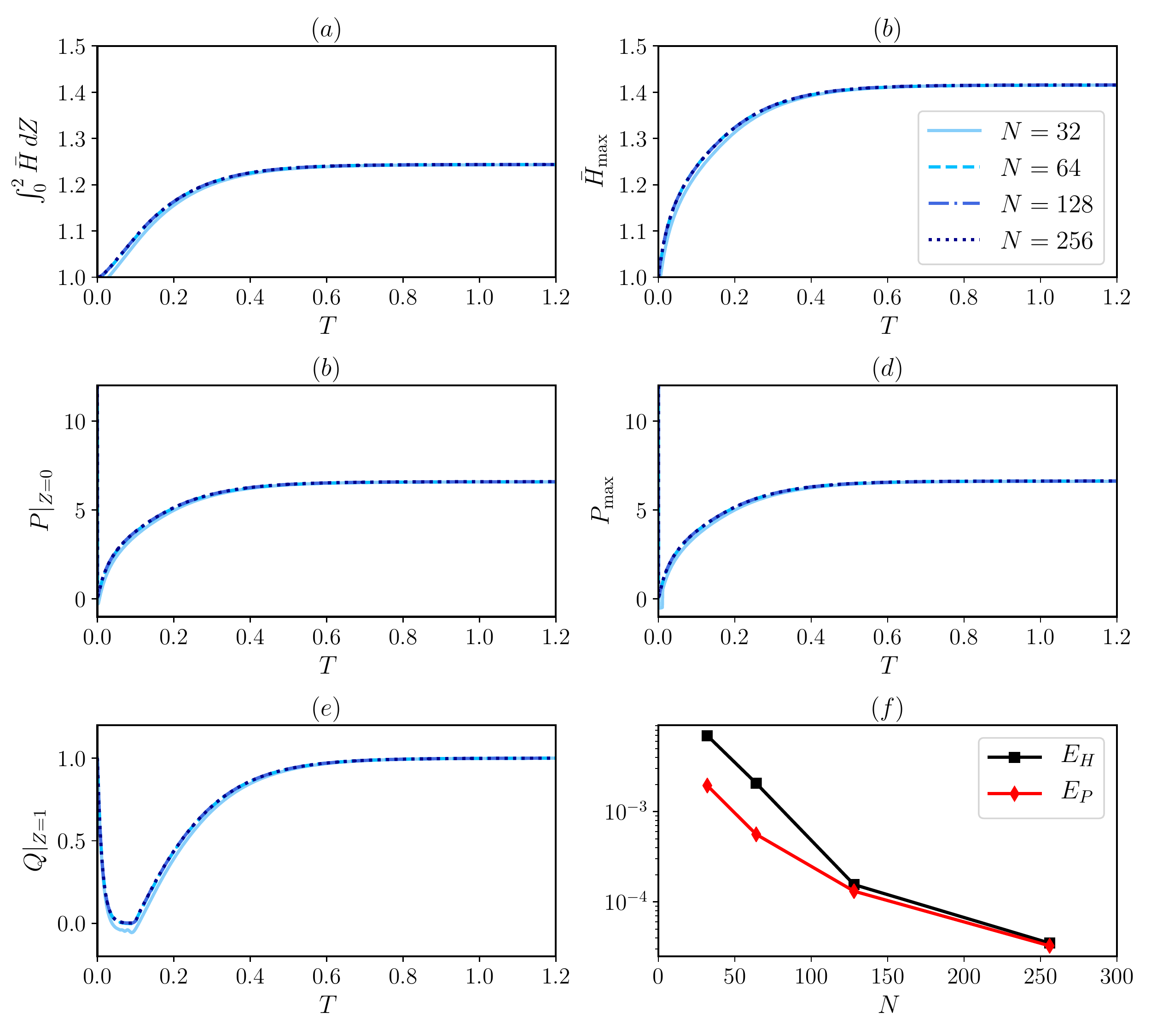}
    \caption{Grid independence study on the dynamic simulations of FSI in the microchannel with $E=2$ MPa under a flow rate of $q=1500~\si{\micro\liter\per\minute}$ (case C4). The time histories of the representative quantities in the system are shown in panels (\textit{a}) to (\textit{e}). Panel (\textit{f}) shows the two-norm of the difference between the simulated steady state solution and the ``exact'' solution, which is computed from the steady simulation with $N=2048$ Gauss--Lobatto points using the scheme described in \S~\ref{subsec:app-b-steady}. The tolerance used in SciPy's {\tt newton\_krylov} was $10^{-8}$. The errors $E_H$ and $E_P$ are computed via equation \eqref{error}.}
    \label{fig:grid-independence}
\end{figure}

\begin{figure}[ht]
    \centering
    \includegraphics[width=\textwidth]{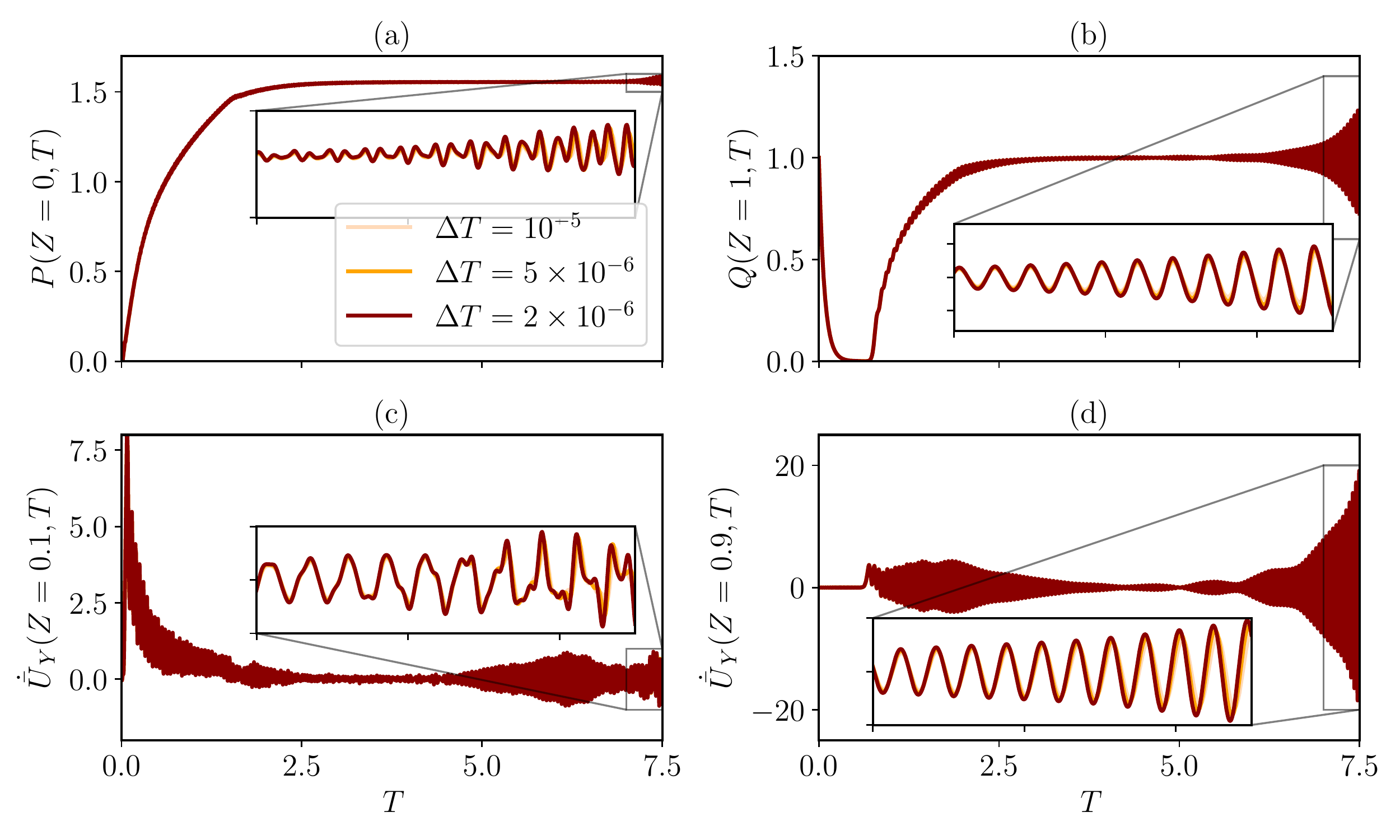}
    \caption{Time histories of (\textit{a}) inlet pressure $P(0,T)$, (\textit{b}) outlet flow rate $Q(1,T)$, the vertical velocity $\dot{\bar{U}}_Y$ of the fluid--solid interface at (\textit{c}) $Z=0.1$ and (\textit{d}) $Z=0.9$. All panels are for case C3 but using different time step sizes $\Delta T$. The spatial grid is fixed to have $N=128$ nodes.}
    \label{fig:sim-diffdt}
\end{figure}

\subsection{Dynamic simulation}
\label{subsec:app-b-dynamic}

The dynamic problem is solved in a similar manner. At each time step, the nonlinear system of equations \eqref{disc-gen1d-cont-eq}, \eqref{disc-gen1d-mom-eq}, \eqref{disc1-solid-eq} and \eqref{disc-interface-eq} must be solved. However, to get the Newmark--$\beta$ time integration \eqref{disc2-solid-eq} involved, the residual, $R_d$, is defined with the pressure as the input, denoted as $P^I$. Then, starting from equation \eqref{disc1-solid-eq}, solved with the scheme \eqref{disc2-solid-eq}, $\bar{U}_Y$ and $\dot{\bar{U}}_Y$ are obtained. Then, $\bar{H}$ and $Q$ are evaluated from equations \eqref{disc-interface-eq} and \eqref{disc-gen1d-cont-eq}, respectively. Equation \eqref{disc-gen1d-mom-eq} gives another evaluation of the pressure, denoted as, $P^O$. The residual is thus evaluated as 
\begin{equation}
    R_d = P^O - P^I.
\end{equation}
At each time step, SciPy's {\tt newton\_krylov} is used to minimize the max-norm of $R_d$. The details of this numerical procedure are summarized in the flow chat in  figure~\ref{fig:scheme-flow-chart}.

Note that at time step $n+1$, $\theta_t$ is evaluated as $\theta_t = \Tilde{\theta}_t\int_0^1 (\boldsymbol{D}^{(1)} \bar{H}^n)^2 \, \rd Z$. The integral is approximated using the Gauss--Lobatto quadrature after interpolating the kernel on the finer grid of $N=1024$. Here, we use $\bar{H}^n$ instead of $\bar{H}^{n+1}$ in order to avoid another nonlinear problem requiring ``internal iterations'' on $\theta_t^{n+1}$. We have verified that the results using $\bar{H}^n$ instead of $\bar{H}^{n+1}$ do not differ from those evaluated based on more involved method using $\bar{H}^{n+1}$.

\subsubsection{Grid independence study}

Next, we verify the grid independence of the numerical results shown in the main text. The case chosen to perform the grid independence study is C4 from table~\ref{table:case-param} (corresponding to $E=2$ \si{\mega\pascal} and $q=1500$ \si{\micro\litre\per\minute}). For case C4, we have shown in \S~\ref{sec:linear-stability} that the inflated steady state is linearly stable to infinitesimal perturbations. If the flat initial condition \eqref{ic} is used, then the system will reach the steady state eventually.

The end time for the simulation is $T_\mathrm{end}=2.0$. However, for the clarity of the presentation, only the results between $T=0$ and $T=1.2$ are shown. Furthermore, the ratio of the smallest grid size to the time step is fixed for every simulation. Since the smallest spacing of the Gauss--Lobatto grid points goes as   $\mathcal{O}(N^{-2})$, then as $N$ is doubled, $\Delta T$ is decreased by a factor $4$ accordingly. The time step for $N=32$ is $\Delta T = 4\times 10^{-4}$. As shown figure~\ref{fig:grid-independence}(\textit{a}) and (\textit{e}), all of the representative quantities agree well with each other as $\Delta T$ is refined, except on the courses grid with $N=32$,

After the simulation has reached $T=T_\mathrm{end}$, the deformed interface shape, $\bar{H}^\mathrm{end}(Z)$, and the pressure distribution within the channel, $P^\mathrm{end}(Z)$ are compared with an ``exact'' steady-state solution. 
The latter are denoted as $\bar{H}^e(Z)$ and $P^e(Z)$, respectively. The ``exact'' solution is taken to be the steady state of the simulation with $N=2048$, and tolerance for SciPy's {\tt newton\_krylov} set to $10^{-8}$. We define two $L^2$-norm based error estimates:
\begin{subequations}\label{error}\begin{align}
    E_H &= \left(\frac{1}{2} \sum_{j=1}^{N-1} (\bar{H}^{\rm end}_j-\bar{H}^e_j) \mathfrak{w} \sqrt{1-\tilde{Z}_j^2}  \right)^{1/2} ,\\
    E_P &= \left(\frac{1}{2} \sum_{j=1}^{N-1} (P^{\rm end}_j-P^e_j) \mathfrak{w} \sqrt{1-\tilde{Z}_j^2} \right)^{1/2},
\end{align}\end{subequations}
which are written in the discrete form using the Gauss--Lobatto quadrature. Here, $\mathfrak{w} = \pi/N$ are the weights, and we choose $N=2048$. Figure~\ref{fig:grid-independence}(\textit{f}) shows that the error decreases with the increase of $N$. The cases of $N=32,\ 64$ and $128$ even display an exponential decay for $E_H$. However, since the ``exact'' solution is not really exact, both error estimates tend to ``saturate'' for $N=256$.

As for the linearly unstable cases, the errors defined in equation \eqref{error} are not applicable because the system will not reach steady state. In these cases, each simulation is tested with different time step sizes, and only the converged results are shown. The spatial grid is typically fixed as $N=128$ for satisfactory accuracy (as shown in figure~\ref{fig:grid-independence}). One example for case C3 is shown in figure~\ref{fig:sim-diffdt}. In panel (\textit{c}) and (\textit{d}), the vertical velocity of the fluid--solid interface is obtained as follows. First, we substitute the simulated $\bar{H}$ into equation \eqref{vel-profile} to obtain $V_Z^{2D}$. Then, we compute $V_Y^{2D}$ based on conservation of mass, \textit{i.e.,} $\partial V_Z^{2D}/\partial Z+\partial V_Y^{2D}/\partial Y=0 $. Lastly, we obtain $\dot{\bar{U}}_Y = \beta^{-1}V_Y^{2D}|_{Y=\bar{H}}$ using equations \eqref{kinematic-bc} and \eqref{Havg}. The actual simulation time is longer than the time window shown for each case. However, it is observed that after a certain $T$, the results with different time step sizes begin to diverge, indicating this nonlinear 1D FSI model's dynamic behavior may be chaotic. Understanding such an interesting possibility is beyond of the scope of the current work.   

\section{A Chebyshev pseudospectral method for the generalized eigenvalue problem}
\label{sec:app-c}

We solve the generalized eigenvalue problem \eqref{linearstab-mat} using the approach proposed by \citet{IWC20}. However, since the boundary conditions for equation \eqref{linearstab-mat} are different from the boundary conditions of \citet{IWC20}, we employ another modified Lagrange polynomial basis, now written as
\begin{equation}\label{eigfunc-basis}
    \widetilde{Q}(\Tilde{Z})\approx(1+\Tilde{Z})\sum_{j=1}^N\widetilde{Q}_j\frac{\ell_j(\Tilde{Z})}{1+\Tilde{Z}_j},\qquad \widetilde{H}(\Tilde{Z})\approx\sum_{j=1}^{N-1}\widetilde{H}_j\ell_j(\Tilde{Z}) + \widetilde{H}_N(1-\Tilde{Z}^2).
\end{equation}
It is easy to check that $\widetilde{Q}(\Tilde{Z}_j)=\Tilde{Q}_j$ and $\widetilde{H}(\Tilde{Z}_j)= \widetilde{H}_j$, except that $\widetilde{H}(\Tilde{Z}_N)=0\neq \widetilde{H}_N$, meaning that  $\widetilde{Q}_j$ and $\widetilde{H}_j$ are the collocated function values; $\widetilde{H}_N$ is introduced ensure the satisfaction of the boundary condition. Also note $j$ starts from $1$ instead of $0$, because equation \eqref{eigfunc-basis} has already satisfied the conditions that $\widetilde{Q}=\rd\widetilde{Q}/\rd\Tilde{Z}=0$ at $\Tilde{Z}=-1$, and $\widetilde{H}=0$ at $\Tilde{Z}=\pm 1$. The two remaining boundary conditions at $\Tilde{Z}=1$ in equation \eqref{linearstab-bc} are enforced manually. 

Next, the generalized eigenvalue problem \eqref{linearstab-mat} is collocated at the Gauss--Lobatto points from $j=1,2,\hdots,N-1$, with the unsatisfied boundary conditions enforced at $j=N$. This approach gives rise to the $2N\times 2N$ matrices $\boldsymbol{A}$ and $\boldsymbol{B}$ in equation \eqref{linearstab-mat}. Due to the imposition of the boundary conditions at $\Tilde{Z}=1$, $\boldsymbol{B}$ is singular. Thus, we invert $\boldsymbol{A}$ to obtain a regular eigenvalue problem,  $\boldsymbol{A}^{-1}\boldsymbol{B}\boldsymbol{\psi} = (\ri\omega_G)^{-1}\boldsymbol{\psi}$, which can be solved using SciPy's {\tt eig}.

Finally, to filter out any spurious modes, each calculation has been performed with $N=600$ and $N=800$ grid points and cross-checked. Only the converged modes are reported in the text.

\end{document}